\documentclass[prd,preprint,tightenlines,floatfix,showpacs,preprintnumbers,nofootinbib,eqsecnum,superscriptaddress]{revtex4}

\usepackage[dvips,final]{graphicx}
 \usepackage{amssymb}
  \usepackage{amsmath}
   \usepackage{amsfonts}
    \usepackage{epsfig}
     \usepackage{bm}% bold math
      \usepackage{multirow}
       \usepackage{tabularx}

\begin{document}

\begin{flushright}
LU TP 13-11\\
October 2013
\end{flushright}
\vspace{1cm}

\title{Chiral-Symmetric Technicolor with Standard Model Higgs boson}

\author{Roman Pasechnik}
\email{Roman.Pasechnik@thep.lu.se} \affiliation{Department of
Astronomy and Theoretical Physics, Lund University, SE-223 62 Lund,
Sweden}

\author{Vitaly Beylin}
\email{vbey@rambler.ru} \affiliation{Research Institute of Physics,
Southern Federal University, 344090 Rostov-on-Don, Russian
Federation}

\author{Vladimir Kuksa}
\email{vkuksa47@mail.ru} \affiliation{Research Institute of Physics,
Southern Federal University, 344090 Rostov-on-Don, Russian
Federation}

\author{Grigory Vereshkov}
\email{gveresh@gmail.com} \affiliation{Research Institute of
Physics, Southern Federal University, 344090 Rostov-on-Don, Russian
Federation} \affiliation{Institute for Nuclear Research of Russian
Academy of Sciences, 117312 Moscow, Russian Federation\vspace{1cm}}

\begin{abstract}
\vspace{0.5cm} Most of the traditional Technicolor-based models are
known to be in a strong tension with the electroweak precision
tests. We show that this serious issue is naturally cured in
strongly coupled sectors with chiral-symmetric vector-like gauge
interactions in the framework of gauged linear $\sigma$-model. We
discuss possible phenomenological implications of such non-standard
chiral-symmetric Technicolor scenario in its simplest formulation
preserving the Standard Model (SM) Higgs mechanism. For this
purpose, we assume the existence of an extra technifermion sector
confined under extra $SU(3)_{\rm{TC}}$ at the energy scales
reachable at the LHC, $\Lambda_{\rm{TC}}\sim 0.1-1$ TeV, and
interacting with the SM gauge bosons in a chiral-symmetric
(vector-like) way. In the framework of this scenario, the SM Higgs
vev acquires natural interpretation in terms of the condensate of
technifermions in confinement in the nearly conformal limit. We
study the influence of the lowest lying composite physical states,
namely, technipions, technisigma and constituent technifermions, on
the Higgs sector properties in the SM and other observables at the
LHC. We found out that the predicted Higgs boson signal strengths in
$\gamma\gamma$, vector-boson $VV^*$ and fermion $f\bar f$ decay
channels can be sensitive to the new strongly-coupled dynamics and
are consistent with the current SM-like Higgs boson observations in
the limit of relatively small Higgs-technisigma mixing. At the same
time, the chiral-symmetric Technicolor provides us with rich
technipion phenomenology at the LHC, and its major implications are
discussed in detail.
\end{abstract}


\maketitle

%----------------------
\section{Introduction}
%----------------------

A complete experimental verification of the Standard Model (SM),
including the discovery of the Higgs boson and precision tests of
its properties, is the most intriguing and challenging task of high
energy particle physics at the moment. Last year, the major LHC
collaborations, ATLAS and CMS \cite{ATLAS,CMS}, have announced the
discovery of a new ``Higgs-like'' particle with the mass of
$125.3\pm0.6$ GeV, which may become the last yet missing piece
predicted within the SM framework -- the Higgs boson. Some evidence
for the Higgs boson has been also seen by CDF and D0 collaborations
at the Tevatron \cite{Aaltonen:2012qt}.

An ultimate proof of the Higgs boson's existence and understanding
of its nature would only be possible after high precision
measurements of its decay parameters which can be sensitive to
details of a particular New Physics scenario. The current situation
with the Higgs boson properties suggests that there are no
significant deviations from the SM (within rather large statistical
and systematical uncertainties) as revealed by the full data set
collected so far at the LHC \cite{latest_LHC} and Tevatron
\cite{latest_Tevatron} (for the most recent comprehensive studies of
the Higgs boson properties, see e.g.
Refs.~\cite{Falkowski:2013dza,Ellis:2013lra,Djouadi:2013qya,Giardino:2013bma}).
Even though the room for New Physics contributions has been greatly
reduced \cite{Ellis:2013lra,Cheung:2013kla}, it is too early to draw
final conclusions about the properties and nature of the newly
discovered particle not only due to large experimental error bars,
but also due to theoretical uncertainties in the SM Higgs production
which are rather high and become dominant
\cite{Djouadi:2013qya,Dittmaier:2011ti}. If the branching ratios
deviate from predictions of the simplest one-doublet SM, even
slightly, this would require a proper extension of the SM and pose a
serious question about theoretical principles such an extension
should be based upon.

Traditionally, ideas of additional to SM strongly-coupled sectors in
confinement were realized in the Technicolor (TC) model which was
one of the strongest alternatives to the Higgs mechanism of the
spontaneous Electroweak Symmetry Breaking (EWSB) \cite{TC}. The
existing Higgs-less TC models with dynamical EWSB (DEWSB) are based
upon the idea that the Goldstone degrees of freedom (technipions)
appearing after the global chiral symmetry breaking $SU(2)_L\otimes
SU(2)_R\to SU(2)_W$ are absorbed by the SM weak gauge bosons which
thereby gain masses. The DEWSB mechanism is then triggered by the
condensate of technifermions in confinement, $\langle {\tilde
Q}\bar{\tilde Q}\rangle\not=0$. Traditional TC models with DEWSB are
faced with the problem of the mass generation of standard fermions,
which was consistently resolved in Extended TC scenarios
\cite{Extended-TC}. However, many of the existing TC models have got
severely constrained or even ruled out by the EW precision data
\cite{Peskin:1990zt} (for a detailed review on the existing TC
models, see e.g. Refs.~\cite{Hill:2002ap,Sannino}). Generally, in
these schemes noticeable contributions to strongly constrained
Flavor Changing Neutral Current (FCNC) processes appear together
with too large contributions to Peskin-Takeuchi (especially, to $S$)
parameters. Further developments of the TC ideas have resulted in
the Walking TC model which succeeded in resolving the
above-mentioned problems and remains a viable model of the DEWSB
\cite{Appelquist:1986an,Sannino:2004qp,Foadi:2007ue}.

Very recently, as was shown in Ref.~\cite{Cheung:2013kla} based on
the latest LHC data, the $1\sigma$ allowed region of the relative to
SM-predicted Higgs-vector-vector fusion $HVV$ coupling is
$0.96^{+0.13}_{-0.15}$, which sets further constraints on the EWSB
models alternative to the SM Higgs mechanism, as well as to
composite Higgs models (see also current bounds on the rescaling of
the SM couplings in Ref.~\cite{Giardino:2013bma,Alanne:2013dra}).
However, even if the newly-discovered particle is indeed the SM
Higgs boson and the Higgs mechanism is experimentally confirmed, all
available LHC and high precision EW data do not completely exclude
the existence of a strongly-coupled fermion sector in confinement,
additional to the SM fermion sector, with a confinement scale, $\sim
0.1 - 1$ TeV, being not very far from the EW scale $M_{\rm EW}\sim
100$ GeV. The main goal of this paper is to prove this statement and
to study a new class of viable realistic models for an extra
strongly-coupled sector assisting the conventional SM Higgs
mechanism at accessible energy scales, along with the study of their
implications to the ongoing New Physics searches at the LHC.

An alternative class of TC models usually referred to as bosonic TC
scenarios include both a Higgs doublet $H$ and a new TC sector
\cite{Simmons:1988fu,Samuel:1990dq,Kagan:1991gh}, without referring
to an origin of the Higgs doublet. Most recent realization of the
bosonic TC is based upon holographic ideas \cite{Carone}, and allows
to explain the existence of recently discovered Higgs-like 125 GeV
particle and its possible non-standard features \cite{Carone-1}. In
this approach, strongly coupled dynamics is defined using the
AdS/CFT correspondence within the holographic approach allowing to
avoid the EW precision constraints
\cite{Agashe:2004rs,Hirn:2006nt,Hong:2006si}. In contrast to
conventional (Extended and Walking) TC models, in bosonic TC models
the mechanism of the EWSB and generation of SM fermions masses is
driven by the Higgs vacuum expectation value (vev) in the standard
way, irrespectively of (elementary or composite) nature of the Higgs
field itself. Due to linear source term in the Higgs potential the
Higgs field $H$ develops vev which in turn is induced by the
technifermion condensate. This means the Higgs mechanism is not the
primary source of the EWSB, but effectively induced by an unknown TC
dynamics at high scales. For more alternatives on TC and
compositeness models, see e.g. Ref.~\cite{Chivukula:2000mb}.

In this work, we start off with the similar ideas about the
existence of an extra Higgs-like scalar field and TC nature of the
SM Higgs vev implemented in the bosonic TC models and study
theoretical and phenomenological opportunities of new possible
strongly coupled sectors with chiral-symmetric (vector-like) gauge
interactions. We further develop these ideas based on the gauged
linear $\sigma$-model \cite{Lee,LSigM,SU2LR} and applied it to new
TC-induced degrees of freedom, in a complete analogy with low-energy
hadron physics applications. In this model, which will further be
referred to as the Chiral-Symmetric (or Vector-Like) Technicolor (in
short, CSTC) scenario, the oblique (Peskin-Takeuchi) parameters and
FCNC corrections turn out to be naturally very small and fully
consistent with the current EW constraints as well as with the most
recent Higgs couplings measurements at the LHC in the limit of small
Higgs-technisigma mixing. Most importantly, this happens naturally
in the standard quantum-field theory framework implemented in
rigorous quark-meson approaches of hadron physics without attracting
any extra holographic or other special arguments from unknown
high-scale physics. For simplicity, we adopt the simplest version of
the Standard Model with one Higgs doublet, and the question whether
it is elementary or composite is not critical for further
considerations. The new heavy physical states of the model
(additional to those in the SM) are the singlet technisigma
$\tilde{\sigma}$, triplet of technipions $\tilde{\pi}_a,\,a=1,2,3$,
and constituent technifermions $\tilde{Q}$ which acquire masses via
the technifermion condensate as an external source and the
technisigma vev (other composite degrees of freedom are usually much
heavier and decoupled from the considering low-energy limit of the
theory). Their possible phenomenological implications and signatures
at the LHC is the subject of our analysis.

Despite the phenomenological advantages mentioned above, the
proposed CSTC scenario, at least, in its simplest form considered
here, does not attempt to resolve the naturalness problem of the SM,
i.e. does not provide a mechanism protecting the Higgs boson mass
itself from becoming arbitrary large. Nevertheless, it points out a
promising path towards a consistent formulation of composite Higgs
models in extended chiral-gauge theories with vector-like UV
completion\footnote{Also, the model does not provide a mechanism for
generation of current (Dirac) technifermion masses which \'a priori
are arbitrary. In analogy to ordinary QCD, however, we consider the
physically interesting conformal limit of the new strongly coupled
dynamics realized in the chiral limit of the theory $m_{U,D}\ll
\Lambda_{\rm TC}$ which leads to an unambiguous determination of the
Higgs vev in terms of the technifermion condensate. The latter means
that the EW symmetry is broken dynamically via the effective Higgs
mechanism in this limit, which makes it particularly interesting.
This statement is stable w.r.t. radiative corrections.}. Indeed, an
existence of composite Higgs-like bosons is often considered as a
primary guideline for Technicolor models. In analogy with hadron
physics, composite bosons can be of two different types: {\it
pseudo-Goldstone collective excitations} (quantum wave of
correlations between non-perturbative technifermion fluctuations in
technivacuum) and {\it techniquarkonia} (a ``bubble'' of
technivacuum stabilized by valence technifermions). After LHC
experiments, the Technicolor models with composite SM-like Higgs
bosons have become favorable. The latter means that the SM-like
Higgs mechanism is indeed realized in Nature even though it can be
treated as an effective one, i.e. the initial fields of collective
excitations or techniquarkonia should be in the fundamental
representation of the EW gauge group with hypercharge $Y=1/2$. In
the CSTC model such objects naturally appear if one extends the
technifermion sector. The simplest extension is such that in
addition to the EW doublet of technifermions $\tilde Q=(U,D)$ one
introduces an extra weak-singlet technifermion $S$. Therefore, a new
composite scalar field appears ${\cal H}=\bar S\tilde Q$ having
transformation properties of the Higgs boson ($SU(2)_{\rm W}$
doublet with $Y=1/2$). In this model, the initial classification
(techniflavor) group is the global chiral group $SU_{\rm
L}(3)\otimes SU_{\rm R}(3)$. A further generalization would be to
consider $SU_{\rm L}(4)\otimes SU_{\rm R}(4)$ giving rise to
effective two Higgs-doublet model. Of course, in such extended
techniflavor models there appears a plenty of new technihadron
states which require a separate lengthy analysis. In analogy to
hadron physics one may expect, however, that the lightest physical
technihadron states which are the most interesting for the LHC
phenomenology in the first place are technipions, technisigma and,
in principle, lightest technibaryons. Therefore, in this paper we
limit ourselves to considering initial (presumably, the {\it
minimal}) techniflavor group $SU_{\rm L}(2)\otimes SU_{\rm R}(2)$
and discuss a simplified model with gauged vector-like subgroup
$SU(2)_{\rm L+R}$ only where the Higgs boson formally (at the
low-energy part of the spectrum of technihadrons) has a status of
the fundamental field, which does not satisfy the naturalness
criterium. An extended techniflavor model $SU_{\rm L}(N_f)\otimes
SU_{\rm R}(N_f)$ with $N_f>2$ will be studied elsewhere.

The paper is organized as follows. The Section II is devoted to
description of theoretical foundations of the CSTC scenario along
with the physical Lagrangian derivation and analysis of the
parameter space. The study of EW constraints (oblique corrections
and FCNC) is performed in Section III. Some basic opportunities for
LHC phenomenology, in particular, in studies of the Higgs sector
properties, as well as in searches for new lightest composites, are
discussed in Section IV. Finally, Section V summarizes the basic
results of the paper.

\section{Chiral-symmetric Technicolor model}

\subsection{Vector-like technifermions vs chiral SM fermions}

Historically, the Nambu-Jona-Lasinio (NJL) model \cite{NJL} based on
the global chiral group $SU(N_f)_{\rm L}\otimes SU(N_f)_{\rm R}$ is
the first model describing dynamical breaking of chiral symmetry in
particles physics (for review on the topic, see e.g.
Ref.~\cite{Vogl:1991qt}). A large interest in the gauged version of
the NJL model (or GNJL) initially proposed in Ref.~\cite{gauged-NJL}
has been stimulated by its importance for constructing extended TC
models and top-quark condensate models (for an extensive review of
the GNJL models and their applications, see
Ref.~\cite{gauged-NJL-intro}). The GNJL approach has fewer
parameters and significantly reduces ambiguities of corresponding
predictions.

As one of the most successful implementation of the GNJL ideas in
hadron physics, the so-called gauged linear $\sigma$-model
(GL$\sigma$M) initially proposed in Ref.~\cite{Lee} and further
elaborated in Refs.~\cite{LSigM,SU2LR} was one of the first models
with local chiral $SU(2)_{\rm R}\otimes SU(2)_{\rm L}$ symmetry,
which incorporates the vector $\rho$ and pseudovector $a_1$ mesons
as corresponding gauge bosons, besides lightest pseudoscalar pion
$\pi$ and scalar $\sigma$ fields. Typically, the local chiral
symmetry is spontaneously broken by the scalar $\sigma$ vev giving
rise to the vector-meson mass terms, constituent light quark masses
\cite{const-quark} and the mass splitting between $\rho$ and $a_1$
mesons.

In what follows, we employ the ideas of the GL$\sigma$M and consider
the {\it global} chiral $SU(N_f)_{\rm L}\otimes SU(N_f)_{\rm R}$
group in the technifermion sector $\tilde{Q}$ in the simplest case
with $N_f=2$, with its subsequent breaking (by the technisigma vev)
down to the vector subgroup $SU(2)_{\rm V\equiv L+R}$ which is then
{\it gauged} at energy scales close to the EWSB scale. Such a
``gauging'', however, does not necessarily mean that one should
introduce extra gauge bosons to the existing theory. The ``gauging''
procedure may also mean that corresponding fundamental
technifermions interact with {\it already existing} gauge bosons in
the SM in the low-energy effective field theory limit, which is a
rather plausible opportunity we wish to explore here. In analogy
with standard QCD and hadron physics, at the scale of the order of
the techniconfinement scale $\Lambda_{\rm TC}$ technifermions
acquire effective non-perturbative constituent masses due to the
chiral symmetry breaking \cite{const-quark}. At lower energies the
initial technifermions condense into technihadron states due to
confinement. This scheme is an analogy of the chiral-invariant
QHD-III model \cite{SU2LR} where the pseudo-Goldstone technipion
fields $\tilde{\pi}_a$ get the same masses (via an external source
term linear in $\tilde{\sigma}$ field) and remain the physical
degrees of freedom, in distinction from many other traditional TC
and compositeness scenarios.

For the sake of simplicity, we consider a possible scenario of the
SM extension by means of an additional chiral-symmetric
(vector-like) technifermion sector confined under $SU(3)_{\rm TC}$
group, which is analogical to the $SU(3)_c$ color group of QCD. Such
an assignment is not unique, of course, but would allow us to use
direct analogies with hadron physics\footnote{For this purpose, one
could choose an extension of the gauge and fermion SM sectors
motivated by a reduction from the grand-unified theories (GUT)
originating from e.g. superstring-inspired $E_8\otimes E_8'$ group
with many appealing features \cite{strings}. In the latter case, one
of the exceptional groups, say, $E_8'$ can exist in confinement and,
possibly, consists of a few unbroken subgroups confined at different
scales, whereas the second $E_8$ gets broken down to the SM gauge
group $G_{\rm SM}\equiv SU(3)_c \otimes SU(2)_{\rm W} \otimes U_{\rm
Y}(1)$ in a straightforward way. As a realistic possibility, one of
the $SU(3)$ subgroups of the original $E_8'$ can be, in principle,
identified with the TC gauge group $SU(3)_{\rm TC}$, which acts only
on new additional technifermion sector, and there are no any
obstacles for it to be confined at relatively low scales being not
very far from the EWSB scale (later it will be shown that the latter
condition is not critical for the TC-induced EWSB).\\}. The
GL$\sigma$M can therefore be efficiently extended to incorporate
constituent technifermion-technimeson interactions as the simplest
way of phenomenological description of the non-perturbative effects
in technihadron dynamics at low energies. We will further refer to
it below as the gauged linear technisigma model, or GLT$\sigma$M. In
the simplest version of this model, the non-perturbative effects are
accounted for by an effective NJL-type theory of constituent
technifermion interactions with the lightest technihadron states
only \cite{const-quark} -- technipions and technisigma. In the
context of GLT$\sigma$M we suggest the following hypothesis, which
will be studied below: {\it the energy scales of the EWSB and
techni-confinement have a common quantum-topological nature and are
determined by a non-perturbative dynamics of the
technifermion-technigluon condensate}. In particular, we would like
to find specific conditions on the model parameters under which the
latter hypothesis is validated. As was noted above, the technipion
d.o.f. $\tilde{\pi}_a$ are the pseudo-Goldstone fields which are
usually considered as collective fluctuations of the
technifermion-technigluon vacuum, while technisigma $\tilde{\sigma}$
is the lightest techniglueball state -- these states are not usual
bound $\tilde{Q}\bar{\tilde{Q}}$ states and thus play a special role
in the GLT$\sigma$M \cite{Lee,LSigM,SU2LR}.

%At the last stage of a GUT symmetry breaking down to the SM gauge
%group $G_{\rm SM}$, we assume the existence of the well-known
%intermediate $LR$-symmetric rank-5 gauge group \cite{Harari}
%supplemented with extra unbroken $SU(3)_{\rm TC}$ group in
%confinement
%\begin{eqnarray}
%SU(3)_c \otimes SU(2)_{\rm L} \otimes SU(2)_{\rm R} \otimes U_{{\rm
%Y}}(1) \otimes SU(3)_{\rm TC}\quad\to\quad G_{\rm SM} \otimes
%SU(3)_{\rm TC}\,. \label{red-TC}
%\end{eqnarray}

From the point of view of the GLT$\sigma$M, the spontaneous breaking
of the {\it global} chiral symmetry group in the technifermion
sector happens in the chiral-symmetric (vector-like) way in a
complete analogy with the chiral symmetry breaking in GNJL models
\cite{gauged-NJL-intro,SU2LR} as follows
\begin{eqnarray}
SU(2)_{\rm L}\otimes SU(2)_{\rm R}\to SU(2)_{\rm V\equiv L+R}\equiv
SU(2)_{\rm W}\,, \label{ident}
\end{eqnarray}
where the subsequent {\it gauging} of the resulting unbroken vector
subgroup $SU(2)_{\rm V}$ and its identification with the weak gauge
group of the SM are performed. Such gauging and identification
procedures are not forbidden theoretically and lead to specific
properties of the technifermion sector, which thereby make it to be
very different from the chiral-nonsymmetric SM fermion sectors. It
therefore means that after the chiral symmetry breaking in the
technifermion sector the left and right components of the original
Dirac technifermion fields can interact with the SM weak $SU(2)_{\rm
W}$ gauge bosons with vector-like couplings, in opposition to
ordinary SM fermions, which interact under $SU(2)_{\rm W}$ by means
of their left-handed components only. Note, analogous vector-like
gauge interactions are rather common and appear e.g. in the chargino
sector of the MSSM.

Note, the above procedure (\ref{ident}) should be understood in
exactly the same way as is done in the QCD hadron physics at low
energies. There, the fundamental gauge group of color $SU(3)_c$ is
vector-like i.e. acts on left-handed $q_L$ and right-handed $q_R$
quarks in exactly the same way, which makes it possible to introduce
the global chiral group $SU(3)_{\rm L}\otimes SU(3)_{\rm R}$. The
latter is typically broken down to the vector-like subgroup
$SU(3)_{\rm V\equiv L+R}$ by the $\sigma$-vev. If one gauges it, one
recovers that its properties are {\it identical} to the color group
$SU(3)_c$ in the low energy limit. This leads to a low-energy
effective field theory where interaction properties of elementary
and composite fields are effectively described by the same gauge
group with renormalized local gauge couplings (as limiting values of
corresponding form factors valid at small momentum transfers).
Similarly, vector-like weak interactions of technifermions make it
possible to introduce the chiral group whose gauged subgroup has
properties identical to the weak isospin group (\ref{ident}). Most
importantly, the latter procedure is valid {\it only} in the
phenomenologically interesting low energy limit of the theory. When
typical momentum transfers become comparable to the
techniconfinement scale or larger $Q^2\gtrsim \Lambda_{\rm TC}^2$
the global chiral symmetry is fully restored, while fundamental EW
gauge interactions of technifermions remain vector-like (similarly
to QCD interactions of quarks in perturbative limit).

So, in this scenario the sector of initial (current) technifermions
transforms according to the local gauge $SU(2)_{\rm W}\otimes U_{\rm
Y}(1)$ symmetry group, and, therefore, interacts only with SM gauge
bosons $B,\,W^a,\,a=1,2,3$, or with $W^{\pm},\,Z^0$ and $\gamma$
after the SM symmetry breaking. Of course, in a complete local
chiral $SU(2)_{\rm L}\otimes SU(2)_{\rm R}$ theory of
technifermion-technimeson-gauge interactions one would need to
include e.g. a mixing of vector technirho $\tilde{\rho}$ with the
elementary SM gauge bosons as is done in local quark-meson
interaction theories \cite{Serot:1979dc,SU2LR} (for review on the
subject, see also Ref.~\cite{Serot:1997xg} and references therein).
However, in this work in what follows we neglect the heavier vector
and pseudovector technimesons, such that only elementary gauge
$B,\,W^a,\,a=1,2,3$ fields remain, and consider only the spectrum of
lightest composite scalar (technisigma $\tilde{\sigma}$) and
pseudoscalar (technipion $\tilde{\pi}_a$) states, relevant for the
LHC measurements. Note, that a reduction scheme to the left-right
(LR) symmetric subgroup $SU(2)_{\rm L+R}$ enables one to introduce
current masses of technifermions directly into the initial
Lagrangian without a need in extra fields which is considered to be
advantageous. While the latter freedom may be regarded as a new form
of the hierarchy problem as there must be a symmetry which protects
the current up ($U$) and down ($D$) technifermion masses $m_{U,D}$
from becoming very large, we take on the phenomenological approach
and consider the chiral limit of the theory with the current masses
being small compared to the techniconfinement energy scale, i.e.
$m_{U,D}\ll\Lambda_{\rm TC}$, in a complete analogy with the chiral
QCD framework. Surely, the latter issue should be addressed in a
high-scale GUT-like theory which incorporates new strongly-coupled
fermion sectors, and this certainly goes beyond the scope of the
present analysis. Additionally, chiral (axial) anomalies do not
appear in this framework; it is {\it anomaly-safe} automatically. We
will further discuss specific consequences of such new vector-like
weak interactions of the additional technifermion sector in
confinement.

One should remember that identification of the local vector subgroup
of the chiral group with the SM weak isospin group (\ref{ident}) is
a purely phenomenological procedure which leads to correct results
in the low energy limit of the theory. In reality, of course, the
global classification techniflavor group $SU_{\rm L}(2)\otimes
SU_{\rm R}(2)$ has nothing to do with the EW gauge group of the SM.
At the first stage, the techniflavor group is used for
classification of {\it composite} technihadrons and, in particular,
predicts the existence of technipions, technisigma and technibaryons
states. At the second stage, one notices that technifermions
entering the composite technihadrons besides technistrong
interactions participate also in the fundamental EW interactions.
One should therefore calculate the {\it EW form factors} of
composite technihadrons. The corresponding EW interactions must then
be also introduced at {\it the fundamental technifermion level}
consistently with those at {\it the composite level technihadron
level}. At the third stage, in the phenomenologically interesting
low-energy limit of the theory the EW form factors approach the
renormalized EW constants (since the technihadron substructure does
not emerge at relatively small momentum transfers). The latter
should be calculated after {\it reclassification} of technihadrons
under the EW group representations. This three-fold generic scheme
will be used below for description of EW interactions of
technihadrons.

According to the standard quark-meson approaches
\cite{Serot:1997xg,const-quark}, constituent quark loops describe
non-perturbative effects at relatively small distances, whereas
meson loops work at larger distances. This scheme should be realized
in the CSTC model under discussion, in a complete analogy with the
standard quark-meson theories, and is valid up to an energy scale of
typical technihadron states. Following to this analogy, we consider
meson (technipions $\tilde{\pi}_a$ and technisigma $\tilde{\sigma}$)
interactions at tree level, and technifermion interactions (with
effective constituent masses) at one-loop level \cite{const-quark}.
At much larger energies, one should turn into the perturbative
techni-QCD framework describing technigluon and technifermion (with
current masses) interactions, in analogy with the standard QCD
approach.

This scenario becomes especially interesting from both theoretical
and phenomenological points of view since it predicts the existence
of the physical technimeson spectrum with relatively light
pseudoscalar $\tilde{\pi}_a$ and scalar $\tilde{\sigma}$
fields\footnote{It is typically assumed that technibaryons, along
with the vector and pseudovector states, are much heavier and thus
likely to be irrelevant for the LHC phenomenology, at least, at the
moment. Although if the techni-confinement scale $\Lambda_{\rm TC}$
is not very far above the EW scale, technibaryon states might emerge
in LHC data as large missing $E_T$ signatures which is a subject for
further studies.}. Latter has quantum numbers identical to the SM
Higgs boson ones. This leads to a mixing of initial $\tilde{\sigma}$
and $H$ fields causing a possible modification of the physical Higgs
boson couplings. Additionally, lightest physical technipion states
$\tilde{\pi}_a$ enrich LHC phenomenology with possible new
observable signatures, to be studied in detail.

\subsection{Gauged linear technisigma model: initial CSTC Lagrangian}

As was shortly discussed in the previous Section, we use the
standard structure of the gauged linear $\sigma$-model for
low-energy TC phenomenology. Let us formulate the CSTC model in
terms of the lightest composite states based on the local weak
isospin symmetry group $SU(2)_{\rm L+R}=SU(2)_{\rm W}$ acting on the
confined technifermion sector. The initial field content of the CSTC
model in its simplest formulation is given by one $LR$-symmetric
doublet of technifermions
 \begin{eqnarray} \label{Tdoub}
 \tilde{Q} = \left(
      \begin{array}{c}
         U \\
         D
      \end{array}
             \right)
 \end{eqnarray}
which forms the fundamental representation of the $SU(2)_{\rm
W}\otimes U(1)_{\rm Y}$ group, the initial scalar technisigma $S$
field which is the singlet representation, and the triplet of
initial technipion fields $P_a,\,a=1,2,3$ which is the adjoint
(vector) representation of $SU(2)_{\rm W}$ (with zeroth $U(1)_{\rm
Y}$ hypercharge). Thus, in terms of the fields introduced above the
GLT$\sigma$M part of the Lagrangian responsible for Yukawa-type
interactions of the technifermions reads
 \begin{eqnarray}
 {\cal L}_Y^{\rm CSTC} = -g_{\rm TC} \bar{\tilde{Q}}(S+i\gamma_5\tau_a
 P_a)\tilde{Q}\,,
 \end{eqnarray}
where $\tau_a,\,a=1,2,3$ are the Pauli matrices. By restricting
ourselves to considering only one technifermion doublet
(\ref{Tdoub}) (the first generation), we imply that other
generations, if exist, are much heavier and split off in the mass
spectrum, based on analogy with the SM, even though such an analogy
is not mandatory.

\begin{figure*}[!h]
\begin{minipage}{0.7\textwidth}
 \centerline{\includegraphics[width=1.0\textwidth]{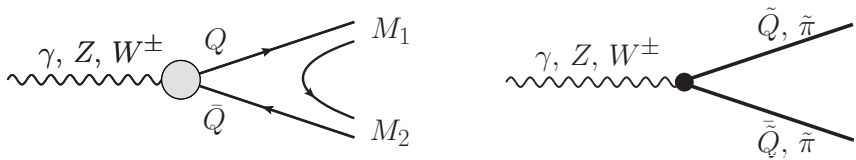}}
\end{minipage}
   \caption{
\small An illustration of the interactions of (techni)fermion and
(techni)meson fields with the SM gauge bosons via
(techni)hadronisation in hadron physics (left panel) and in the
point-like approximation adopted in the considered CSTC scenario
(right panel).}
 \label{fig:fig-1}
\end{figure*}

In the SM, the gauge boson interactions with usual hadrons are
typically introduced by means of hadronisation effects (see
Fig.~\ref{fig:fig-1} (left)).
In our case, such an effect is strongly suppressed by large
constituent masses of technifermions $\sim \Lambda_{\rm TC}$.
Instead, the interactions of $\tilde{Q}$ and $P_a$ fields with
initial SM gauge fields $B_{\mu},\,V_{\mu}^a$ can be introduced via
the local approximation which is illustrated in Fig.~\ref{fig:fig-1}
(right). Generally speaking, these interactions should be written in
terms of nonlocal form factors since both technimesons and dressed
(constituent) technifermions are the objects delocalized at energy
scales exceeding the scale of non-perturbative technigluon
fluctuations. We assume, however, that the latter scale is large
compared to the EWSB scale and corresponding effects can be
neglected at experimentally accessible energy scales. Thus, in the
first approximation one can replace the form factors by point-like
couplings as is usually done in the local quantum-field theory
approach\footnote{In a more rigorous analysis this approximation can
be easily lifted by introducing the Pauli form factors, although in
this very first analysis of the CSTC we work in the point-like
approximation for the sake of simplicity.}. The coupling constants
of $\tilde{Q}$ and $P_a$ with gauge fields can be taken the same as
in the SM, but calculated via the Renormalisation Group evolution at
corresponding scales. Since this evolution is logarithmic and rather
weak, whereas $\Lambda_{\rm TC}$ is assumed to be in a vicinity of
the EW scale, in the leading-order numerical analysis below we fix
all the relevant couplings at the $M_Z$ scale.

The vector-like gauge interactions can be introduced via covariant
derivatives over the local $SU(2)_{\rm W}\otimes U(1)_{\rm Y}$ group
in the same form as the SM gauge interactions, i.e. the additional
(to the SM) kinetic terms have the following form
 \begin{eqnarray}
  {\cal L}_{kin}^{\rm CSTC} = \frac12 \partial_{\mu} S\, \partial^{\mu} S + \frac12 D_{\mu} P_a\, D^{\mu} P_a +
  i \bar{\tilde{Q}}\hat{D}\tilde{Q}\,, \label{LG}
 \end{eqnarray}
where the covariant derivatives of the $\tilde{Q}$ and $P_a$ fields
reads
 \begin{eqnarray}
    \hat{D}\tilde{Q} = \gamma^{\mu} \left( \partial_{\mu}
       - \frac{iY_{\tilde Q}}{2}\, g'B_{\mu} - \frac{i}{2}\, g W_{\mu}^a \tau_a \right)\tilde{Q},\,
       \qquad D_{\mu} P_a = \partial_{\mu} P_a + g
\epsilon_{abc} W^b_{\mu} P_c\,, \label{DQ}
 \end{eqnarray}
respectively. Further, we wish to employ analogies with the SM and,
in particular, with QCD as much as possible, so for the sake of
convenience and simplicity in actual calculations we fix the
hypercharge of the technifermion doublet (\ref{Tdoub}) to be the
same as that of the quark doublet in the SM, i.e.
$Y_{\tilde{Q}}=1/3$, unless noted otherwise. Certainly, the
hypercharge $Y_{\tilde{Q}}$, the number of technifermion
generations, the respective properties of interactions, etc. should
be ultimately constrained in extended chiral-gauge or grand-unified
theories incorporating extra technifermion sectors, which is a
subject of further studies.

In Eqs.~(\ref{LG}) and (\ref{DQ}) we notice two key differences of
the CSTC scenario from traditional TC-based models (cf.
Ref.~\cite{Peskin:1990zt,Chivukula:2000mb}) -- the existence of
physical technisigma and technipion states, introduced via the
GLT$\sigma$M approach, and the equivalence of left and right
technifermion chiralities in their interactions with weak gauge
bosons, following from the gauging of the initial chiral group of
the linear $\sigma$ model. Along with the absence of chiral
anomalies, the CSTC scenario under discussion can be considered as a
solid theoretically motivated basis for the whole new class of more
elaborated TC-based extensions of the SM and their phenomenological
tests.

Next, let us consider the potential part of the CSTC model
Lagrangian giving rise to (pseudo)scalar self-interactions and
$\tilde{\pi}$, $\tilde{\sigma}$ masses after the chiral symmetry
breaking and the EWSB. As was mentioned in the Introduction, in the
simplest formulation of the CSTC model developed in this work we
keep the SM Higgs mechanism of the EWSB and the one-Higgs-doublet SM
untouched, and simply add extra technifermion sector (\ref{Tdoub})
in confinement. As an essential part of the CSTC model, we introduce
the interaction terms between the standard Higgs doublet
$\mathcal{H}$, and the new $P_a$ and $S$ states which are allowed by
the local $SU(2)_W$ symmetry. As will be demonstrated below, such
extra terms lead to a mixing between the scalar Higgs and
technisigma fields. The most general form of the Lagrangian
corresponding to the scalar self-interactions including $\mu$-terms
as follows \cite{Serot:1997xg}
\begin{eqnarray}
 {\cal L}_{U,\,\rm self}^{\rm CSTC} = \frac12\mu^2_{\rm S}(S^2+P^2)+\mu_{\rm H}^2{\mathcal{H}}^2
- \frac14\lambda_{\rm TC}(S^2+P^2)^2-\lambda_{\rm H}{\mathcal{H}}^4
+ \lambda{\mathcal{H}}^2(S^2+P^2)\,, \label{L-U-self}
\end{eqnarray}
and the extra linear ``source'' term which appears after averaging
over the technifermion vacuum fluctuations and describes
interactions of the scalar singlet $S$ field with scalar modes of
the technifermion condensate, i.e.
\begin{eqnarray}
 {\cal L}_{U,\,\rm source}^{\rm CSTC} &=& -g_{\rm{TC}}\,S\,\langle\bar{\tilde{Q}}\tilde{Q}\rangle\,. \label{L-U-source}
\end{eqnarray}
The potential part of the GLT$\sigma$M Lagrangian is then given by
\begin{eqnarray}
 {\cal L}_{U}^{\rm CSTC} = {\cal L}_{U,\,\rm self}^{\rm CSTC} + {\cal L}_{U,\,\rm source}^{\rm CSTC}\,. \label{L-U}
\end{eqnarray}
In Eq.~(\ref{L-U-self}) we defined $P^2\equiv \sum_a
P_aP_a=\tilde{\pi}^{0}\tilde{\pi}^{0}+2\tilde{\pi}^+\tilde{\pi}^-$,
whereas gauge-Higgs interaction terms are the same as in the SM.
\begin{figure*}[!h]
 \centerline{\includegraphics[width=1.0\textwidth]{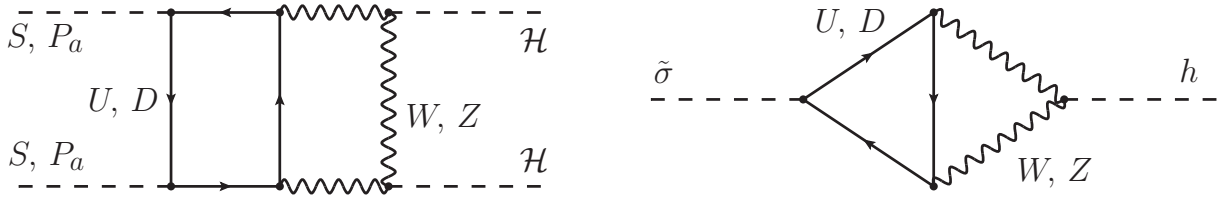}}
   \caption{
\small Typical radiative corrections to the quartic Higgs-TC
coupling $\lambda$ (in particular, giving rise to the
$h\tilde\sigma$-mixing) before the EWSB (left) and after the EWSB
(right).}
 \label{fig:operators}
\end{figure*}

The mixing between the Higgs boson and scalar technisigma fields is
governed by the quartic Higgs-TC coupling $\lambda$ in
Eq.~(\ref{L-U-self}). Such a mixing is one of the characteristic
effects of the chiral-symmetric Technicolor. In a sense, this effect
is indeed one of the motivations of the model under discussion. It
has to be taken into consideration if the precision LHC measurements
uncover possibly small deviations of the Higgs-like $126$ GeV boson
(especially, in the $\gamma\gamma$ decay channel) from the standard
Higgs boson. The quartic coupling $\lambda$ controls such a mixing
and \'a priori is allowed by the gauge symmetry of the initial
Lagrangian, thus, cannot be identically equal to zero. Indeed, any
terms which are allowable by the initial symmetry of the model, even
being equal to zero at the tree level, necessarily appear in
divergent radiative corrections. In order to renormalize such
divergencies one has to introduce corresponding counterterms. So if
at a given scale $\mu_0$ the coupling $\lambda(\mu_0)\to0$ vanishes
it will reappear at another scale. In particular, before the
spontaneous EW symmetry breaking the operator
$\sim{\mathcal{H}}^2(S^2+P^2)$ is supported by the two-loop box-box
diagram illustrated in Fig.~\ref{fig:operators} (left) with incoming
initial $S$ and $P_a$ fields and outgoing initial Higgs field ${\cal
H}$. This operator thus contributes to remormalization of $\lambda$
coupling. After the EWSB, the resulting physical $h\tilde\sigma$
mixing is renormalized by two-loop triangle-triangle diagram shown
in Fig.~\ref{fig:operators} (right)\footnote{In addition, there is
an extra one-loop contribution to the $h\tilde\sigma$-mixing which
is going via a technipion loop. The latter correction exists for
non-zeroth tree-level $\lambda_{\rm tree}\not=0$ only.}. In extended
$SU_{\rm L}(N_f)\otimes SU_{\rm R}(N_f)$ models mentioned above the
corresponding quartic Higgs-TC operator which mixes physical $h$ and
$\tilde\sigma$ appears automatically from the main invariant of the
linear $\sigma$-model and cannot be eliminated.

In order to provide the EWSB and the chiral symmetry breaking in the
simplest way, the Higgs $\mathcal{H}$ and technisigma $S$ fields get
vevs and corresponding physical scalar degrees of freedom are mixed
up, i.e.
\begin{eqnarray} \nonumber
 \mathcal{H} &=& \frac{1}{\sqrt{2}}\left(
\begin{array}{c}
\sqrt{2}i\phi^-  \\
H+i\phi^0
\end{array}\right)\,,\quad
H=v+h c_\theta-\tilde{\sigma}s_\theta\,,\quad \langle \mathcal{H}
\rangle = \frac{1}{\sqrt{2}}\left(
\begin{array}{c}
0  \\
v
\end{array}
\right)\,,\\ v &=& \frac{2M_W}{g} \simeq 246\,{\rm GeV}\,, \quad
 S = u + h s_\theta+\tilde{\sigma} c_\theta\,,\quad \langle S \rangle = u
 \gtrsim v \,,
\label{shifts}
\end{eqnarray}
where $M_W$ is the $W$ boson mass, $v,\ u$ are the Higgs boson and
technisigma $\tilde{\sigma}$ vevs; $h,\ \tilde{\sigma}$ are the
corresponding physical fields with positively definite masses $M_h,\
M_{\tilde \sigma}$, respectively; $c_{\theta}\equiv \cos\theta$,
$s_{\theta}\equiv \sin\theta$, and $\theta$ is the mixing angle,
which diagonalizes the respective scalar mass form. We therefore end
up with the physical Lagrangian which describes new types of
interactions, namely, between Higgs boson, technipions and
technisigma, Yukawa technifermion interactions, as well as mixing
effects between the Higgs boson and technisigma fields, relevant for
the LHC phenomenology.

As it is well-known, in the SM framework we deal with two energy
scales of a completely different nature. The first one is the scale
of quark-gluon condensate which has a quantum-topological nature.
The second one given by the amplitude of the constant Higgs field
(vev) has classical (non-quantum) origin. In the framework of the
CSTC model we suggest another {\it interpretation} of the classical
Higgs mechanism in which the nature of all energy scales (including
the Higgs vev) is quantum-topological, in the essence of original TC
and compositeness models of the DEWSB. The simplest way to realize
this idea is to introduce into the scalar potential an ``external
source'' term (the first term in Lagrangian (\ref{L-U}) linear in
$S$ field) which describes interactions between technifermion
condensate with the singlet scalar $S$ field \cite{Serot:1997xg}. As
will be demonstrated below, in the framework of the CSTC model this
term leads to a close connection between the Higgs and technifermion
condensates. A possible experimental verification of the CSTC model
at the LHC relies on our assumption that both EW and TC scales are
relatively close to each other, within the LHC energy scales.
Indeed, in this case it is natural to assume that the Higgs and
technifermion condensates ($v$ and $u$, respectively) may have the
same origin. Our specific goal is to study possible observable
effects of such a phenomenon related, in particular, to the Higgs
boson properties as well as to lightest technihadron phenomenology
at the LHC energy scales.

\subsection{Parameter space of the CSTC model}

As was mentioned above, in the framework of CSTC scenario it is
assumed that the EWSB in the SM sector (via ordinary Higgs mechanism
by the Higgs vev, $v$) and the chiral symmetry breaking in the TC
sector (via the scalar technisigma field vev, $u$) may happen at
energy scales relatively close to each other, i.e. $u\sim
\Lambda_{\rm TC}\sim 0.1-1$ TeV. In what follows, we adopt this
limiting case where one may expect possible specific signatures of
the chiral-symmetric strongly coupled sectors potentially observable
at the LHC.

Minimizing the potential (\ref{L-U}) using expressions
(\ref{shifts}) one arrives at the set of tadpole equations for the
vacuum expectation values
\begin{equation}
\begin{array}{c}
\displaystyle \langle\delta {\cal L}_U^{\rm
CSTC}/\delta\mathcal{H}\rangle=v\left(\mu_{\rm H}^2-\lambda_{\rm
H}v^2+\lambda u^2\right)=0\ ,
\\[5mm]
\displaystyle \langle\delta {\cal L}_U^{\rm CSTC}/\delta
S\rangle=u\left(\mu_{\rm S}^2-\frac{g_{\rm TC}\langle\bar
QQ\rangle}{u}-\lambda_{\rm TC}u^2+\lambda v^2\right)=0\,.
\end{array}
 \label{tadpole}
\end{equation}
The solution of the above equations with respect to scalar fields
vevs has the following form
\begin{equation}
\begin{array}{c}
\displaystyle v^2=\frac{\lambda_{\rm TC}\mu_{\rm
H}^2+\lambda(\mu_{\rm S}^2+m_{\tilde{\pi}}^2)}{\lambda_{\rm
TC}\lambda_{\rm H}-\lambda^2}\ ,
\\[5mm]
\displaystyle u^2=\frac{\lambda_{\rm H}(\mu_{\rm
S}^2+m_{\tilde{\pi}}^2)+\lambda\mu_{\rm H}^2}{\lambda_{\rm
TC}\lambda_{\rm H}-\lambda^2}\ ,
\end{array}
 \label{tadsols}
\end{equation}
where
\begin{eqnarray}
m_{\tilde{\pi}}^2=-\frac{g_{\rm TC}\langle{\bar {\tilde
Q}}\tilde{Q}\rangle}{u}\,, \qquad \langle{\bar {\tilde
Q}}\tilde{Q}\rangle<0\,, \qquad g_{\rm TC}>0 \label{pion-mass}
\end{eqnarray}
is the technipion mass squared proportional to the (negative-valued)
technifermion condensate $\langle{\bar {\tilde Q}}\tilde{Q}\rangle$,
similarly to that in low-energy hadron physics. The vacuum stability
is ensured by the minimum of the potential $U=-{\cal L}_U^{\rm
CSTC}$ (\ref{L-U}), i.e. by
\begin{eqnarray*}
\Delta\equiv\Bigl\langle\frac{\delta^2 {\cal L}_U^{\rm
CSTC}}{\delta\mathcal{H}\delta
S}\Bigr\rangle^2-\Bigl\langle\frac{\delta^2 {\cal L}_U^{\rm
CSTC}}{\delta\mathcal{H}^2}\Bigr\rangle \Bigl\langle\frac{\delta^2
{\cal L}_U^{\rm CSTC}}{\delta S^2}\Bigr\rangle <0\,,\quad
\Bigl\langle\frac{\delta^2 {\cal L}_U^{\rm
CSTC}}{\delta\mathcal{H}^2}\Bigr\rangle<0\,,\quad
\Bigl\langle\frac{\delta^2 {\cal L}_U^{\rm CSTC}}{\delta
S^2}\Bigr\rangle<0\,,
\end{eqnarray*}
leading to
\begin{eqnarray}
\lambda_{\rm TC}>-\frac{m_{\tilde{\pi}}^2}{2u^2}\,,\qquad
\lambda_H>0\,, \label{minimum}
\end{eqnarray}
which are automatically satisfied for the positively defined scalar
mass form, i.e. for $M_{\tilde{\sigma}}^2>0$ and $M_h^2>0$.
\begin{figure*}[!b]
\begin{minipage}{0.4\textwidth}
 \centerline{\includegraphics[width=1.0\textwidth]{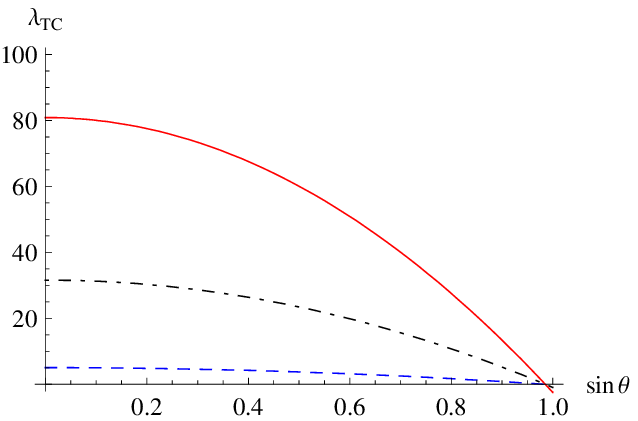}}
\end{minipage}
\begin{minipage}{0.4\textwidth}
 \centerline{\includegraphics[width=1.0\textwidth]{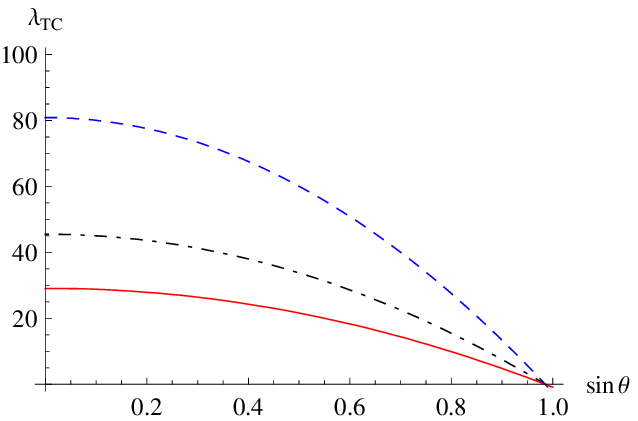}}
\end{minipage}
\begin{minipage}{0.4\textwidth}
 \centerline{\includegraphics[width=1.0\textwidth]{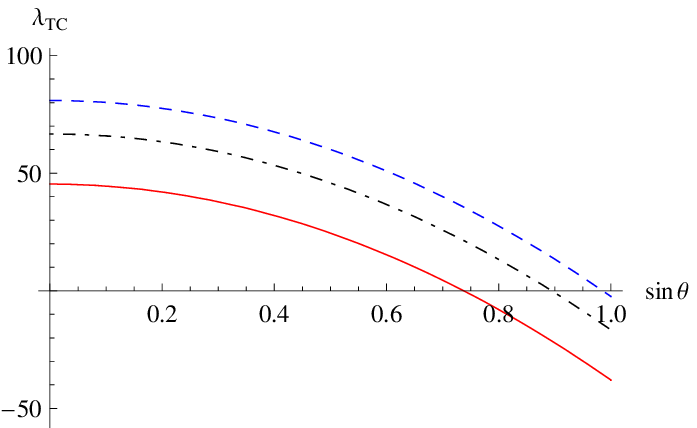}}
\end{minipage}
\begin{minipage}{0.4\textwidth}
 \centerline{\includegraphics[width=1.0\textwidth]{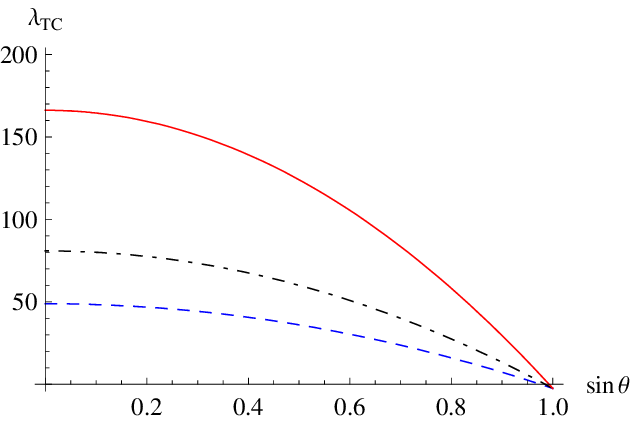}}
\end{minipage}
   \caption{
\small Dependence of the quartic TC self-coupling $\lambda_{\rm TC}$
on the $h\tilde{\sigma}$-mixing $s_\theta$ with dashed, dash-dotted
and solid lines corresponding to (1) $g_{\rm TC}=2,5,8$,
$M_{\tilde{Q}}=300$ GeV, $m_{\tilde{\pi}}=150$ GeV, and
$M_{\tilde{\sigma}}=500$ GeV; (2) $g_{\rm TC}=8$,
$M_{\tilde{Q}}=300,400,500$ GeV, $m_{\tilde{\pi}}=150$ GeV, and
$M_{\tilde{\sigma}}=500$ GeV; (3) $g_{\rm TC}=8$,
$M_{\tilde{Q}}=300$ GeV, $m_{\tilde{\pi}}=150,250,350$ GeV, and
$M_{\tilde{\sigma}}=500$ GeV; (4) $g_{\rm TC}=8$,
$M_{\tilde{Q}}=300$ GeV, $m_{\tilde{\pi}}=150$ GeV, and
$M_{\tilde{\sigma}}=400,500,700$ GeV, in each plot from top to
bottom and left to right, respectively. Here and below, $M_h=125$
GeV. The coupling $\lambda_{\rm TC}$ is symmetric w.r.t.
$s_\theta\to-s_\theta$.}
 \label{fig:fig-lamTC}
\end{figure*}

Notice that in the limiting case of $\mu_{{\rm S},{\rm H}}\ll
m_{\tilde \pi}$ which, in principle, is not forbidden (while origin
of $\mu$-terms is generally unclear in the SM theory) and even can
be motivated in the nearly conformal limit of new strongly coupled
dynamics (see below), both vevs $v$ and $u$ are expressed in terms
of the technifermion condensate, having thereby the same dynamical
origin. The extra confined TC sector is now responsible for the EWSB
in the CSTC model, so the role of extra $\mu$-terms, which are
usually required for the classical Higgs mechanism in the rigorous
SM formulation, is taken over by the technifermion condensate. This
observation thus supports the above argument about the common
quantum-topological nature of the EWSB and the chiral symmetry
breaking mechanisms in the considering CSTC model. In what follows,
we discuss both cases. In the first case, for the sake of
generality, we keep the scalar $\mu$-terms permitted by the gauge
symmetry as free independent parameters. In the second theoretically
motivated limiting case $\mu_{{\rm S},{\rm H}}\ll m_{\tilde \pi}$,
we will also consider the minimal CSTC model neglecting the small
$\mu$-terms below.

In the general case, the mass form of the scalar fields can be
diagonalized and represented in the form
\begin{equation}
\begin{array}{c}
\displaystyle \Delta {\cal L}^{\rm
CSTC}_{sc}=-\frac12[m_{\tilde{\pi}}^2(2\tilde{\pi}^+\tilde{\pi}^-+\tilde{\pi}^0\tilde{\pi}^0)+M_{\tilde
\sigma}^2{\tilde \sigma}^2+M_h^2h^2]\,,
\end{array}
 \label{scmass}
 \end{equation}
where the technipion mass squared expressed in terms of vevs and
scalar self-couplings is
\begin{eqnarray}
m_{\tilde{\pi}}^2=\lambda_{\rm TC} u^2 - \lambda v^2 - \mu_{\rm
S}^2\,, \label{pion-mass-2}
\end{eqnarray}
the technisigma and Higgs boson masses squared are
\begin{equation}
\begin{array}{c}
\displaystyle M_h^2=\frac12\left[2\lambda_{\rm
TC}u^2+m_{\tilde{\pi}}^2+2\lambda_{\rm H}v^2- \sqrt{(2\lambda_{\rm
TC}u^2+m_{\tilde{\pi}}^2-2\lambda_{\rm
H}v^2)^2+16\lambda^2u^2v^2}\right]\,,
\\[5mm]
\displaystyle M_{\tilde \sigma}^2=\frac12\left[2\lambda_{\rm
TC}u^2+m_{\tilde{\pi}}^2+2\lambda_{\rm H}v^2+ \sqrt{(2\lambda_{\rm
TC}u^2+m_{\tilde{\pi}}^2-2\lambda_{\rm
H}v^2)^2+16\lambda^2u^2v^2}\right]\,,
\end{array}
 \label{SHmasses}
\end{equation}
respectively. Finally, the expression for the
$h\tilde{\sigma}$-mixing angle reads
\begin{equation}
\begin{array}{c}
\displaystyle \tan 2\theta=\frac{4\lambda uv}{2\lambda_{\rm
TC}u^2+m_{\tilde{\pi}}^2-2\lambda_{\rm H}v^2}\,,
\end{array}
 \label{theta}
 \end{equation}
whereas the sign of $s_\theta$ is given by
\begin{equation}
\begin{array}{c}
\displaystyle {\rm sign}(s_\theta)={\rm sign}\Bigl(\frac{\lambda
uv}{2\lambda_{\rm H}v^2-M_h^2}\Bigr) \,.
\end{array}
 \end{equation}
\begin{figure*}[!h]
\begin{minipage}{0.32\textwidth}
 \centerline{\includegraphics[width=1.0\textwidth]{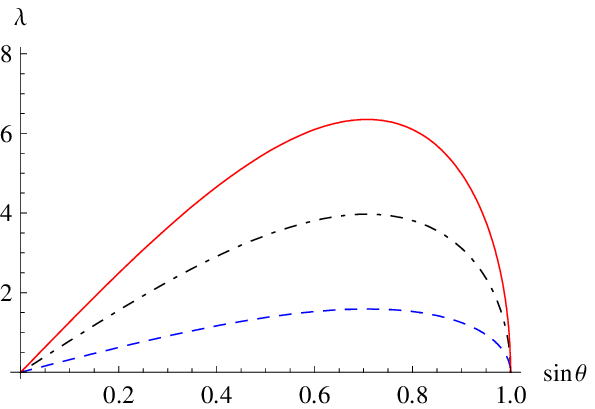}}
\end{minipage}
\begin{minipage}{0.32\textwidth}
 \centerline{\includegraphics[width=1.0\textwidth]{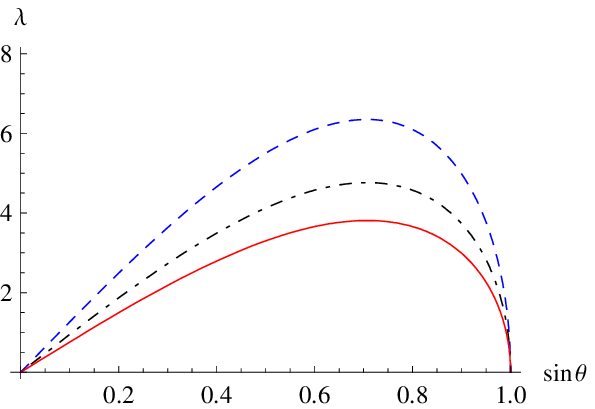}}
\end{minipage}
\begin{minipage}{0.32\textwidth}
 \centerline{\includegraphics[width=1.0\textwidth]{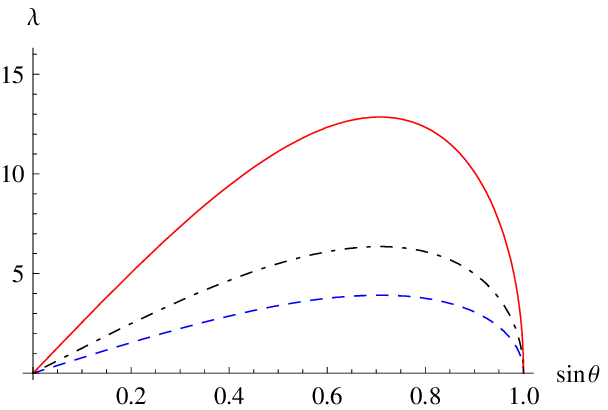}}
\end{minipage}
   \caption{
\small Dependence of the quartic Higgs-TC coupling $\lambda$ on the
$h\tilde{\sigma}$-mixing $s_\theta$ with dashed, dash-dotted and
solid lines corresponding to (1) $g_{\rm TC}=2,5,8$,
$M_{\tilde{Q}}=300$ GeV, and $M_{\tilde{\sigma}}=500$ GeV; (2)
$g_{\rm TC}=8$, $M_{\tilde{Q}}=300,400,500$ GeV, and
$M_{\tilde{\sigma}}=500$ GeV; (3) $g_{\rm TC}=8$,
$M_{\tilde{Q}}=300$ GeV, and $M_{\tilde{\sigma}}=400,500,700$ GeV,
in each plot from left to right, respectively. It does not depend on
$m_{\tilde{\pi}}$. The coupling $\lambda$ is antisymmetric w.r.t.
$s_\theta\to-s_\theta$.}
 \label{fig:fig-lam}
\end{figure*}

In general, the additional sector of the Lagrangian under discussion
together with the modified SM Higgs sector contains seven parameters
in total, namely
\begin{equation}
\begin{array}{c}
\displaystyle \mu_{\rm H}^2\ ,\qquad \mu_{\rm S}^2\ ,\qquad
\lambda_{\rm H}\ ,\qquad \lambda_{\rm TC}\ ,\qquad \lambda\ ,\qquad
 g_{\rm TC}\ ,\qquad \langle{\bar {\tilde Q}}\tilde{Q}\rangle\ .
\end{array}
 \label{inipars}
 \end{equation}
In phenomenological studies, it can be convenient to turn to
mathematically equivalent set of other independent physical
parameters, namely,
\begin{equation}
\begin{array}{c}
\displaystyle M_h\ ,\qquad M_{\tilde \sigma}\ ,\qquad
m_{\tilde{\pi}}\ ,\qquad M_W\ ,\qquad M_{\tilde Q}\ ,\qquad g_{\rm
TC}\ ,\qquad s_\theta\ .
\end{array}
 \label{physpars}
 \end{equation}
where $M_{\tilde Q}=g_{\rm TC}u$ is the constituent technifermion
mass. For this purpose, the following relations between scalar
self-couplings and physical quantities (\ref{physpars}) following
directly from Eqs.~(\ref{pion-mass-2}), (\ref{SHmasses}) and
(\ref{theta}) can be useful:
\begin{equation}
\begin{array}{c}
\displaystyle 2\lambda_{\rm TC}u^2=-m_{\tilde{\pi}}^2+M_{\tilde
\sigma}^2c^2_\theta+M_{h}^2s^2_\theta\ ,
\\[5mm]
\displaystyle 2\lambda_{\rm H}v^2=M_{\tilde
\sigma}^2s^2_\theta+M_{h}^2c^2_\theta\ ,
\\[5mm]
\displaystyle 2\lambda uv=\pm(M_{\tilde \sigma}^2-M_{h}^2)c_\theta
s_\theta\ .
\end{array}
 \label{parrels}
 \end{equation}

In reality, two mass parameters in Eq.~(\ref{physpars}) can be fixed
by the SM phenomenology, namely, $M_W\simeq 80.4$ GeV and $M_h\simeq
125.3$ GeV, so effectively only five-dimensional parameter space
remains to be analyzed. Apparently, two phenomenologically
interesting cases are possible: the lightest observed scalar
particle is indeed the Higgs boson, then $M_h<M_{\tilde \sigma}$, or
the technisigma is the lightest one $M_{\tilde \sigma}<M_h$. In
Eq.~(\ref{parrels}) we restrict ourselves to the first solution for
$\lambda$, with ``plus'' sign, and fix $\cos\theta>0$, such that the
sign of $\lambda$ is the same as the sign of $s_\theta$ for
$M_{\tilde \sigma}>M_{h}$, opposite to the sign of $s_\theta$ for
reversed hierarchy $M_{\tilde \sigma}<M_{h}$. In what follows, we
work with the direct mass hierarchy with the lightest Higgs boson in
the scalar sector of the model $M_{\tilde \sigma}>M_{h}$, unless
noted otherwise.
\begin{figure*}[!h]
\begin{minipage}{0.4\textwidth}
 \centerline{\includegraphics[width=1.0\textwidth]{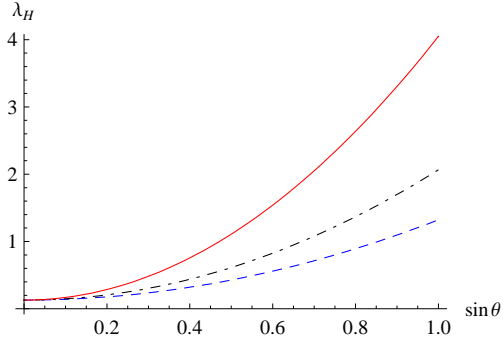}}
\end{minipage}
   \caption{
\small Dependence of the quartic Higgs boson self-coupling
$\lambda_{\rm H}$ on the $h\tilde{\sigma}$-mixing $s_\theta$ with
dashed, dash-dotted and solid lines corresponding to
$M_{\tilde{\sigma}}=400,500,700$ GeV, respectively. It does not
depend on other free parameters of the CSTC model. The coupling
$\lambda_{\rm H}$ is symmetric w.r.t. $s_\theta\to-s_\theta$.}
 \label{fig:fig-lamH}
\end{figure*}

In Fig.~\ref{fig:fig-lamTC} we represent dependence of the quartic
TC self-coupling $\lambda_{\rm TC}$ on the $h\tilde{\sigma}$-mixing
angle, or more precisely $s_\theta$, over reasonable ranges of
$g_{\rm TC}$, $M_{\tilde Q}$, $m_{\tilde \pi}$ and $M_{\tilde
\sigma}$ parameters. One notices that $\lambda_{\rm TC}$ vanishes in
the maximal $h\tilde{\sigma}$-mixing limit $s_\theta\to1$ for any
$g_{\rm TC}$, $M_{\tilde Q}$, $M_{\tilde \sigma}$ values and for
small $m_{\tilde \pi}\sim 150$ GeV. For small mixing angles and
rather large $M_{\tilde \sigma}\gtrsim 700$ GeV and $g_{\rm
TC}\gtrsim 8$, it can become very large $\lambda_{\rm TC}\sim 100$,
where the non-linear non-perturbative effects turn out to be
important, and applicability of the corresponding GLT$\sigma$M may
be restricted. This has to be taken into consideration in analysis
of the available parameter space of the model and possible
phenomenological signatures.

Similarly, the quartic Higgs-TC coupling $\lambda$ and the quartic
Higgs boson self-coupling $\lambda_{\rm H}$ w.r.t. $s_\theta$ are
given in Figs.~\ref{fig:fig-lam} and \ref{fig:fig-lamH},
respectively. The $\lambda$ coupling does not depend on the
technipion mass, and vanish in both limits $s_\theta\to 1$ and
$s_\theta\to 0$ limits as is seen in figures. The $\lambda_H$
coupling depends only on the $s_\theta$ and $M_{\tilde \sigma}$, and
both $\lambda$ and $\lambda_{\rm H}$ are generally constrained
$\lambda,\,\lambda_{\rm H}\lesssim 10$.

In our analysis for the sake of simplicity and transparency we wish
to employ an analogy with QCD and hadron physics as long as
possible, which is reasonable (even though not necessary) since the
TC confinement group and technifermion hypercharge are assumed to be
the same as for standard quarks. If such an analogy is indeed
realized in nature, one would need to pay attention to other
possible similarities e.g. in properties of QCD and techni-QCD
vacuum subsystems. The QCD vacuum at scales $\Lambda_{\rm QCD}\sim
200$ MeV is formed by gluon and quark condensates
\cite{Shifman:1978bx}:
 \begin{eqnarray} \nonumber
&&\langle 0| \frac{\alpha_s}{\pi} \hat{G}_{\mu\nu}\hat{G}^{\mu\nu}
|0\rangle = (365\pm 20 \, {\rm MeV})^4\simeq (2\Lambda_{\rm QCD})^4\,,\\
&& \langle 0|\bar{u}u |0\rangle=\langle 0|\bar{d}d |0\rangle=-l_g
\langle 0| \frac{\alpha_s}{\pi} \hat{G}_{\mu\nu}\hat{G}^{\mu\nu}
|0\rangle=-(235\pm 15 \, {\rm MeV})^3\,, \label{QCD-vac}
 \end{eqnarray}
where $\Lambda_{\rm QCD}^{-1}\simeq 10^{-13}$ sm is the
characteristic hadron size, whereas the correlation length
$l_g\simeq (1500\,\rm{MeV})^{-1}$ is the characteristic length scale
of the non-perturbative gluon field fluctuations. In the meson
spectrum, the lightest states are pions with mass $m_{\pi}\simeq
140$ MeV (the pseudo-Goldstone modes of the quark condensate
excitations) and $\sigma$-meson $\sigma=f_0(500)$ with mass
$m_{\sigma}\simeq 500$ MeV (the lightest glueball as a collective
excitation of the gluon condensate). In the framework of the
hypothesis about the technicolor nature of the Higgs vacuum
$v\sim200$ GeV, it is natural to assume that the second techni-QCD
vacuum subsystem is formed by condensate of technigluons and light
technifermions at a nearby scale $\Lambda_{\rm TC}\gtrsim 200$ GeV,
being therefore, at least, a thousand times higher than
$\Lambda_{\rm QCD}$ scale. Then, a reasonable order-of-magnitude
estimate leads to
 \begin{eqnarray*}
&&\langle 0| \frac{\alpha_{\rm TC}}{\pi}
\hat{F}_{\mu\nu}\hat{F}^{\mu\nu}
|0\rangle \sim (2\Lambda_{\rm TC})^4\,,\\
&& \langle 0|\bar{U}U |0\rangle=\langle 0|\bar{D}D |0\rangle\sim
-l_{\rm TC} (2\Lambda_{\rm TC})^4\,.
 \end{eqnarray*}
If the current technifermion masses obey the same hierarchy as that
of usual quarks, the lightest technihadron excitations in the
bosonic spectrum are technipions $\tilde{\pi}^{0,\pm}$ and
technisigma meson $\tilde{\sigma}$, whereas in the fermion spectrum
-- techninucleons $P$ and $N$. Such a dynamical similarity between
color and technicolor enables us to estimate characteristic masses
of the lightest technihadrons and constituent technifermions through
the scale transformation of ordinary hadron states via scale factor
$\zeta=\Lambda_{\rm TC}/\Lambda_{\rm QCD}\gtrsim 1000$, i.e.
 \begin{eqnarray}
m_{\tilde{\pi}}\gtrsim 140\,{\rm GeV}\,,\quad
M_{\tilde{\sigma}}\gtrsim 500\,{\rm GeV}\,,\quad
M_{\tilde{Q}}\gtrsim 300\,{\rm GeV}\,,\quad M_{P}\simeq M_{N}\gtrsim
1\,{\rm TeV}\,, \label{QCD-an}
 \end{eqnarray}
which imply that $m_{\tilde{\pi}}>M_h$, $M_{\tilde{\sigma}}>M_h$,
$M_{\tilde{\sigma}}>m_{\tilde{\pi}}$, and $u\gtrsim 100$ GeV for
$g_{\rm TC}\simeq 3$. Also, with respect to interactions with known
particles at typical 4-momentum squared transfers $Q^2\ll l_{\rm
TC}^{-2}\gtrsim 2.3\,{\rm TeV}^2$, the lightest technihadrons behave
as elementary particles, which participate in electroweak
interactions only. The technipions are then treated as being in the
adjoint representation of the $SU(2)_{\rm W}$ with hypercharge equal
to zero, thus justifying what was done above, whereas techninucleons
can be included as the fundamental representation of the electroweak
group $SU(2)_{\rm W}\otimes U(1)_{\rm Y}$ with hypercharge $Y_{\rm
TN}=1$ along with the constituent technifermion doublet
(\ref{Tdoub}). Heavy techninucleons, however, are likely to be
irrelevant for the LHC phenomenology, but can play an important role
in astrophysics as a plausible candidate for the Dark Matter. The
conditions (\ref{QCD-an}) following from the analogy of QCD and
techni-QCD will be used below in phenomenological studies in the
CSTC framework.

\subsection{The physical Lagrangian of the CSTC model}

In this Section, we consider the principal part of the physical CSTC
Lagrangian relevant for studies of the basic phenomenological
processes in the CSTC model e.g. corrections to EW precision
observables, as well as Higgs, technipion and technisigma production
and decays, discussed below.

The vector-like interactions $\bar{\tilde{Q}}\tilde{Q}V$ of
technifermions and gauge bosons $V=Z^0,\,W^{\pm},\,\gamma$ are given
by
\begin{eqnarray}
 L_{\bar{\tilde{Q}}\tilde{Q}V}&=&\frac{1}{\sqrt{2}}\,g\bar{U}\gamma^{\mu}D\cdot W_{\mu}^+ +
\frac{1}{\sqrt{2}}\,g\bar{D}\gamma^{\mu}U\cdot W_{\mu}^- \nonumber \\
&+& \frac{g}{c_W}\, Z_{\mu} \sum_{f=U,D}
\bar{f}\gamma^{\mu}\bigl(t_3^f-q_f\,s^2_W\bigr)f +
e\sum_{f=U,D}q_f\,\bar{f}\gamma^{\mu}A_{\mu}f\,, \label{L-QV}
\end{eqnarray}
where $e=g s_W$ is the electron charge, $t_3^f$ is the weak isospin
($t_3^U=1/2$, $t_3^D=-1/2$), $q_f=Y_{\tilde Q}/2+t_3^f$ is the
technifermion charge. As agreed above, we choose $Y_{\tilde Q}=1/3$
in analogy to the SM, thus $q_U=2/3$ and $q_D=-1/3$.

The Yukawa-type interactions
$\bar{\tilde{Q}}\tilde{Q}h+\bar{\tilde{Q}}\tilde{Q}\tilde{\sigma}+\bar{\tilde{Q}}\tilde{Q}\tilde{\pi}$
of constituent technifermions with scalar ($h$ and $\tilde{\sigma}$)
and pseudoscalar ($\pi^{0,\pm}$) fields are driven by
\begin{eqnarray} &&L_{\bar{\tilde{Q}}\tilde{Q}h}+L_{\bar{\tilde{Q}}\tilde{Q}\tilde{\sigma}}+
L_{\bar{\tilde{Q}}\tilde{Q}\tilde{\pi}}=- g_{\rm
TC}\,(c_{\theta}\tilde{\sigma}+s_{\theta}h)\cdot (\bar{U}U +
\bar{D}D) \nonumber \\ && - i\sqrt{2}g_{\rm
TC}\,\tilde{\pi}^+\bar{U}\gamma_5 D - i\sqrt{2}g_{\rm
TC}\,\tilde{\pi}^-\bar{D}\gamma_5 U - ig_{\rm TC}\,
\tilde{\pi}^0(\bar{U}\gamma_5 U - \bar{D}\gamma_5 D)\,.
\label{L-QhSpi}
\end{eqnarray}
As was advocated above, at relatively low energies $\sim 0.1$ TeV
close to the $M_{\rm EW}$ scale the Lagrangians of the technifermion
interactions (\ref{L-QV}) and (\ref{L-QhSpi}) should be used in the
loop-induced processes with constituent quarks propagating inside
loops only.

The interactions of technipions with gauge bosons which will be used
in further calculations are defined as follows
\begin{eqnarray}
L_{\tilde{\pi}\tilde{\pi}V} &=&
 ig{W^{\mu}}^+ \cdot (\tilde{\pi}^0\tilde{\pi}_{,\mu}^- -
\tilde{\pi}^-\tilde{\pi}_{,\mu}^0) +
 ig{W^{\mu}}^- \cdot (\tilde{\pi}^+\tilde{\pi}_{,\mu}^0 -
\tilde{\pi}^0\tilde{\pi}_{,\mu}^+) \nonumber \\
 &+& ig(c_W Z_{\mu} + s_W A_{\mu})\cdot
 (\tilde{\pi}^-\tilde{\pi}_{,\mu}^+ -
 \tilde{\pi}^+\tilde{\pi}_{,\mu}^-) \nonumber \\
 &+& g^2\, W_{\mu}^+ {W^{\mu}}^- \cdot (\tilde{\pi}^0\tilde{\pi}^0
 + \tilde{\pi}^+\tilde{\pi}^-) + g^2\, (c_W Z_{\mu} + s_W A_{\mu})^2
 \cdot \tilde{\pi}^+\tilde{\pi}^- +\,...\,,  \label{L-piV}
\end{eqnarray}
where $\tilde{\pi}_{,\mu} \equiv \partial_{\mu}\tilde{\pi}$. All
triple and quartic interactions, which are necessary in calculations
of technipion contributions to the gauge bosons self-energies, are
written down here.

The Yukawa interactions $\bar{f}fh+\bar{f}f\tilde{\sigma}$ of the
ordinary fermions get modified compared to the SM
\begin{eqnarray}
L_{\bar{f}fh}+L_{\bar{f}f\tilde{\sigma}}=-g(c_{\theta}h-s_{\theta}\tilde{\sigma})\cdot
\frac{m_f}{2M_W}\bar{f}f\,. \label{Yukawa}
\end{eqnarray}

The Lagrangians of the $h\tilde{\pi}\tilde{\pi}$ and $hWW+hZZ$
interactions are
\begin{equation}
\begin{array}{c}
\displaystyle L_{h\tilde{\pi}\tilde{\pi}}=-(\lambda_{\rm
TC}u\,s_\theta-\lambda vc_\theta)\,h(\tilde{\pi}^0\tilde{\pi}^0 +
2\tilde{\pi}^+\tilde{\pi}^-)=
-\frac{M_h^2-m_{\tilde{\pi}}^2}{2M_{\tilde Q}}\,g_{\rm TC}s_\theta\,
h(\tilde{\pi}^0\tilde{\pi}^0 + 2\tilde{\pi}^+\tilde{\pi}^-)\ ,
\\[5mm]
\displaystyle L_{hWW}+L_{hZZ}=gM_W c_\theta \,hW_\mu^+{W^{\mu}}^-
+\frac12(g^2+g_1^2)^{1/2}M_Zc_\theta\, hZ_\mu Z^\mu\ .
\end{array}
 \label{L-hVVpipi}
 \end{equation}
The Lagrangians of the $\tilde{\sigma} \tilde{\pi}\tilde{\pi}$ and
$\tilde{\sigma} WW+\tilde{\sigma} ZZ$ interactions are
\begin{equation}
\begin{array}{c}
\displaystyle L_{\tilde{\sigma}
\tilde{\pi}\tilde{\pi}}=-(\lambda_{\rm TC}u c_\theta+\lambda v
s_\theta)\,\tilde{\sigma} (\tilde{\pi}^0\tilde{\pi}^0 +
2\tilde{\pi}^+\tilde{\pi}^-)= -\frac{M_{\tilde
\sigma}^2-m_{\tilde{\pi}}^2}{2M_{\tilde Q}}\,g_{\rm TC}c_\theta\,
\tilde{\sigma} (\tilde{\pi}^0\tilde{\pi}^0 +
2\tilde{\pi}^+\tilde{\pi}^-)\ ,
\\[5mm]
\displaystyle L_{\tilde{\sigma} WW}+L_{\tilde{\sigma} ZZ}=-gM_W
s_\theta\, \tilde{\sigma}
W_\mu^+{W^{\mu}}^--\frac12(g^2+g_1^2)^{1/2}M_Z s_\theta\,
\tilde{\sigma} Z_\mu Z^\mu\ .
\end{array}
 \label{L-SVVpipi}
 \end{equation}
The Lagrangian of quartic scalar-gauge $(\tilde{\sigma}/h)^2VV$
interactions is given by
\begin{eqnarray}
L_{(\tilde{\sigma}/h)^2VV}&=&
\frac{1}{4}(c_{\theta}h-s_{\theta}\tilde{\sigma})^2 \cdot
\left(g^2W_{\mu}^+{W^{\mu}}^- + \frac12
(g^2+g_1^2)Z^{\mu}Z_{\mu}\right)\,. \label{L-quart-VV}
\end{eqnarray}

\subsection{Nearly conformal limit: the minimal CSTC}

In the SM, the arbitrary quadratic terms with ``wrong'' sign in the
Higgs potential are usually required for the classical (non-quantum)
Higgs mechanism of the EWSB. As we have noticed above, in the
framework of the CSTC model there is a possibility for another
interpretation of the Higgs mechanism in which the nature of all
energy scales (including the Higgs vev) is quantum-topological. Let
us look into the latter possibility in detail.

In the rigorous QCD framework, there is not any fundamental scalar
sector and thus scalar $\mu$-terms do not appear. In the theory of
non-perturbative QCD vacuum all the scale parameters have
quantum-topological nature and are expressed through the gluon
condensate $\langle GG\rangle$ and the correlation length $l_g$,
whereas the quark condensate $\langle q\bar{q}\rangle$ is induced by
the gluon one (\ref{QCD-vac}). Clearly, low energy hadron physics
based upon the effective GL$\sigma$M should reproduce the
non-perturbative QCD predictions. On the other hand, it is
well-known that in the limit of small current quark masses $m_q\to
0$ (the chiral limit), the QCD Lagrangian restores the conformal
symmetry. Similarly, the $\sigma$-model as an effective model of
non-perturbative QCD should obey the conformal symmetry in the
chiral QCD limit. In this case, the $\mu_S$-term corresponding to
the $\sigma$ field is forbidden by the conformal symmetry. In a
realistic case, the conformal symmetry in QCD is broken due to
non-zeroth current quark masses. However, the current up- and
down-quark masses are small compared to the value of the quark
condensate $\langle q\bar{q}\rangle$ or, equivalently, the pion
mass, i.e. $m_{u,d}\ll m_{\pi}$, so it is meaningful to assume that
an induced $\mu_S$-term, if exists, should also be small $\mu_S\ll
m_{\pi}$. In this case, since $\langle GG\rangle$, $\langle
q\bar{q}\rangle$ and small current masses $m_{u,d}\ll m_{\pi}$ are
the only physical parameters in non-perturbative QCD, the $\sigma$
vev $u\sim m_{\pi}$ has quantum-topological nature, so it should be
expressed only through these parameters and given by e.g. $\langle
q\bar{q}\rangle$ or, equivalently, $m_{\pi}$. Of course, this logic
is rather naive since the $\sigma$-model does not have status of a
fundamental theory, but rather serves as an effective low-energy
phenomenological model with its own limitations and constraints.
Note, a dynamical theory of the QCD vacuum does not exists yet, and
our understanding of non-perturbative effects is very limited and
one cannot make any strong claims here.

The above line of naive arguments can be naturally extended to the
technifermion sector in confinement adopting a direct analogy
between non-perturbative QCD and techni-QCD. Looking at the
Eqs.~(\ref{shifts}) we notice that for not very large scalar
self-couplings $|\lambda|,\,|\lambda_{\rm TC}|,\,\lambda_{\rm H}\sim
0.1 - 10$ in the potential (\ref{L-U}), the technisigma vev $u$ can
be expressed through the technifermion condensate, or $m_{\tilde
\pi}$, for small $\mu_S\ll m_{\tilde \pi}$ which can be valid in the
nearly conformal limit of chiral techni-QCD $m_{U,D}\ll m_{\tilde
\pi}$ if and only if the Higgs boson vev is also small compared to
the techni-confinement scale, i.e. $\mu_{\rm H}\ll m_{\tilde \pi}$.
The latter means that both the vacua, the Higgs and technisigma
vevs, have the same quantum-topological nature and completely
determined by the technifermion condensate. This theoretically
appealing scenario would be rigorous and strictly valid in the exact
chiral techni-QCD limit with vanishing current technifermion masses
$m_{U,D}\to 0$. In the nearly-conformal limit there is a weak or no
running of the strong techni-QCD coupling. This is in accordance
with the analytic QCD (see e.g. Ref.~\cite{Shirkov:1997wi}) or other
phenomenological approaches predicting a rather slow bounded or even
``frozen'' behavior of the strong QCD coupling in the infrared
domain while non-perturbative QCD contributions are strongly
dominated over the perturbative ones in the constituent quark-meson
interactions at small $Q^2$. To this end, in the nearly-conformal
limit all the $\mu$-terms can be neglected in the Lagrangian
(\ref{L-U}) without affecting the SM Higgs mechanism itself, which
then would be triggered completely by the technifermion condensate,
giving rise to even more restricted parameter space of the model.
Let us look into this non-trivial possibility, which is simply a
particular case of the more general CSTC model described above, in
some more detail.

The solutions of the two tadpole equations (\ref{tadpole}) can then
be written w.r.t vevs as follows
 \begin{eqnarray}
u=\left(\frac{\lambda_{\rm H}}{\delta}\right)^{1/3}
\bar{g}_{\rm{TC}}^{1/3}\,, \qquad
v=\left(\frac{\xi\lambda}{\lambda_{\rm H}}\right)^{1/2}
\left(\frac{\lambda_{\rm H}}{\delta}\right)^{1/3}
\bar{g}_{\rm{TC}}^{1/3}\,, \label{u-v-min}
 \end{eqnarray}
where $\delta=\lambda_{\rm H}\lambda_{\rm{TC}} - \lambda^2$,
$\bar{g}_{\rm{TC}}=g_{\rm{TC}}|\langle\bar{\tilde{Q}}\tilde{Q}\rangle|>0$
and the sign factor $\xi={\rm
sign}(M_{\tilde{\sigma}}^2-3m_{\tilde{\pi}}^2)$ such that
$\xi\lambda\equiv |\lambda|\geq 0$ and $\lambda_H>0$ always. From
relations (\ref{u-v-min}) it follows that both vevs (and hence both
the EWSB and the chiral symmetry breaking) are induced by the
technifermion condensate since $u,\, v \sim
|\langle\bar{\tilde{Q}}\tilde{Q}\rangle|^{1/3}$. So, our choice of
the potential part of the TC Lagrangian $L_U$ (\ref{L-U}) provides
physically interesting interpretation of the Higgs vacuum condensate
as triggered by the technifermion condensate $\langle
\bar{\tilde{Q}}\tilde{Q}\rangle\not=0$ at low scales $\sim 0.1$ TeV.

It is convenient to redefine yet unknown parameters, the technisigma
vev, $u$, and $\bar{g}_{\rm{TC}}$ in terms of the Higgs vev, $v$,
and scalar self-couplings $\lambda,\,\lambda_{\rm
H},\,\lambda_{\rm{TC}}$ as follows
 \begin{eqnarray}
u=v\cdot \left(\frac{\lambda_{\rm H}}{\xi\lambda}\right)^{1/2}\,,
\qquad \bar{g}_{\rm{TC}} = v^3\left(\frac{\lambda_{\rm
H}\lambda_{\rm{TC}}}{\lambda} -
\lambda\right)\cdot\left(\frac{\lambda_{\rm
H}}{\xi\lambda}\right)^{1/2}\,. \label{uvTC}
 \end{eqnarray}
The technipion mass is given by
 \begin{eqnarray}
 m_{\tilde{\pi}}^2=v^2\left(\frac{\lambda_H\lambda_{\rm{TC}}}
 {\lambda}-\lambda\right)\,, \qquad m_{\tilde{\pi}}\sim v \,. \label{pion-mass-spec}
 \end{eqnarray}
Note, in the limit $M_{\tilde{\sigma}}\to \sqrt{3}m_{\tilde{\pi}}$,
we have $\delta\sim \lambda \to0$, whereas $\bar{g}_{\rm{TC}}\sim u
\sim M_{\tilde{Q}} \sim 1/\sqrt{|\lambda|}\to \infty$ at finite
$m_{\tilde{\pi}}$ and $v$. Also, $s_\theta\to0$ in this case, so $h$
and $\tilde{\sigma}$ do not mix (``no $h\tilde{\sigma}$-mixing''
limit). This peculiar limit physically corresponds to decoupling of
the technifermion condensate (and hence the techniconfinement scale
$\Lambda_{\rm TC}$) up to very high scales, while providing light
technipions and technisigma in the spectrum and the TC-induced EWSB
mechanism in the usual way. Of course, the formal mathematical
singularities corresponding to a very large techniconfinement scale
$\Lambda_{\rm TC}$, or equivalently, large $u$ and
$|\langle\bar{\tilde{Q}}\tilde{Q}\rangle|$ (see
Fig.~\ref{fig:fig-gTCbar-vevS} below), should be regularized by yet
unknown high-scale TC physics, and thus vicinities of these special
points are to be excluded from the current consideration.
Interestingly enough, the Higgs boson turns out to be absolutely
standard close to the singular points -- its properties are not
affected by the extra TC degrees of freedom, since corresponding new
TC-induced couplings vanish in this case at $M_{\tilde{\sigma}}\to
\sqrt{3}m_{\tilde{\pi}}$. While physically possible, this peculiar
situation, however, is not realized if one adopts the naive scaling
between the QCD and techni-QCD considered in this analysis. Absence
of any deviations from the SM in the measured Higgs boson
properties, from the point of view of the minimal CSTC discussed
here, would then mean physically that the ``no
$h\tilde{\sigma}$-mixing'' scenario is realized in Nature, but this
does not rule out the TC-induced EWSB mechanism (see below).

The mass form of the physical scalars, $h$ and $\tilde{\sigma}$
fields, can be represented by the following matrix
 \begin{eqnarray}
 M_{h\tilde{\sigma}} = \left(
      \begin{array}{ccc}
         \phantom{.} 3m_{\tilde{\pi}}^2 + 2\lambda v^2 & \phantom{...} & -2v^2\sqrt{\xi\lambda\lambda_H} \phantom{.} \\
         & \\
         -2v^2\sqrt{\xi\lambda\lambda_H}      &                 & 2\lambda_H v^2
      \end{array}
             \right)\,. \label{mass-matr}
 \end{eqnarray}
The diagonalization of this matrix leads to masses of the physical
states scalar states, i.e.
 \begin{eqnarray}
M_{\tilde{\sigma},\,h}^2=\frac12 v^2 \left\{
\Big(2\lambda_H+2\lambda+3\frac{m_{\tilde{\pi}}^2}{v^2}\Big) \pm
\sqrt{\Big(2\lambda_H+2\lambda+3\frac{m_{\tilde{\pi}}^2}{v^2}\Big)^2+16\lambda\lambda_H}\right\}
\label{scmasses-spec}
 \end{eqnarray}
Then, the $h\tilde{\sigma}$-mixing angle is given by
\begin{eqnarray}
 c_{\theta} =
 \left(1+\frac{(M_{\tilde{\sigma}}^2-m_{11})^2}{m_{12}^2}\right)^{-1/2}, \qquad
 s_{\theta}=\xi\sqrt{1-c_{\theta}^2}\,,
\end{eqnarray}
where $m_{11}=(M_{h\tilde{\sigma}})_{11}$,
$m_{12}=(M_{h\tilde{\sigma}})_{12}$ are the elements of the mass
matrix (\ref{mass-matr}). In analysis of the parameter space it is
again convenient to express free scalar self-couplings
$\{\lambda,\,\lambda_{\rm H},\,\lambda_{\rm{TC}}\}$ through the
physical masses $\{m_{\tilde{\pi}}^2,\, M_{\tilde{\sigma}}^2,\,
M_h^2\}$:
\begin{eqnarray}
\lambda=\frac{3m_{\tilde{\pi}}^2(M_{\tilde{\sigma}}^2+M_h^2)-
M_{\tilde{\sigma}}^2M_h^2-9m_{\tilde{\pi}}^4}{6v^2m_{\tilde{\pi}}^2}\,,\quad
\lambda_{\rm
H}=\frac{M_{\tilde{\sigma}}^2M_h^2}{6v^2m_{\tilde{\pi}}^2}\,,\quad
\lambda_{\rm TC}=\frac{\lambda}{\lambda_{\rm
H}}\left(\lambda+\frac{m_{\tilde{\pi}}^2}{v^2}\right)\,.
\label{spec-lam}
\end{eqnarray}
By fixing the Higgs boson mass to its recently measured value
$M_h\simeq 125$ GeV, one further reduces the freedom down to three
free parameters only, $\{m_{\tilde{\pi}},\, M_{\tilde{\sigma}},\,
M_{\tilde{Q}}\}$, compared to five parameters in the non-minimal
case (cf. Section II.C). Note, the scalar self-couplings and the
mixing angle $\theta$ depend only on two parameters
$\{m_{\tilde{\pi}},\, M_{\tilde{\sigma}}\}$, whereas $M_{\tilde{Q}}$
can be used to define $g_{\rm TC}$ or $\langle
\bar{\tilde{Q}}\tilde{Q}\rangle$.

\begin{figure*}[!h]
\begin{minipage}{0.4\textwidth}
 \centerline{\includegraphics[width=1.0\textwidth]{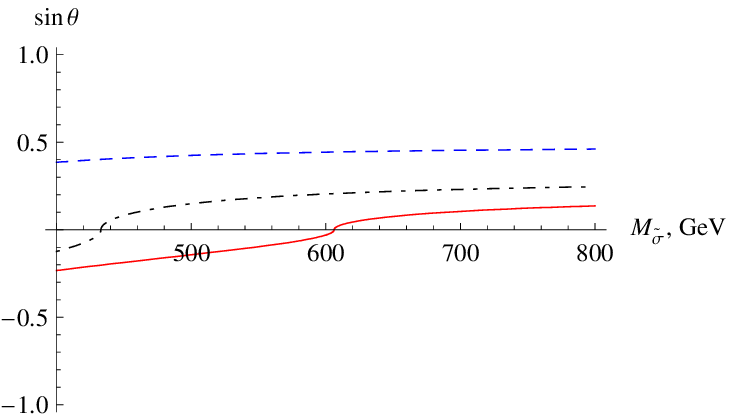}}
\end{minipage}
\begin{minipage}{0.41\textwidth}
 \centerline{\includegraphics[width=1.0\textwidth]{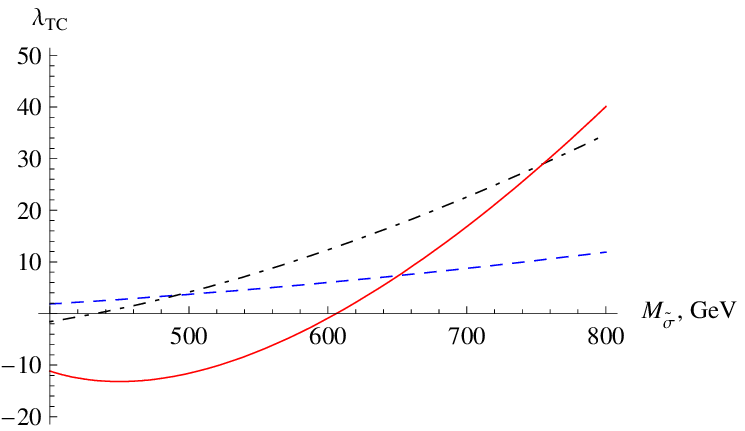}}
\end{minipage}
\begin{minipage}{0.4\textwidth}
 \centerline{\includegraphics[width=1.0\textwidth]{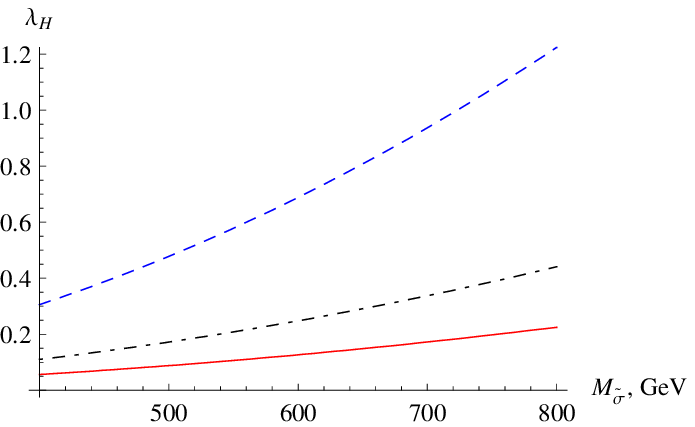}}
\end{minipage}
\begin{minipage}{0.4\textwidth}
 \centerline{\includegraphics[width=1.0\textwidth]{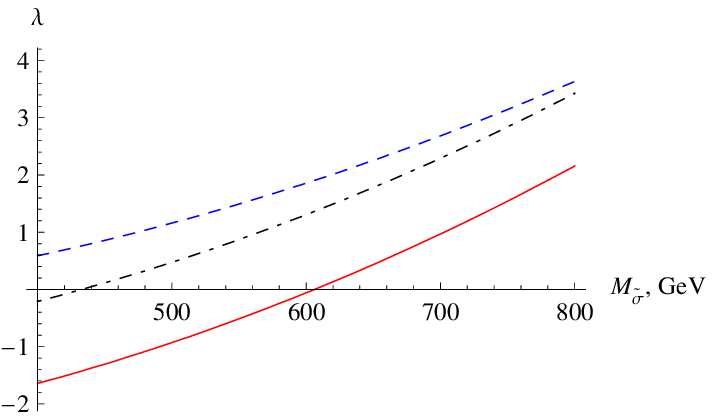}}
\end{minipage}
   \caption{
\small Dependence of the $h\tilde{\sigma}$-mixing $s_\theta$ and the
Higgs/TC self-couplings, $\lambda_{\rm TC},\,\lambda_{\rm
H}\,,\lambda$, on $M_{\tilde{Q}}$ in the minimal CSTC scenario with
dashed, dash-dotted and solid lines corresponding to
$m_{\tilde{\pi}}=150,250,350$ GeV, respectively. The ``no
$h\tilde{\sigma}$-mixing'' limit corresponds to zeros of the curves
at $M_{\tilde{\sigma}}=\sqrt{3}m_{\tilde{\pi}}$.}
 \label{fig:fig-spec}
\end{figure*}

In Fig.~\ref{fig:fig-spec} we have presented plots of sine of the
mixing angle $s_{\theta}=s_{\theta}(m_{\tilde{\pi}},\,
m_{\tilde{\sigma}})$, and scalar self-couplings --
Higgs-(pseudo)scalar coupling $\lambda=\lambda(m_{\tilde{\pi}},\,
m_{\tilde{\sigma}})$, quartic Higgs self-coupling $\lambda_{\rm
H}=\lambda_{\rm H}(m_{\tilde{\pi}},\, m_{\tilde{\sigma}})$ and
(pseudo)scalar self coupling $\lambda_{\rm TC}=\lambda_{\rm
TC}(m_{\tilde{\pi}},\, m_{\tilde{\sigma}})$. At relatively large
technipion masses $m_{\tilde{\pi}}\gtrsim 250$ GeV the
$h\tilde{\sigma}$-mixing becomes rather small, $s_{\theta}\lesssim
0.2$, while it does not strongly depend on the technisigma mass,
away from ``no-mixing'' points. As was noticed above, the condition
$\lambda=0$ (or $s_{\theta}=0$) corresponds to ``no-mixing'' limit
and is represented by a relation on masses,
$M_{\tilde{\sigma}}=\sqrt{3}m_{\tilde{\pi}}$. In the considering
ranges of masses, the values of $\lambda$ and $\lambda_{\rm H}$ do
not exceed a few units, so they are of the order of strong (``fat'')
couplings in usual hadron dynamics (e.g. $g_{\rho\pi\pi}\sim 5-6$)
and gradually increase at large $M_{\tilde{\sigma}}$. The
(pseudo)scalar self-coupling $\lambda_{\rm TC}$ can reach larger
values $\sim 100$ at large values of $M_{\tilde{\sigma}}\gtrsim
800\,{\rm GeV}$ restricting the allowable region of physical
parameters and applicability of the GLT$\sigma$M under
consideration. An experimental information on the scalar
self-couplings $\lambda,\,\lambda_{\rm TC}$ would shed light on the
true origin of the Higgs mechanism making it possible to determined
which minimal or non-minimal CSTC scenario is realized in Nature.
\begin{figure*}[!h]
\begin{minipage}{0.4\textwidth}
 \centerline{\includegraphics[width=1.0\textwidth]{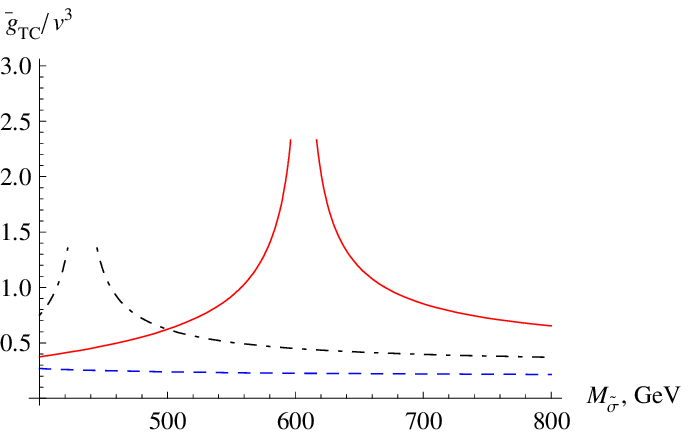}}
\end{minipage}
\begin{minipage}{0.41\textwidth}
 \centerline{\includegraphics[width=1.0\textwidth]{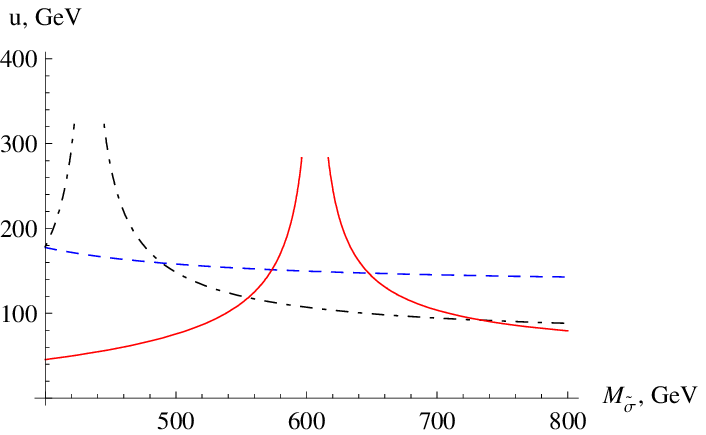}}
\end{minipage}
   \caption{
\small The dimensionless $\bar{g}_{\rm TC}/v^3$ (left) and
$\tilde{\sigma}$ vev $u$ (right) parameters with respect to
$M_{\tilde{\sigma}}$ in the minimal CSTC scenario with dashed,
dash-dotted and solid lines corresponding to
$m_{\tilde{\pi}}=150,250,350$ GeV, respectively. The ``no
$h\tilde{\sigma}$-mixing'' limit corresponds to positions of the
singularities on the curves at
$M_{\tilde{\sigma}}=\sqrt{3}m_{\tilde{\pi}}$, vicinity of those are
excluded from the plots.}
 \label{fig:fig-gTCbar-vevS}
\end{figure*}

In Fig.~\ref{fig:fig-gTCbar-vevS} we show the dimensionless
$\bar{g}_{\rm TC}/v^3$ (left) and $\tilde{\sigma}$ vev $u$ (right)
parameters with respect to $M_{\tilde{\sigma}}$ for different values
of $m_{\tilde \pi}$ in the minimal CSTC scenario. The technisigma
vev $u=u(m_{\tilde{\pi}},\, m_{\tilde{\sigma}})$, can be smaller
than the Higgs vev $v$,  $u\lesssim v$, almost in all physically
favorable regions of parameter space where $m_{\tilde{\pi}} \sim v$,
except for vicinities of ``no $h{\tilde \sigma}$-mixing'' points
$M_{\tilde{\sigma}}\simeq\sqrt{3}m_{\tilde{\pi}}$ where $u$ can be
larger or even much larger than the Higgs vev $v$. The latter case
can be interesting both theoretically and phenomenologically in case
of absence of any deviations of Higgs boson properties from the SM
predictions at the LHC. Then, the only source of new information
about the TC sector can only come from measurements of the Higgs
boson scalar self-couplings and possible technipion/technisigma
phenomenology.

One should notice here that if the small $h\tilde\sigma$-mixing
scenario with $s_\theta\to0$ and $\lambda\to0$ is realized in
Nature, we have the Technicolor decoupling regime with large $u\gg
v$ and hence $\Lambda_{\rm TC}\sim M_{\tilde Q}\gg M_{\rm EW}$,
while the Higgs boson, technipions and technisigma remain at the EW
scale according to the tree-level mass formulae of the model.
Remarkably enough, the Higgs vev is still expressed in terms of the
technifermion condensate by Eq.~(\ref{u-v-min}) for vanishingly
small but non-zeroth $\lambda\not=0$ preserving the dynamical nature
of the EWSB (or effective Higgs) mechanism.

\section{Electro-weak constraints on the CSTC}

\subsection{Oblique corrections}

The effects of heavy New Physics (NP) particles of various types
onto $Z^0$ and $W^\pm$ observables (e.g. masses, widths) typically
emerging through extra loop contributions to $Z^0$, $W^\pm$ and
$\gamma$ (diagonal and non-diagonal) self-energies can be
parameterized by means of the so-called oblique corrections or
Peskin-Takeuchi (PT) parameters \cite{Peskin:1990zt}. The first
three of these parameters $S,\,T,\,U$ are normally introduced in the
limiting case of large masses of new particles compared to the EW
scale, i.e. in the limit $M_{\rm EW}/M_{\rm NP}\ll 1$ ($M_{\rm NP}$
is the NP scale). If one relaxes this assumption, the $S,\,T,\,U$
parameters get somewhat modified, and additional three independent
parameters denoted as $V,\,W,\,X$ are introduced (see e.g.
Refs.~\cite{Maksymyk:1993zm,Burgess:1993mg}). The oblique
corrections are rather strongly constrained by the EW precision
measurements \cite{PDG12}
\begin{eqnarray} \label{STU-constr}
 S=0.00^{+0.11}_{-0.10},\quad T=0.02^{+0.11}_{-0.12},\quad U=0.08\pm 0.11
\end{eqnarray}
and must be respected by realistic NP models (for existing
constraints on higher $V,\,W,\,X$ parameters see e.g.
Ref.~\cite{Barbieri}). The extensive studies of these constraints
are very critical for all existing TC models. In particular, some of
the traditional TC scenarios are currently being ruled out or are in
a considerable tension with constraints on $S,\,T,\,U$ parameters
\cite{Peskin:1990zt} (see also Refs.~\cite{Hill:2002ap,Sannino}).
Let us analyze these constraints in the suggested CSTC scenario.

The analysis we present further in this Section does not depend on
whether one includes $\mu_{S,H}$-terms or not; the difference
between these non-minimal and minimal versions of the CSTC model can
only be crucial for processes with (pseudo)-scalar self-couplings,
which can be important e.g. for Higgs and technipion phenomenology.

In the earlier Sections, we have established the phenomenologically
reasonable intervals for masses and couplings of new TC particles
(technipions, technisigma and constituent technifermions) based on
analogies with ordinary QCD and hadron physics together with the
relative proximity of the new TC scale $\Lambda_{\rm TC}\sim 0.1-1$
TeV. In what follows, these regions of parameter space will be
tested against the EW precision constraints given by
Eq.~(\ref{STU-constr}).

The generic definitions of the PT parameters are given in terms of
corrections to the vacuum polarization functions $\delta\Pi_{\rm
XY}(q^2)$ of the gauge bosons ($\rm{X,\,Y}=W,\,Z,\,\gamma$) coming
either from new particles, additional to those in the SM, $\Pi_{\rm
XY}^{\rm new}(q^2)$, or from a modification of the SM parameters due
to NP effects, $\Pi_{\rm XY}^{\rm SM'}(q^2)$, i.e.
\begin{eqnarray}
\delta\Pi_{\rm XY}(q^2)\equiv \Pi_{\rm XY}^{\rm NP}(q^2)-\Pi_{\rm
XY}^{\rm SM}(q^2)\,,\qquad \Pi_{\rm XY}^{\rm NP}(q^2)=\Pi_{\rm
XY}^{\rm new}(q^2)+\Pi_{\rm XY}^{\rm SM'}(q^2)\,. \label{deltaPi}
\end{eqnarray}

The expressions for the $S,\,T,\,U$ parameters in terms of generic
polarisation functions $\delta\Pi_{\rm XY}(q^2)$ and their
derivatives $\delta\Pi'_{\rm XY}(q^2)=d\delta\Pi/dq^2$ calculated
beyond the linear approximation in $q^2$ variable read
\cite{Maksymyk:1993zm,Burgess:1993mg}
\begin{eqnarray}
\frac{\alpha}{4s_W^2c_W^2}\,S &=&
\frac{\delta\Pi_{ZZ}(M_Z^2)-\delta\Pi_{ZZ}(0)}{M_Z^2} -
\frac{c_W^2-s_W^2}{c_Ws_W}\delta\Pi'_{Z\gamma}(0)-
\delta\Pi'_{\gamma\gamma}(0)\,, \nonumber \\
\alpha\,T &=& \frac{\delta\Pi_{WW}(0)}{M_W^2} -
\frac{\delta\Pi_{ZZ}(0)}{M_Z^2}\,, \label{STU} \\
\frac{\alpha}{4s_W^2}\,U &=&
\frac{\delta\Pi_{WW}(M_W^2)-\delta\Pi_{WW}(0)}{M_W^2}
-c_W^2\,\frac{\delta\Pi_{ZZ}(M_Z^2)-\delta\Pi_{ZZ}(0)}{M_Z^2} \nonumber  \\
&&\qquad -\, s_W^2 \delta\Pi'_{\gamma\gamma}(0) - 2c_Ws_W
\delta\Pi'_{Z\gamma}(0)\,. \nonumber
\end{eqnarray}
Note, in the limit $\zeta=M_{\rm EW}/\Lambda_{\rm TC}\ll 1$ we have
\begin{eqnarray}
\frac{\delta\Pi_{WW}(M_Z^2)-\delta\Pi_{WW}(0)}{M_Z^2}&=&
\frac{\delta\Pi_{WW}(M_W^2)-\delta\Pi_{WW}(0)}{M_W^2}+{\cal
O}(M_{\rm EW}^4/\Lambda_{\rm TC}^4)\,,\\
\frac{\delta\Pi_{\rm XY}(q^2)-\delta\Pi_{\rm
XY}(0)}{q^2}&=&\delta\Pi'_{\rm XY}(0)+{\cal O}(q^4/\Lambda_{\rm
TC}^4)\,, \label{linq2}
\end{eqnarray}
which are equivalent to working in the linear order in $q^2$ in
power expansions of the polarisation functions $\delta\Pi_{\rm
XY}(q^2)$. In fact, applying approximate relation (\ref{linq2}) to
expressions (\ref{STU}) at $q^2=M_Z^2$ and having in mind that
$\delta\Pi_{Z\gamma}(0)=\delta\Pi_{\gamma\gamma}(0)=0$ in a
realistic case, one arrives at the Particle Data Group formulas (see
Eq.~(10.65b,c) in Ref.~\cite{PDG12}). We, however, do not assume
smallness of $\zeta$ in calculations (unless noted otherwise) since
the new TC scale $\Lambda_{\rm TC}$ can be rather close to the
electroweak scale $M_{\rm EW}$ since they may have the same physical
nature in the considering CSTC scenario, and therefore rigorous
definitions (\ref{STU}) should be applied.

Other three parameters which appear beyond the linear order in $q^2$
in addition to the $S,\,T,\,U$ are defined as follows
\cite{Maksymyk:1993zm,Burgess:1993mg}
\begin{eqnarray}\nonumber
\alpha V &=& \delta\Pi'_{ZZ}(M_Z^2)-\frac{\delta\Pi_{ZZ}(M_Z^2)-
\delta\Pi_{ZZ}(0)}{M_Z^2}\,,\\
\alpha W &=& \delta\Pi'_{WW}(M_W^2)-\frac{\delta\Pi_{WW}(M_W^2)-
\delta\Pi_{WW}(0)}{M_W^2}\,, \label{VWX} \\
\alpha X &=&
-s_Wc_W\,\Bigl[\frac{\delta\Pi_{Z\gamma}(M_Z^2)}{M_Z^2}-
\delta\Pi'_{Z\gamma}(0)\Bigr]\,. \nonumber
\end{eqnarray}

\begin{figure*}[!h]
\begin{minipage}{0.7\textwidth}
 \centerline{\includegraphics[width=1.0\textwidth]{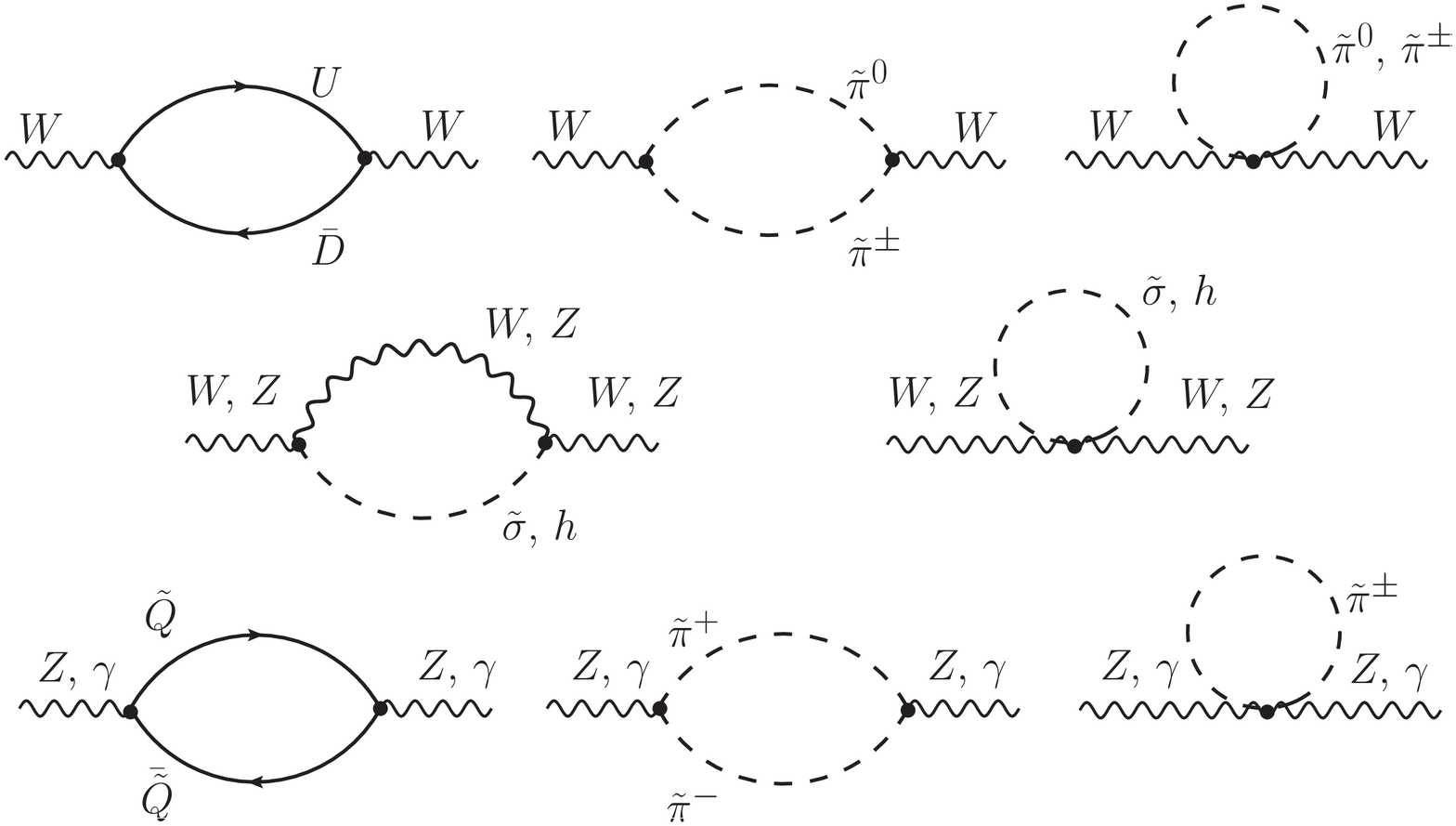}}
\end{minipage}
   \caption{
\small The additional new (via $\tilde{\pi}$, $\tilde{Q}$ and
$\tilde{\sigma}$) and modified (via Higgs boson $h$) contributions
to the gauge bosons $Z^0,\,W^\pm$ and $\gamma$ vacuum polarisation
functions $\delta\Pi_{\rm XY}(q^2)$.}
 \label{fig:fig-pols}
\end{figure*}

In the framework of the CSTC model, the new contributions to $W,\,Z$
and $\gamma$ vacuum polarizations come from technipion, constituent
technifermions and technisigma loops, i.e.
\begin{eqnarray}
\Pi_{\rm XY}^{\rm new}(q^2)=\Pi_{\rm XY}^{\tilde{\pi}}(q^2)+\Pi_{\rm
XY}^{\tilde{Q}}(q^2)+\Pi_{\rm XY}^{\tilde{\sigma}}(q^2)
\end{eqnarray}
while the SM modified contributions come only from the Higgs boson
due to modified $hVV$ couplings, $\Pi_{\rm XY}^{h}(q^2)$, whereas
other SM couplings are not changed in the CSTC model, thus we have
\begin{eqnarray}
\Pi_{\rm XY}^{\rm SM'}(q^2)=\Pi_{\rm XY}^{h}(q^2)\,.
\end{eqnarray}
The corresponding diagrams are presented in Fig.~\ref{fig:fig-pols}.

Note that the modified Higgs contribution to the gauge bosons
polarisation functions $\Pi_{\rm XY}^{h}(q^2,M_h^2)$ can be obtained
by multiplying the corresponding SM result presented many times in
the literature (see e.g. Ref.~\cite{Kniehl:1993ay}), $\Pi_{\rm
XY}^{\rm{SM},h}(q^2,M_h^2)$, by a factor of $c_\theta^2$. Also, the
extra contribution due to $\tilde{\sigma}$ meson, $\Pi_{\rm
XY}^{\tilde{\sigma}}(q^2,M_{\tilde{\sigma}}^2)$, can be easily
obtained from the Higgs boson one, $\Pi_{\rm XY}^{h}(q^2,M_h^2)$, by
a replacement $c_\theta\to s_\theta$ and $M_h\to M_{\tilde{\sigma}}$
in corresponding polarisation functions (cf. Eqs.~(\ref{L-hVVpipi})
and (\ref{L-SVVpipi})). Therefore, the total contribution of the
scalar states $\delta\Pi_{\rm XY}^{\rm sc}(q^2)$ to the total
$\delta\Pi_{\rm XY}(q^2)$ defined in Eq.~(\ref{deltaPi}) reads
\begin{eqnarray} \label{deltaPi-1}
\delta\Pi_{\rm XY}(q^2)&=&\delta\Pi_{\rm XY}^{\rm sc}(q^2)+\Pi_{\rm
XY}^{\tilde{\pi}}(q^2,m_{\tilde{\pi}}^2)+\Pi_{\rm XY}^{\tilde{Q}}(q^2,M_{\tilde{Q}}^2)\,,\\
\delta\Pi_{\rm XY}^{sc}(q^2)&=&\Pi_{\rm
XY}^{\tilde{\sigma}}(q^2,M_{\tilde{\sigma}}^2)+\Pi_{\rm
XY}^{h}(q^2,M_h^2)-\Pi_{\rm XY}^{SM,h}(q^2,M_h^2) \nonumber \\
&=& s_\theta^2\,\Pi_{\rm
XY}^{\rm{SM},h}(q^2,M_{\tilde{\sigma}}^2)-s_\theta^2\,\Pi_{\rm
XY}^{\rm{SM},h}(q^2,M_h^2)\,. \label{scPi}
\end{eqnarray}
Apparently, $\delta\Pi_{\rm XY}^{sc}(q^2)\to 0$ and hence the
corresponding contributions to the oblique corrections (\ref{STU})
and (\ref{VWX}) turn to zero in the limit of degenerated
$\tilde{\sigma}$ and $h$ masses, $M_{\tilde{\sigma}}\to M_h$. Also,
the function $\delta\Pi_{\rm XY}^{sc}(q^2)$ vanish in the ``no
$\tilde{\sigma}h$-mixing'' limit, when $s_\theta\to0$, so the
corresponding oblique corrections can be very small and fit the EW
precision data without a significant tension.

Finally, consider the new contributions coming from $\tilde{\pi}$
and $\tilde{Q}$ loops. For illustration, below we show analytical
results for the limiting ``no $h\tilde{\sigma}$-mixing'' scenario
and degenerated technifermions implying that their constituent
masses are equal $M_U=M_D\equiv M_{\tilde{Q}}$, while forthcoming
numerical results and figures will be presented also for the general
case with $M_U\not=M_D$ and arbitrary mixing angle. Note that if one
employs an analogy with hadron physics, where the non-perturbative
QCD contribution to the constituent masses of up and down quarks is
much larger than their current masses, the approximate degeneracy
$M_U\simeq M_D$ (or, more precisely, $\Delta M_{\tilde Q}\equiv
M_D-M_U\ll M_U,\,M_D$) is physically reasonable and justified.
\begin{figure*}[!h]
\begin{minipage}{0.45\textwidth}
 \centerline{\includegraphics[width=1.0\textwidth]{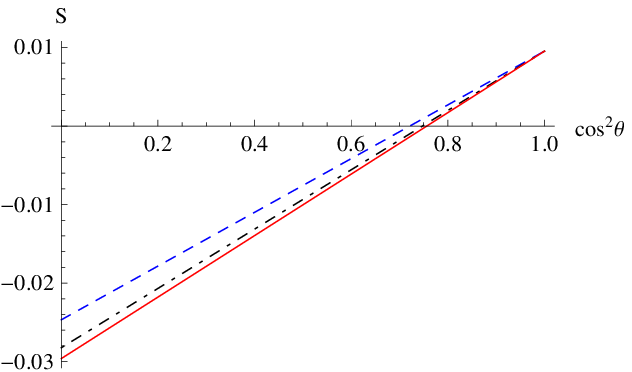}}
\end{minipage}
\hspace{9mm}\vspace{5mm}
\begin{minipage}{0.45\textwidth}
 \centerline{\includegraphics[width=1.0\textwidth]{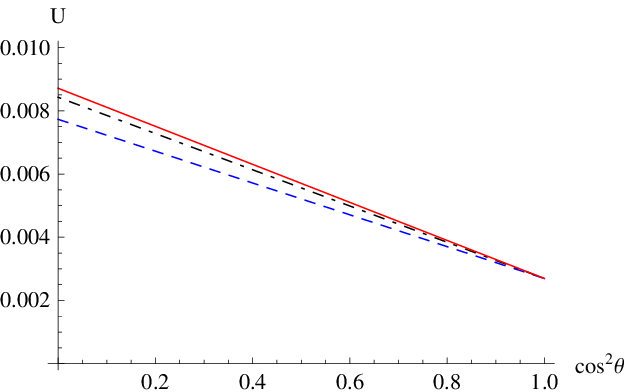}}
\end{minipage}
\begin{minipage}{0.45\textwidth}
 \centerline{\includegraphics[width=1.0\textwidth]{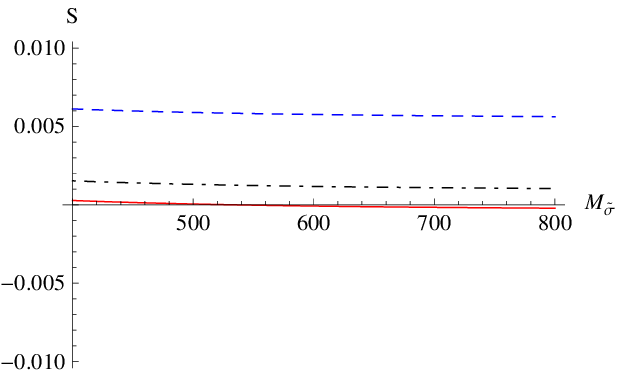}}
\end{minipage}
\hspace{9mm}\vspace{5mm}
\begin{minipage}{0.45\textwidth}
 \centerline{\includegraphics[width=1.0\textwidth]{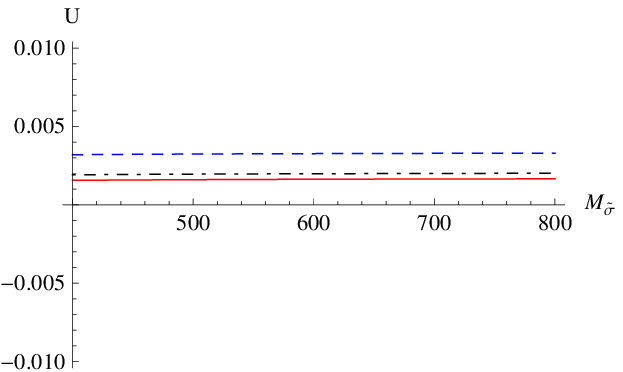}}
\end{minipage}
\begin{minipage}{0.45\textwidth}
 \centerline{\includegraphics[width=1.0\textwidth]{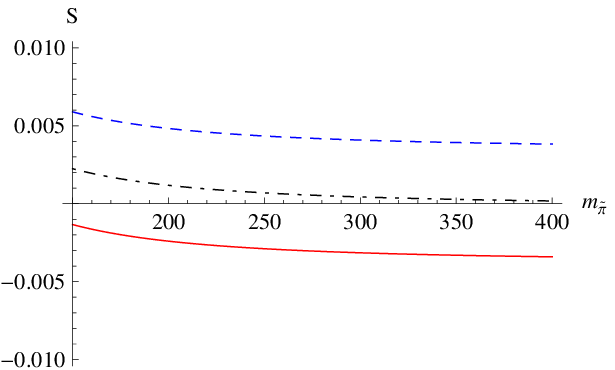}}
\end{minipage}
\hspace{9mm}
\begin{minipage}{0.46\textwidth}
 \centerline{\includegraphics[width=1.0\textwidth]{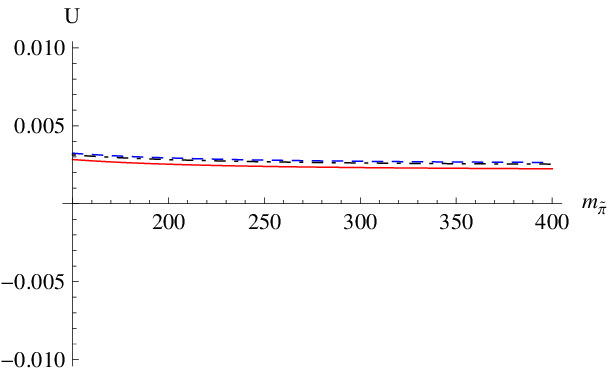}}
\end{minipage}
   \caption{\small The complete $S$ and $U$ parameters in the CSTC scenario
   (the non-minimal case with $\mu_S$ and $\mu_H$ included) as functions of
   (1) $\cos^2\theta$ for fixed $m_{\tilde \pi}=150$ GeV, $M_{\tilde Q}\equiv M_U=M_D=300$ GeV,
   and $M_{\tilde \sigma}=400,600,800$ GeV, corresponding to dashed,
   dash-dotted and solid lines, respectively (first row);
   (2) $M_{\tilde \sigma}$ for fixed $m_{\tilde \pi}=150$ GeV,
   $\cos^2\theta=0.9$, and $M_{\tilde Q}=300,500,700$ GeV, corresponding
   to dashed, dash-dotted and solid lines, respectively (second row);
   and (3) $m_{\tilde \pi}$ for fixed $M_{\tilde \sigma}=500$ GeV,
   $\cos^2\theta=0.9$, $M_{\tilde Q}=300$ GeV and $\Delta M_{\tilde Q}\equiv
   M_D-M_U=0,5,10$ GeV, corresponding to dashed, dash-dotted and
   solid lines, respectively (third row). Also, here and for other PT parameters below
   the sine of the mixing angle due to symmetry is chosen to be
   positive,
   $s_\theta>0$.}
 \label{fig:fig-PT-SU}
\end{figure*}

Then, the technipion and technifermion contributions can be
represented in the following generic form:
\begin{eqnarray}
 \Pi_{\rm XY}^{\tilde{\pi}}(q^2,m_{\tilde{\pi}}^2)=\frac{g^2}{24\pi^2}\,
 K_{\rm XY}\, F_{\tilde{\pi}}(q^2,m_{\tilde{\pi}}^2)\,, \quad
 \Pi_{\rm XY}^{\tilde{Q}}(q^2,M_{\tilde{Q}}^2)=\frac{g^2N_c}{24\pi^2}\,
 K_{\rm XY}\,\kappa_{\rm XY}\,F_{\tilde{Q}}(q^2,M_{\tilde{Q}}^2)\,. \label{Pi-TQpi-XY}
\end{eqnarray}
where $N_{\rm TC}=3$ is the number of technicolors, coefficients
$K_{\rm XY}$ and $\kappa_{\rm XY}$ are shown for two different cases
with $Y_{\tilde Q}=0$ and $Y_{\tilde Q}=1/3$ in
Table~\ref{tab-coeff}, and momentum-dependent parts are defined as
\begin{eqnarray*}
&&F_{\tilde{\pi}}(q^2,m_{\tilde{\pi}}^2)=\frac13 q^2 -
2m_{\tilde{\pi}}^2+2A_0(m_{\tilde{\pi}}^2)+\frac12
(q^2-4m_{\tilde{\pi}}^2)B_0(q^2,m_{\tilde{\pi}}^2,m_{\tilde{\pi}}^2)\,,\\
&&F_{\tilde{Q}}(q^2,M_{\tilde{Q}}^2)=-\frac13 q^2 +
2M_{\tilde{Q}}^2-2A_0(M_{\tilde{Q}}^2)+
(q^2+2M_{\tilde{Q}}^2)B_0(q^2,M_{\tilde{Q}}^2,M_{\tilde{Q}}^2)\,,
\end{eqnarray*}
where $A_0(m^2)$ and $B_0(q^2,m^2,m^2)$ are the standard one- and
two-point functions \cite{PV}, respectively. Furthermore, one
evaluates these functions and their derivatives for a given set of
arguments and substitutes them into Eq.~(\ref{Pi-TQpi-XY}) and then
to Eq.~(\ref{deltaPi-1}).
\begin{table}[h!]
\caption{\small\sf Summary of coefficients $K_{\rm XY}$ and
$\kappa_{\rm XY}$ in gauge bosons self-energies ${\rm
X,Y}=Z^0,\,W^\pm,\,\gamma$ coming from $\tilde{\pi}$ and $\tilde{Q}$
loops (\ref{Pi-TQpi-XY}). Two different cases for technifermion
hypercharges are considered.}
\begin{center}\label{tab-coeff}
\begin{tabular}{ |c|c|c|c|c| }
\hline   $\quad K,\,\kappa\quad$ & $\quad WW\quad$ & $\quad ZZ\quad$ & $\quad \gamma\gamma\quad$ & $\quad Z\gamma\quad$ \\
 \hline\hline
 $K_{\rm XY}$   & $1$  & $c_W^2$ & $s_W^2$ & $c_Ws_W$  \\
\hline
                 $\kappa_{\rm XY},\,Y_{\tilde{Q}}=0$   & $1$  & $1$     & $1$     & $1$       \\ \hline
                 $\kappa_{\rm XY},\,Y_{\tilde{Q}}=1/3$ & $1$  & $1+s_W^4/9c_W^4$  & $10/9$  & $1-s_W^2/9c_W^2$  \\ \hline
\end{tabular}
\end{center}
\end{table}
Using the relations
\begin{eqnarray*}
B_0(0,m^2,m^2)=\frac{A_0(m^2)}{m^2}-1\,,\qquad
A_0(m^2)=m^2\Bigl(\frac{1}{\bar{\varepsilon}}+1-\ln\frac{m^2}{\mu^2}\Bigr)\,,
\end{eqnarray*}
it can be checked directly that
$F_{\tilde{\pi}}(0,m_{\tilde{\pi}}^2)=0$ and
$F_{\tilde{Q}}(0,M_{\tilde{Q}}^2)=0$ which means that technipions
and degenerated technifermions do not contribute to the
$T$-parameter, i.e. $T^{\tilde{\pi}}=T^{\tilde{Q}}=0$ automatically.
The only contribution to the $T$-parameter comes from the scalar
sector of the theory: $\tilde{\sigma}$ loops and modified Higgs
loops given by Eq.~(\ref{scPi}).

The $S$ and $U$ parameters calculation becomes especially
transparent if one works in the linear order in $q^2$ power
expansion and applies an approximate relation (\ref{linq2}). For
this purpose, let us consider the simplest case of degenerated
technifermion sector with $Y_{\tilde{Q}}=0$. Then, having $\Pi_{\rm
XY}^{\tilde{\pi}}(0,m_{\tilde{\pi}}^2)=0$ and $\Pi_{\rm
XY}^{\tilde{Q}}(0,m_{\tilde{Q}}^2)=0$ for any ${\rm X,Y}$ we observe
that the $\tilde{\pi}$ and $\tilde{Q}$ contributions to $S$ and $U$
parameters also vanish for $Y_{\tilde{Q}}=0$ in the linear order in
$q^2$. Indeed, using the corresponding $K_{\rm XY}$ and $\kappa_{\rm
XY}$ coefficients from Table~\ref{tab-coeff}, we explicitly see that
\begin{eqnarray*}
\frac{\alpha\,S^{\tilde{\pi}+\tilde{Q}}}{4s_W^2c_W^2}&=&
f(M_Z^2,m_{\tilde{\pi}}^2,M_{\tilde{Q}}^2)
\cdot\left[c_W^2-\frac{c_W^2-s_W^2}{c_Ws_W}\cdot
c_Ws_W-s_W^2\right]=0,\\
\frac{\alpha\,U^{\tilde{\pi}+\tilde{Q}}}{4s_W^2}
&=&f(M_Z^2,m_{\tilde{\pi}}^2,M_{\tilde{Q}}^2)
\cdot\left[1-c_W^4-s_W^4-2c_W^2s_W^2\right]=0\,,
\end{eqnarray*}
where $f(m_1^2,m_2^2,m_3^2)$ is some finite regular function of the
respective mass scales.
\begin{figure*}[!h]
\begin{minipage}{0.43\textwidth}
 \centerline{\includegraphics[width=1.0\textwidth]{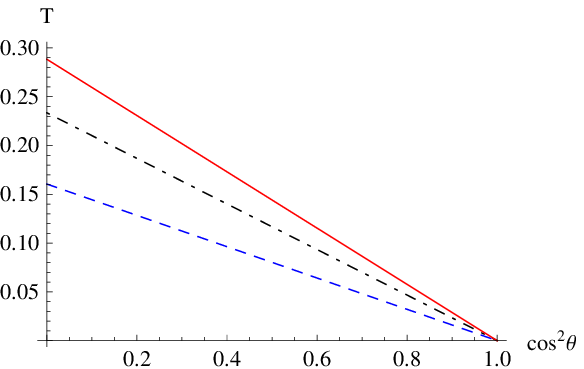}}
\end{minipage}
\hspace{9mm}\vspace{5mm}
\begin{minipage}{0.43\textwidth}
 \centerline{\includegraphics[width=1.0\textwidth]{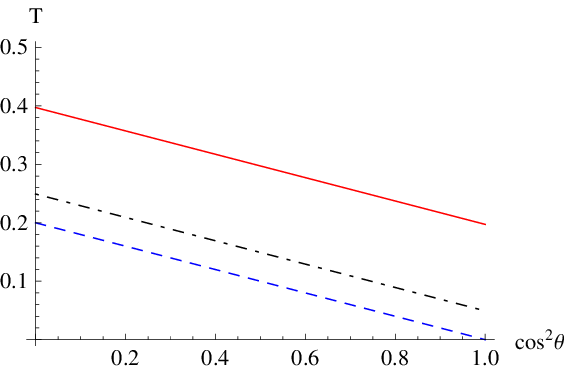}}
\end{minipage}
   \caption{
\small  The complete $T$ parameter in the CSTC scenario
   (the non-minimal case with $\mu_S$ and $\mu_H$ included) as function of
   $\cos^2\theta$ for two different cases: (1) $\Delta M_{\tilde
   Q}=0$ and
   $M_{\tilde \sigma}=400,600,800$ GeV, corresponding to dashed,
   dash-dotted and solid lines, respectively (left panel); (2)
   $M_{\tilde \sigma}=500$ GeV,
   and $\Delta M_{\tilde Q}\equiv M_D-M_U=0,5,10$ GeV,
   corresponding to dashed, dash-dotted and solid lines,
   respectively (right panel). The $T$ parameter does not depend
   on degenerated $M_{\tilde Q}\equiv M_U=M_D$ mass and $m_{\tilde \pi}$.}
 \label{fig:fig-PT-T}
\end{figure*}
We summarize that the only contribution to the $S,\,T,\,U$
parameters (in the simplest scenario with $Y_{\tilde{Q}}=0$ and in
the linear order in $q^2$) comes from scalar loops given by
Eq.~(\ref{scPi}). This result is different from traditional TC-based
scenarios with chiral-nonsymmetric weak interactions, where
$S$-parameters does not vanish and is equal to a relatively large
constant, even in the limit of infinitely heavy technifermions
\cite{Peskin:1990zt}. In the considering CSTC scenario this problem
does not appear at all.

The calculations in more elaborated case with the SM-like
technifermion hypercharge $Y_{\tilde{Q}}=1/3$ are less transparent
and more cumbersome. Remarkably enough, in this case
$S^{\tilde{\pi}+\tilde{Q}}$ and $U^{\tilde{\pi}+\tilde{Q}}$ are not
zeroth any longer, but still strongly suppressed. Since in the
first, linear, order in $q^2$ power expansion technipions and
technifermions do not contribute or contribute very little, it is
worth to go beyond this approximation, also incorporating
$V,\,W,\,X$ parameters into the analysis. Keeping the degeneracy
condition $\Delta M_{\tilde Q}=0$, we have
$T^{\tilde{\pi}+\tilde{Q}}=0$, as shown above, and other parameters
read
\begin{eqnarray*}
S^{\tilde{\pi}+\tilde{Q}}&=&\frac{2c_W^4}{3\pi}\Bigl\{\frac13-
\beta_Z^{\tilde{\pi}}(1-\phi_Z^{\tilde{\pi}})+N_{\rm
TC}\Bigl(1+\frac{s_W^4}{9c_W^4}\Bigr)
\Bigl[-\frac13+(3+\beta_Z^{\tilde{Q}})(1-\phi_Z^{\tilde{Q}})\Bigr]\Bigr\}\,,\\
U^{\tilde{\pi}+\tilde{Q}}&=&\frac{2}{3\pi}\Bigl\{\frac13(1-c_W^4)
-\beta_W^{\tilde{\pi}}(1-\phi_W^{\tilde{\pi}})+c_W^4\beta_Z^{\tilde{\pi}}
(1-\phi_Z^{\tilde{\pi}}) \nonumber \\ &+& N_{\rm
TC}\Bigl[-\frac13\Bigl(1-c_W^4-\frac19s_W^4\Bigr)
+(3+\beta_W^{\tilde{Q}})(1-\phi_W^{\tilde{Q}})-\Bigl(c_W^4+\frac19s_W^4\Bigr)
(3+\beta_Z^{\tilde{Q}})(1-\phi_Z^{\tilde{Q}})\Bigr]\Bigr\}\,,
\end{eqnarray*}
\begin{eqnarray*}
V^{\tilde{\pi}+\tilde{Q}}&=&\frac{c_W^2}{6\pi s_W^2} \Bigl\{-\frac12
M_Z^2\,\beta_Z^{\tilde{\pi}}\,B_0'(M_Z^2,m_{\tilde{\pi}}^2,m_{\tilde{\pi}}^2)
+(\beta_Z^{\tilde{\pi}}+1)(1-\phi_Z^{\tilde{\pi}}) \nonumber \\
&+&N_{\rm TC}\Bigl(1+\frac{s_W^4}{9c_W^4}\Bigr) \Bigl[\frac12
M_Z^2\,(3+\beta_Z^{\tilde{Q}})\,B_0'(M_Z^2,M_{\tilde{Q}}^2,M_{\tilde{Q}}^2)-
(1+\beta_Z^{\tilde{Q}})(1-\phi_Z^{\tilde{Q}})\Bigr]\Bigr\}\,,\\
W^{\tilde{\pi}+\tilde{Q}}&=&\frac{1}{6\pi s_W^2} \Bigl\{-\frac12
M_W^2\,\beta_W^{\tilde{\pi}}\,B_0'(M_W^2,m_{\tilde{\pi}}^2,m_{\tilde{\pi}}^2)
+(\beta_W^{\tilde{\pi}}+1)(1-\phi_W^{\tilde{\pi}}) \nonumber \\
&+&N_{\rm TC}\Bigl[\frac12
M_W^2\,(3+\beta_W^{\tilde{Q}})\,B_0'(M_W^2,M_{\tilde{Q}}^2,M_{\tilde{Q}}^2)-
(1+\beta_W^{\tilde{Q}})(1-\phi_W^{\tilde{Q}})\Bigr]\Bigr\}\,,\\
X^{\tilde{\pi}+\tilde{Q}}&=&\frac{c_W^2}{6\pi}\Bigl\{-\frac13+
\beta_Z^{\tilde{\pi}}(1-\phi_Z^{\tilde{\pi}})+N_{\rm
TC}\Bigl(1-\frac{s_W^2}{9c_W^2}\Bigr)
\Bigl[\frac13-(3+\beta_Z^{\tilde{Q}})(1-\phi_Z^{\tilde{Q}})\Bigr]\Bigr\}\,,
\end{eqnarray*}
where
\begin{eqnarray*}
&&\beta_{Z,W}^{\tilde{\pi}}=\frac{4m_{\tilde{\pi}}^2}{M_{Z,W}^2}-1>0\,,\qquad
\beta_{Z,W}^{\tilde{Q}}=\frac{4M_{\tilde{Q}}^2}{M_{Z,W}^2}-1>0\,,\\
&&\phi_{Z,W}^{\tilde{\pi},\tilde{Q}}=\Bigl(\beta_{Z,W}^{\tilde{\pi},\tilde{Q}}\Bigr)^{1/2}
\arctan\Bigl(\beta_{Z,W}^{\tilde{\pi},\tilde{Q}}\Bigr)^{-1/2}\,,\quad
B_0'(M^2,m^2,m^2)=\int_0^1 dx\,\frac{x(1-x)}{m^2-M^2\,x(1-x)}\,.
\end{eqnarray*}
\begin{figure*}[!h]
\begin{minipage}{0.45\textwidth}
 \centerline{\includegraphics[width=1.0\textwidth]{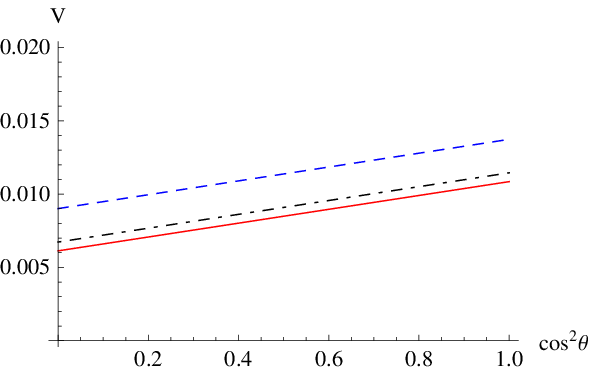}}
\end{minipage}
\hspace{6mm}\vspace{5mm}
\begin{minipage}{0.45\textwidth}
 \centerline{\includegraphics[width=1.0\textwidth]{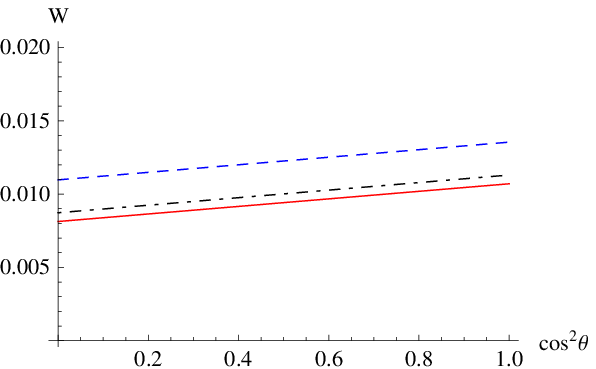}}
\end{minipage}
\begin{minipage}{0.45\textwidth}
 \centerline{\includegraphics[width=1.0\textwidth]{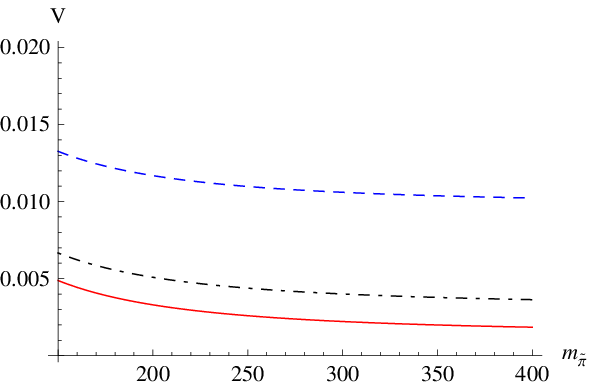}}
\end{minipage}
\hspace{5mm}
\begin{minipage}{0.45\textwidth}
 \centerline{\includegraphics[width=1.0\textwidth]{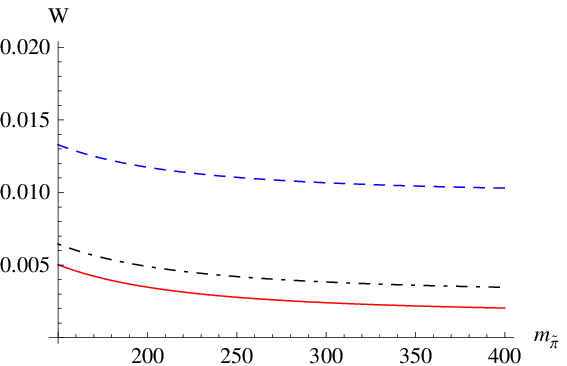}}
\end{minipage}
   \caption{\small The complete $V$ and $W$ parameters in the CSTC scenario
   (the non-minimal case with $\mu_S$ and $\mu_H$ included) as functions of
   (1) $\cos^2\theta$ for fixed $M_U=M_D=300$ GeV,
   and $m_{\tilde \pi}=150,250,350$ GeV, corresponding to dashed,
   dash-dotted and solid lines, respectively (first row);
   and (2) $m_{\tilde \pi}$ for fixed
   $\cos^2\theta=0.9$, and $M_{\tilde Q}=300,500,700$ GeV,
   corresponding to dashed, dash-dotted and solid lines,
   respectively (second row). Both $V$ and $W$ parameters
   do not depend on $M_{\tilde \sigma}$ and $\Delta M_{\tilde Q}$.}
 \label{fig:fig-PT-VW}
\end{figure*}

\begin{figure*}[!h]
\begin{minipage}{0.43\textwidth}
 \centerline{\includegraphics[width=1.0\textwidth]{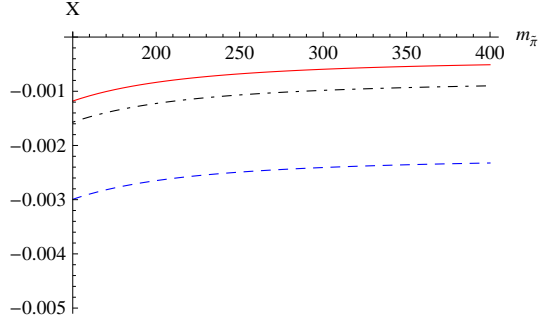}}
\end{minipage}
   \caption{
\small The $X$ parameter $m_{\tilde \pi}$ for fixed $M_{\tilde
Q}=300,500,700$ GeV, corresponding to dashed, dash-dotted and solid
lines, respectively. It does not depend $\cos^2\theta$, $M_{\tilde
\sigma}$ and $\Delta M_{\tilde Q}$.}
 \label{fig:fig-PT-X}
\end{figure*}

In order to constrain the viability of the CSTC model, let us look
at the complete EW precision PT ($S,\,T,\,U,\,V,\,W,\,X$) parameters
in general case, appearing due to both the modifications in the
scalar sector and the new states propagating in loops, as well as at
their dependence on the physical parameters of the model. These are
demonstrated in Figs.~\ref{fig:fig-PT-SU}, \ref{fig:fig-PT-T},
\ref{fig:fig-PT-VW} and \ref{fig:fig-PT-X}. In particular, we see
that the $S$-parameter is always restricted by $|S|<0.03$, and can
even turn to zero for small mixing angles $\sin^2\theta\sim 0.2$,
moderate values of $\Delta M_{\tilde Q}\sim 5$ GeV and large values
of $M_{\tilde Q}\gtrsim 600$ GeV, and this is weakly dependent on
$M_{\tilde \sigma}$ (see Fig.~\ref{fig:fig-PT-SU}). So, we conclude
that in the CSTC there is not such a big issue to satisfy the
constraints on the $S$ parameter (\ref{STU-constr}): the predictions
fit well with $|S^{data}|\lesssim 0.1$ for the whole physically
reasonable parameter space.

Does this fortunate conclusion persist also for other PT parameters?
The $U$ parameter is strongly suppressed too, and never exceeds
0.01, while being rather weakly dependent on all the physical
parameters except for the mixing angle, however, it never turns into
zero exactly $U\gtrsim 0.002$ (see Fig.~\ref{fig:fig-PT-SU}). Thus,
both $S$ and $U$ parameters cannot be used for an efficient
constraining the model parameter space at the current level of data
uncertainties (\ref{STU-constr}). The same holds true for associated
oblique corrections beyond the linear $q^2$ power expansions, given
in terms of $V,\,W,\,X$ parameters (\ref{VWX}). In particular, $V$
and $W$ parameters remain of the same order of magnitude as the $S$
and $U$ parameters. They belong to the interval $0.002\lesssim
V,W\lesssim0.01$ and are weakly dependent on physical parameters
(see Fig.~\ref{fig:fig-PT-VW}), whereas the $X$ parameter is even
stronger suppressed, $|X|\sim0.001$ (see Fig.~\ref{fig:fig-PT-X}).
In general, this situation is not noticeably affected by having more
than one generation of technifermions or other $N_{\rm TC}$
different from three.

The strongest bounds to the CSTC parameter space actually come from
the $T$ parameter (see Fig.~\ref{fig:fig-PT-T}). The EW precision
constraints to the $T$ parameter encoding the vector isospin
breaking effects (\ref{STU-constr}) are satisfied only for a
relatively small $h\tilde{\sigma}$ mixing $\sin^2\theta\lesssim 0.3$
and a small splitting between current technifermion masses $\Delta
M_{\tilde Q}\lesssim 5$ GeV. The latter is natural since similarly
the relatively small splitting between the current up and down quark
masses compared to their constituent masses applies also for usual
QCD. In the degenerated case with $\Delta M_{\tilde Q}=0$ and in the
``no $h{\tilde \sigma}$-mixing'' limit $\cos^2\theta\to 1$, the
$T$-parameter vanishes identically, $T\to 0$. So, the CSTC model has
enough room to fit with the EW precision data, together with tight
constraints to the Higgs sector properties.

Note, the $S,\,U,\,V,\,W,\,X$ parameters are always UV finite. The
$T$ parameter is finite in the degenerated case when $M_{U}=M_{D}$,
whereas in general case it has logarithmic divergence proportional
to the technifermion mass difference, i.e. ${\rm div}(T)\sim
(M_U-M_D)^2/\varepsilon$ with a constant coefficient. Note also that
the EW constraints put much stronger limits on parameter space in
the case of inverse mass hierarchy in the scalar sector of the
theory, i.e. assuming that technisigma is the lightest scalar
particle observed at the LHC $M_{\tilde \sigma}<M_{h}$. In this
case, the $h\tilde{\sigma}$-mixing angle has to be much closer to
being vanished in order to satisfy the constraints on the
corresponding oblique corrections.

\subsection{Qualitative remarks on FCNC constraints}

Another source of (less) stringent constraints onto TC models comes
from the FCNC-induced processes (see e.g.
Ref.~\cite{Fukano:2009zm}). In particular, here one would be
interested in constraints coming from such processes as mixing in
system of neutral mesons $M^0-\bar{M}^0$, as well as from rare
leptonic decays of neutral mesons $M^0\to l\bar{l}$, etc. The
semi-leptonic decays are presumably more uncertain theoretically due
to larger contributions from poorly known hadronic form factors thus
making it rather hard to set definite constraints to NP
contributions. The flavor constraints can be very relevant for
phenomenological tests of the TC models with relatively light spin-1
resonances with the same quantum numbers as the SM gauge bosons. In
the considering CSTC model under discussion adopting the QCD-like
mass hierarchy in the technihadron spectrum there are no light
spin-1 particles; heavy vector $\tilde{\rho}$ and axial-vector
$\tilde{a}_1$ states are considered to be decoupled from the
lightest technipion and technisigma states and do not participate in
processes at low momentum transfers. This is, of course, a valid
approximation motivated by advances of the usual hadron physics. An
extended theory which supposedly includes heavy states should then
be quantitatively tested against the flavor constraints according to
Ref.~\cite{Fukano:2009zm}, in particular, setting up the low bounds
on masses of heavy (pseudo)vector particles. However, this analysis
will only be reasonable after the lightest (pseudo)scalar states
have been discovered experimentally.
\begin{figure*}[!h]
\begin{minipage}{0.6\textwidth}
 \centerline{\includegraphics[width=1.0\textwidth]{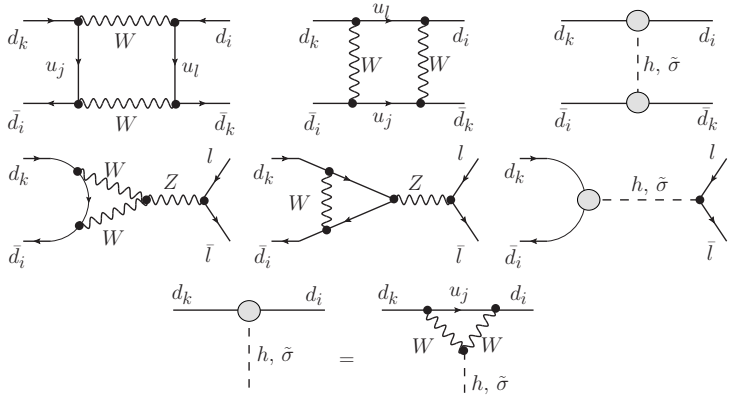}}
\end{minipage}
   \caption{
\small Typical FCNC contributions in the CSTC model. The rightmost
diagrams with scalar exchanges are the only weakly affected
contributions due to a small $h\tilde{\sigma}$-mixing and additional
$\tilde{\sigma}$ meson, which are however negligibly small (see main
text).}
 \label{fig:FCNC}
\end{figure*}

In Fig.~\ref{fig:FCNC} we illustrate new contributions (besides
those in the gauge bosons polarisations) to typical FCNC processes
(rightmost diagrams), along with the standard part (first two
diagrams on the left hand side). These diagrams describe the
short-distance contributions, which dominate the FCNC observables
for heavy flavor mesons (for instance, $B^0_d,\,B^0_s$). In the
framework of the CSTC model, an additional effect comes only from
the $h\tilde{\sigma}$-mixing, whereas technipions and technifermions
can only contribute to the gauge bosons polarisation functions
inside the loop propagators.

The qualitative analysis of these contributions reveals that these
contributions are strongly suppressed due to the following
arguments:
\begin{itemize}
\item the typical contributions from two-loop FCNC effects with the Higgs
boson in the $t$-channel in neutral mesons $M^0-\bar{M}^0$ mixing is
extremely small, and usually neglected in the SM calculations. An
additional (small) mixing with the heavy $\tilde{\sigma}$ field can
not change this situation noticeably;
\item in the case of rare (semi)leptonic decays of the Higgs boson, as well as
$\tilde{\sigma}$ meson, Yukawa couplings to leptons are usually very
small ($\sim gm_l/M_W$), and the corresponding contributions are
also neglected;
\item in all cases the $\tilde{\sigma}$ contributions are additionally
suppressed by a large technisigma mass compared to vector boson
masses, $M_{\tilde{\sigma}}\gg M_{W,Z}$;
\item an extra (double) suppression in the limit of small
$h\tilde{\sigma}$-mixing by $s_{\theta}^2\ll 1$ factor in the
amplitude;
\item the higher-loop effect from the technipions and technifermions contributing
only to the gauge bosons polarisation functions inside the loop
propagators vanish at small loop momentum $q^2\to 0$, but otherwise
is expected to be extremely small.
\end{itemize}

We conclude, that the most stringent constraints on the parameter
space in the considering CSTC scenario come from the $T$-parameter
which sets the upper bound to the $h\tilde{\sigma}$-mixing (see
previous Section). Now, we turn to a discussion of the
phenomenological consequences of the CSTC.

\section{Collider phenomenology of the CSTC}

\subsection{Higgs boson production and decay}

The properties of the Higgs sector in the SM are the subject of
intensive studies and discussions in the light of the latest data
from the LHC \cite{ATLAS,CMS,latest_LHC}. The Higgs couplings are
expected to be rather sensitive to NP contributions, and could
therefore serve as a good probe of physics beyond the SM. However,
it is important to notice here that even though the Higgs boson may
look standard according to the latest observations and studies, this
does not totally exclude possible role of NP in the EWSB and, in
particular, in its underlined dynamical reasons. Further in this
subsection we will examine basic possible signatures of the CSTC in
Higgs boson observables.

Consider first the simplest $s$-channel subprocess of the Higgs
boson production with subsequent decay into final states, i.e.
$ab\to h\to XY$. Typically, the initial states of this subprocess
are $ab=gg,ZZ,WW$ and the final states are $XY=f\bar{f}$, $WW^*$,
$ZZ^*$, $\gamma\gamma$, and $\gamma Z$. As is seen from the physical
Lagrangian of the Higgs boson interactions (\ref{Yukawa}) and
(\ref{L-hVVpipi}), the standard tree-level $hVV$ and $hf\bar{f}$
couplings are modified by a common factor $c_\theta$ only caused by
a mixing with heavy technisigma state.

For $ab=gg$, the $ggh$ and $gg\tilde{\sigma}$ couplings are
loop-induced via heavy quarks, and there are no additional loop
diagrams can contribute here in the framework of CSTC. Hence, in the
Higgs boson production amplitude there always comes an extra factor
$c_\theta$ compared to the corresponding SM amplitude. Further, the
first three Higgs decay channels $XY=f\bar{f}$, $WW^*$, $ZZ^*$ are
the tree-level ones, with another factor $c_\theta$ in the
amplitude, so the corresponding amplitudes $VV\to h \to f\bar{f},\,
WW^*,\, ZZ^*$ can only be different w.r.t. the SM ones by a factor
of $c^2_\theta$ only (or a factor of $s^2_\theta$ in the case of
intermediate $\tilde{\sigma}$ meson). But this is true only if one
considers the $s$-channel production process far from the resonance,
$\hat{s}^{\rm res}=M_h^2$ (or $\hat{s}^{\rm res}=M_{\tilde
\sigma}^2$ for the intermediate $\tilde{\sigma}$ meson). However, in
the resonance region the modifications of the SM amplitudes can be
different from mere mixing factor multiplication.
\begin{figure*}[!h]
\begin{minipage}{0.4\textwidth}
 \centerline{\includegraphics[width=1.0\textwidth]{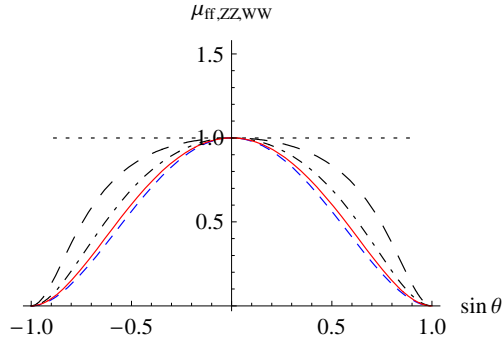}}
\end{minipage}
   \caption{
\small The Higgs boson signal strength in the tree-level $f\bar{f}$,
$ZZ^*$ and $WW^*$ channels $\mu_{fbar{f},WW,ZZ}$ as a function of
$s_{\theta}$ calculated according to Eq.~(\ref{mu-exp}) for $\delta
E=0$ (dotted line), $\delta E=\Gamma^{h,\rm SM}_{tot}\simeq 4.03$
MeV (dash-dotted line), $\delta E=\Gamma^{h,\rm SM}_{tot}/2$
(long-dashed line), $\delta E=2\Gamma^{h,\rm SM}_{tot}$ (solid
line), and $c_\theta^4=(1-s_\theta^2)^2$ curve is also shown for
reference (dashed line).}
 \label{fig:mu_ffWWZZ}
\end{figure*}

In order to calculate the $s$-channel cross section for the scalar
Higgs boson (and $\tilde{\sigma}$ meson) production with
two-particle final states one starts from the universal factorized
formula which reproduces the well-known narrow-width approximation
formula and has been proven to be exact in the framework of the
unstable particles model with smeared mass shell (see e.g.
Ref.~\cite{VI})
\begin{eqnarray}
\sigma(ab\to h(q),\tilde{\sigma}(q)\to XY)=\frac{16\pi
k_{h,\tilde{\sigma}}}{k_ak_b\overline{\lambda}^2(m_a,m_b;q)}\frac{\Gamma(h(q),\tilde{\sigma}(q)\to
ab)\Gamma(h(q),\tilde{\sigma}(q)\to
XY)}{[q^2-M_{h,\tilde{\sigma}}^2]^2+[q\Gamma_{h,\tilde{\sigma}}^{tot}(q)]^2}\,.
\label{sig-hS}
\end{eqnarray}
where $q=p_a+p_b$ is the total $s$-channel 4-momentum, $k_a=2J_a+1$
is the number of polarisation states, and $J_a$ is the spin of
particle $a$ (i.e. $k_{h,\tilde{\sigma}}$=1), and
\begin{eqnarray}
\overline{\lambda}^2(m_a,m_b;q)=1-2\frac{m_a^2+m_b^2}{q^2}+\frac{(m_a^2-m_b^2)^2}{q^4}
\label{Kallen}
\end{eqnarray}
is the normalized K\"allen function. A good estimate of
modifications in $h,\,\tilde{\sigma}$ couplings in the resonance
region where $q^2\simeq M_{h,\tilde{\sigma}}^2$ can thus be obtained
from the formula
\begin{eqnarray}
\sigma(ab\to h,\tilde{\sigma}\to XY)\simeq
\frac{16\pi}{k_ak_b\overline{\lambda}^2(m_a,m_b;m_{h,\tilde{\sigma}})M_{h,\tilde{\sigma}}^2}\,{\rm
Br}(h,\tilde{\sigma}\to ab)\cdot{\rm Br}(h,\tilde{\sigma}\to XY)\,.
\label{peak-CS}
\end{eqnarray}
\begin{figure*}[!h]
\begin{minipage}{0.9\textwidth}
 \centerline{\includegraphics[width=1.0\textwidth]{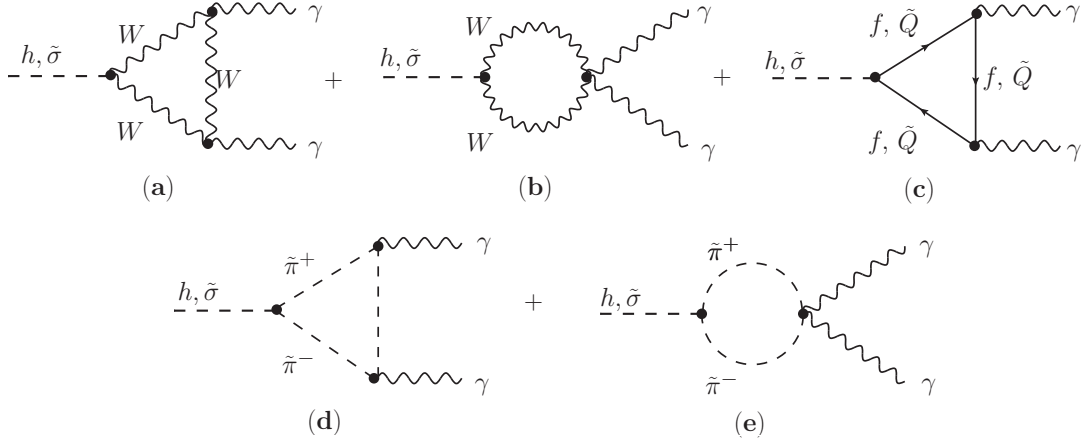}}
\end{minipage}
   \caption{
\small Typical one-loop contributions to the
$h,\,\tilde{\sigma}\to\gamma\gamma$ decay channel in the CSTC.}
 \label{fig:scalar-to-GG}
\end{figure*}

As was mentioned above, the Higgs couplings to SM fermions and
vector bosons in the considering scenario contain extra $c_{\theta}$
factor compared to the SM ones, so in the resonance region we have
for decay widths and branching fractions to a good accuracy
\begin{eqnarray}
\frac{\Gamma^{h,mod}_{tot}}{\Gamma^{h,\rm SM}_{tot}}\simeq
c_{\theta}^2,\qquad \frac{{\rm Br}^{mod}(h\to XY)}{{\rm Br}^{\rm
SM}(h\to XY)}\simeq 1,\qquad XY=f\bar{f},\,WW^*,\,ZZ^*\,,
\end{eqnarray}
i.e. for all Born-level Higgs/technisigma decays which strongly
dominate in the total decay width. This reveals the fact that the
Higgs branching ratios, in fact, in the SM and in the considering
CSTC scenario are the same. Thus, according to Eq.~(\ref{peak-CS})
the ratio between the resonant cross sections in the considering
model to the SM one is close to unity
\begin{eqnarray} \label{mu-res}
\mu^{\rm res}_{f\bar{f},\,ZZ,\,WW}=\frac{\sigma^{mod}(VV\to
h(q)\to\bar{f}f,\,ZZ^*,\,WW^*)}{\sigma^{\rm{SM}}(VV\to
h(q)\to\bar{f}f,\,ZZ^*,\,WW^*)}\simeq 1,\quad q^2\simeq M_h^2\,.
\end{eqnarray}
which are essentially the Higgs boson signal strengths in respective
channels which were measured earlier at the LHC and no significant
deviations from the SM have been found.
\begin{figure*}[!h]
\begin{minipage}{0.45\textwidth}
 \centerline{\includegraphics[width=1.0\textwidth]{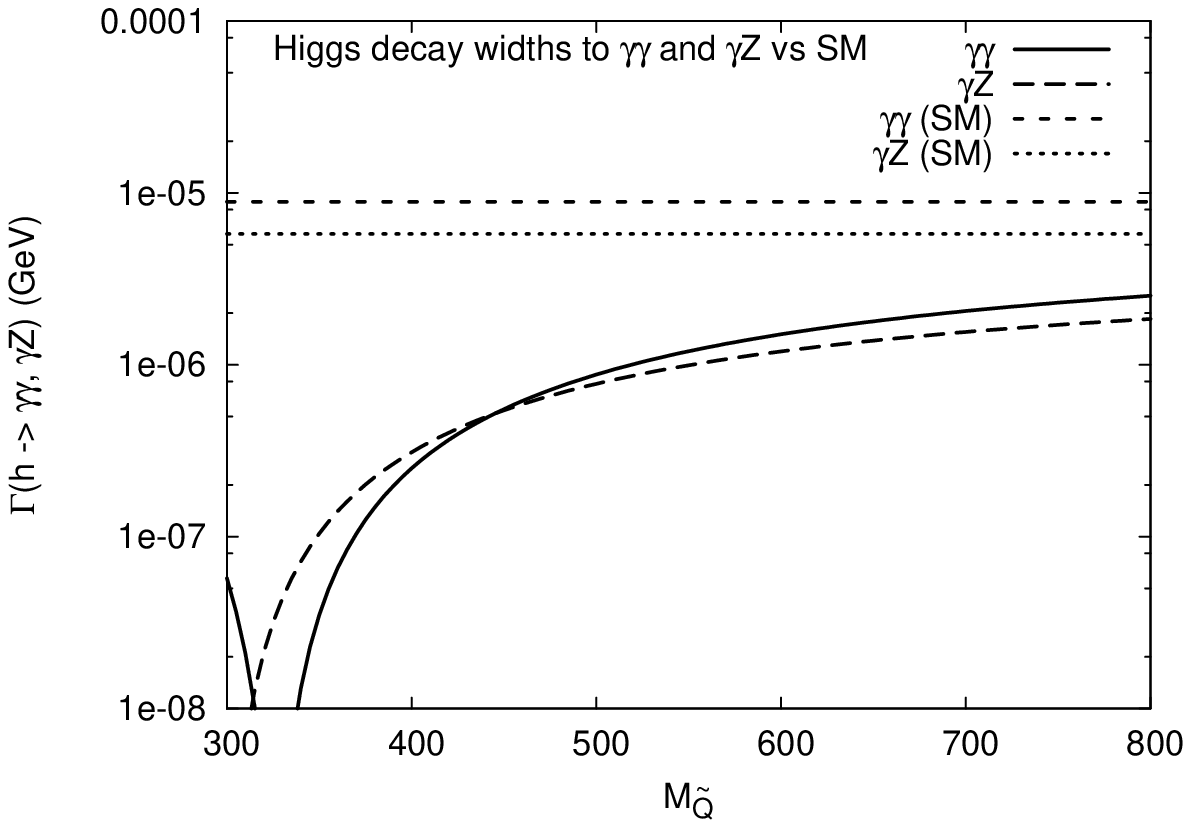}}
\end{minipage}
\begin{minipage}{0.45\textwidth}
 \centerline{\includegraphics[width=1.0\textwidth]{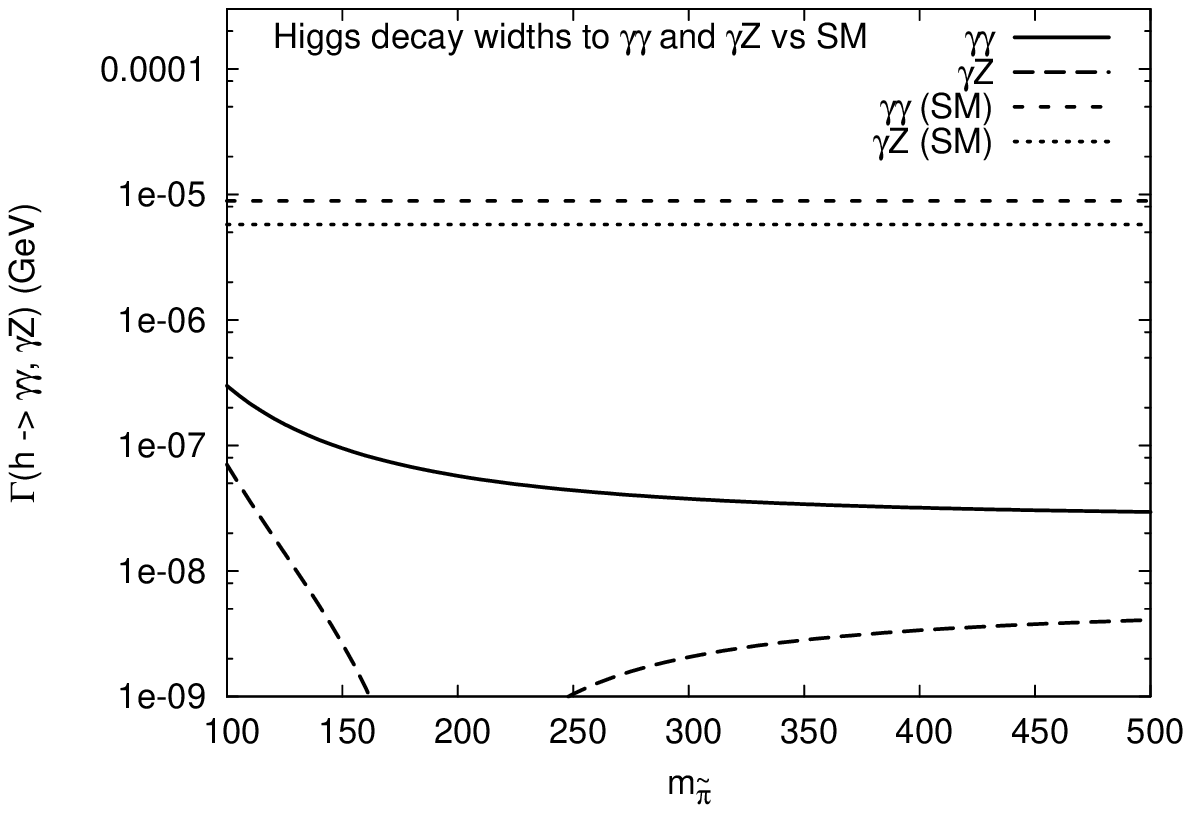}}
\end{minipage}
\begin{minipage}{0.45\textwidth}
 \centerline{\includegraphics[width=1.0\textwidth]{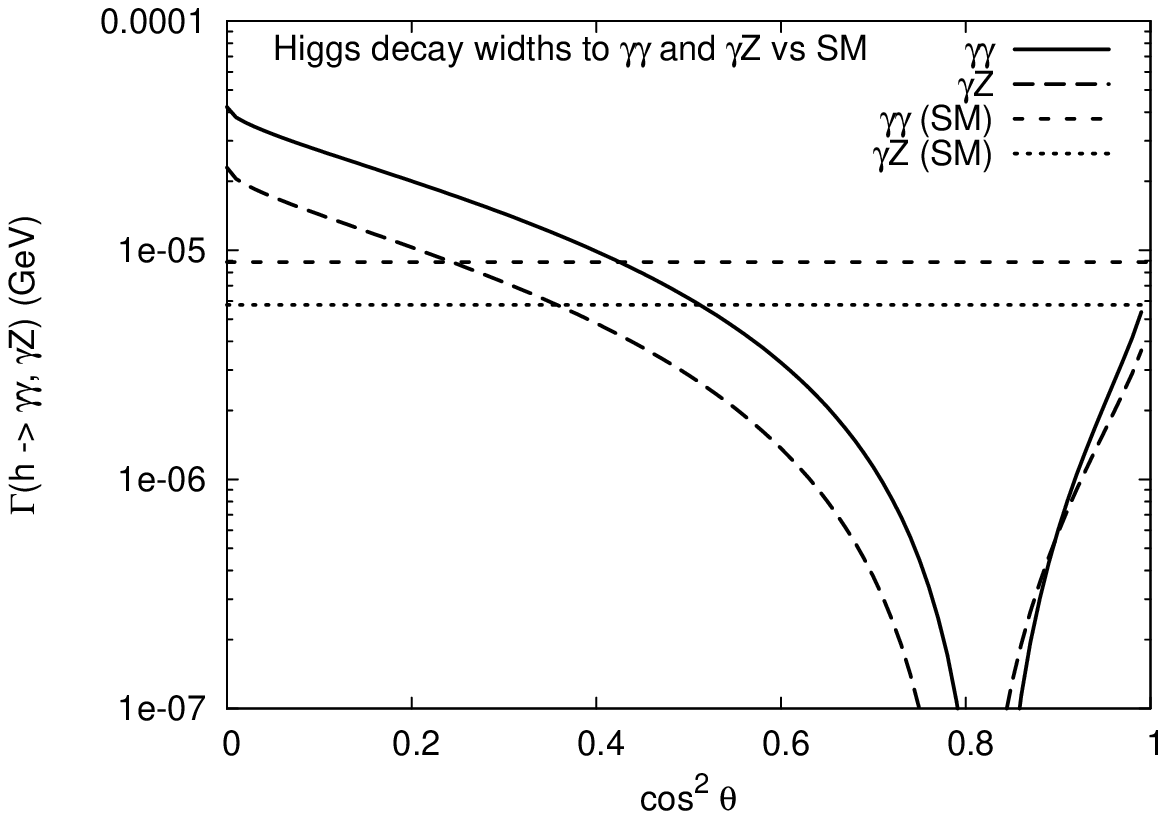}}
\end{minipage}
\begin{minipage}{0.45\textwidth}
 \centerline{\includegraphics[width=1.0\textwidth]{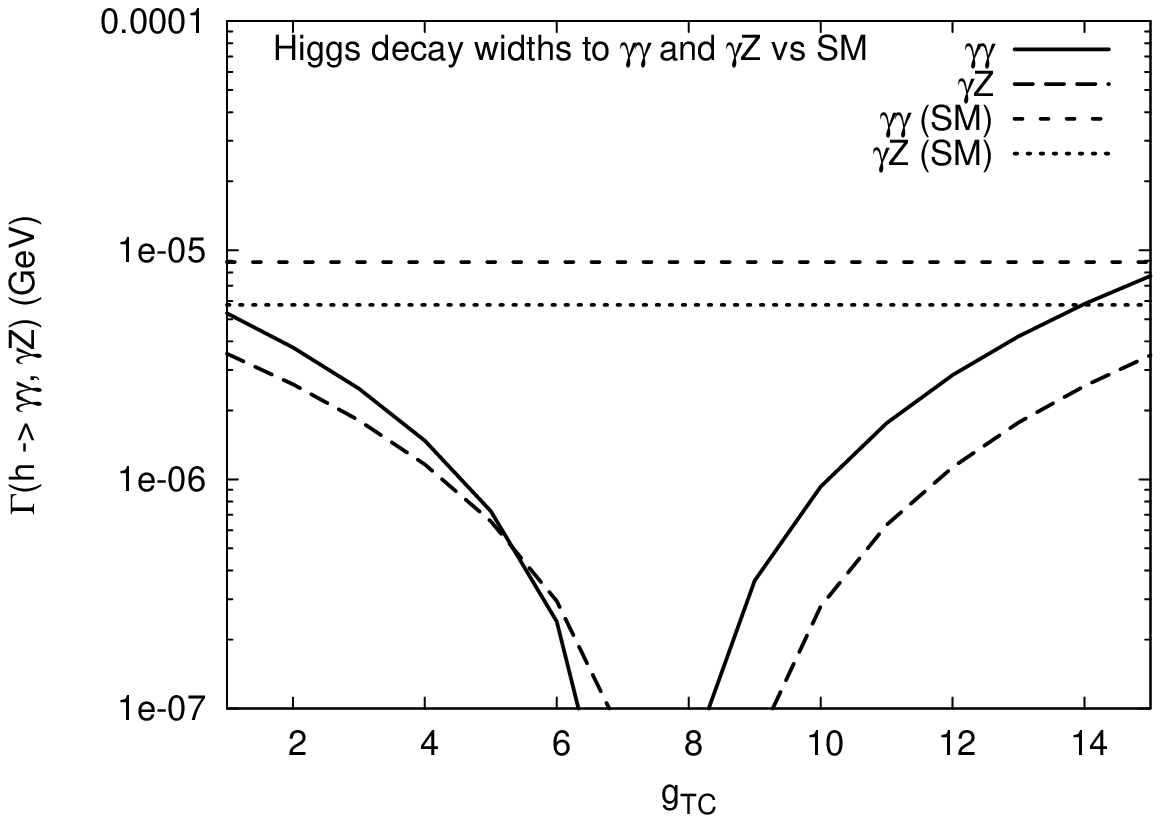}}
\end{minipage}
   \caption{
\small The Higgs boson decay widths in the loop-induced
$\gamma\gamma$ and $\gamma Z$ channels in the non-minimal CSTC (with
scalar $\mu_{S,H}$-terms included) as functions of physical
parameters of the model. The corresponding SM predictions are shown
for comparison. The parameters in each figure are set as follows:
(top-left) $m_{\tilde \pi}=200$ GeV, $c_\theta^2=0.8$, and $g_{\rm
TC}=8$; (top-left) $M_{\tilde Q}=300$ GeV, $c_\theta^2=0.8$, and
$g_{\rm TC}=8$; (bottom-left) $M_{\tilde Q}=300$ GeV, $m_{\tilde
\pi}=200$ GeV, and $g_{\rm TC}=8$; (bottom-right) $M_{\tilde Q}=300$
GeV, $m_{\tilde \pi}=200$ GeV, and $c_\theta^2=0.8$. These results
do not depend on $M_{\tilde \sigma}$, and the positive sign of the
mixing angle, or $s_\theta>0$, is fixed here.}
 \label{fig:Gamma-HGG-VV}
\end{figure*}

In fact, experimentally one never measures events exactly at the
resonance peak position $q^2 = M_h^2$, but one rather has a smearing
of the resonance by e.g. detector conditions. In this case, a more
precise estimation of the Higgs boson signal strength is given by
the ratio of the cross sections integrated (or averaged) over the
energy resolution of an experiment $\delta E$ which can be
comparable or exceeds the small Higgs boson decay width in the SM,
$\delta E \geq \Gamma^{h,\rm SM}_{tot}\simeq 4.03$ MeV (at
$M_h\simeq 125$ GeV) \cite{Dittmaier:2011ti}, i.e.
\begin{eqnarray} \nonumber
\mu_{\rm XY}(\delta E)=\frac{\int_{M_h-\delta E}^{M_h+\delta
E}\sigma^{mod}_{\rm XY}(q)dq}{\int_{M_h-\delta E}^{M_h+\delta
E}\sigma^{\rm{SM}}_{\rm XY}(q)dq} &\simeq& \frac{\Gamma^{mod}(h\to
ab)\Gamma^{mod}(h\to XY)}{\Gamma^{\rm SM}(h\to ab)\Gamma^{\rm
SM}(h\to XY)} \\ &\times& \frac{\int_{M_h-\delta E}^{M_h+\delta
E}[(q^2-M_h^2)^2+q^2(\Gamma_{tot}^{h,\rm SM})^2]dq}{\int_{M_h-\delta
E}^{M_h+\delta E}[(q^2-M_h^2)^2+q^2(\Gamma_{tot}^{h,mod})^2]dq}\,,
\label{mu-exp}
\end{eqnarray}
whose values have to be compared to the measured ones. The last part
of the formula above is fulfilled approximately and valid to a good
accuracy for $\delta E \gg \Gamma^{h,\rm SM}_{tot}$ which is the
case in actual measurements. Clearly, the formula (\ref{mu-exp})
turns into the Eq.~(\ref{mu-res}) in the limit of very narrow
$\delta$-shaped resonance, i.e. when $\delta E\ll \Gamma^{h,\rm
SM}_{tot}$.

In Fig.~\ref{fig:mu_ffWWZZ} we show the dependence of the
$\mu_{f\bar{f},WW,ZZ}(\delta E)$ on the mixing $s_\theta$ for
different values of the peak smearing $\delta E=0$ (short-dashed
line), $\delta E=\Gamma^{h,\rm SM}_{tot}\simeq 4.03$ MeV
(dash-dotted line), $\delta E=2\Gamma^{h,\rm SM}_{tot}$ (solid
line), and $c_\theta^4=(1-s_\theta^2)^2$ curve is also shown for
reference (dashed line). No smearing case with $\delta E=0$
corresponds precisely to the resonance formula (\ref{mu-res}) with
the unit strength, while an increase in the peak smearing quickly
approaches to the off-resonance result with
$\mu_{f\bar{f},WW,ZZ}\sim c_\theta^4$. Clearly, an influence of the
peak smearing vanishes in the no mixing limit $s_\theta\to 0$.
\begin{figure*}[!h]
\begin{minipage}{0.35\textwidth}
 \centerline{\includegraphics[width=1.0\textwidth]{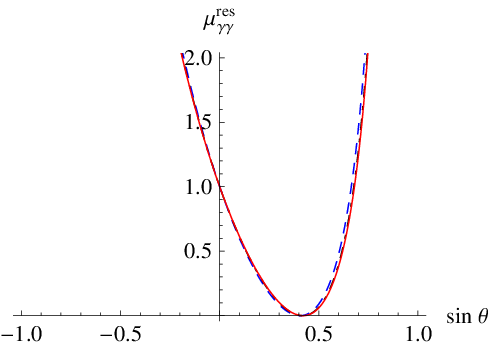}}
\end{minipage}
\hspace{9mm}\vspace{5mm}
\begin{minipage}{0.35\textwidth}
 \centerline{\includegraphics[width=1.0\textwidth]{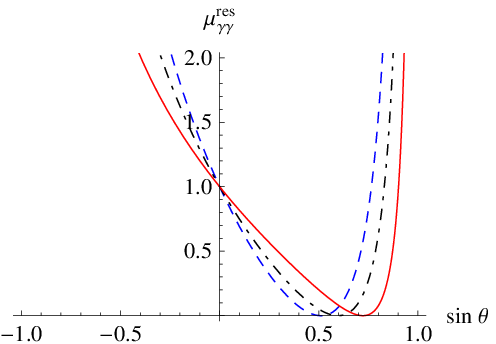}}
\end{minipage}
\begin{minipage}{0.35\textwidth}
 \centerline{\includegraphics[width=1.0\textwidth]{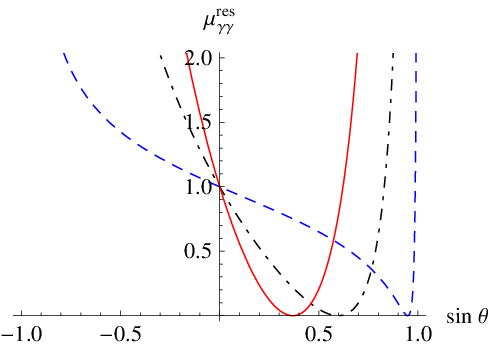}}
\end{minipage}
\hspace{9mm}
\begin{minipage}{0.35\textwidth}
 \centerline{\includegraphics[width=1.0\textwidth]{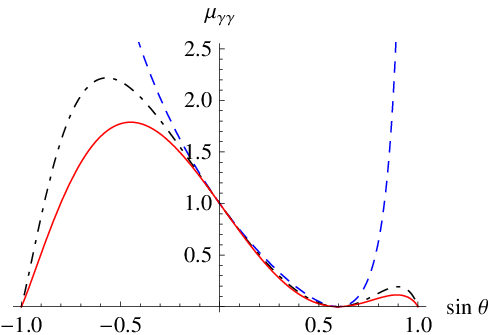}}
\end{minipage}
   \caption{
\small Dependence of the Higgs boson signal strength in the
resonance given by Eq.~(\ref{mu-res-GG}) in the non-minimal case of
the CSTC model (with scalar $\mu_{S,H}$-terms included),
$\mu_{\gamma\gamma}^{\rm res}$, on $s_\theta$ for different sets of
the physical parameters: (top-left) $g_{\rm TC}=8$, $M_{\tilde
Q}=300$ GeV, and $m_{\tilde \pi}=150,250,350$ GeV, corresponding to
dashed, dash-dotted and solid lines, respectively; (top-right)
$g_{\rm TC}=8$, $m_{\tilde \pi}=150$ GeV, and $M_{\tilde
Q}=400,500,700$ GeV, corresponding to dashed, dash-dotted and solid
lines, respectively; (bottom-left) $m_{\tilde \pi}=150$ GeV,
$M_{\tilde Q}=500$ GeV, and $g_{\rm TC}=2,8,15$, corresponding to
dashed, dash-dotted and solid lines, respectively. Finally,
bottom-right figure corresponds to smeared
$\mu_{\gamma\gamma}(\delta E)$ given by Eq.~(\ref{mu-exp}) as a
function of $s_\theta$ for fixed $m_{\tilde \pi}=150$ GeV,
$M_{\tilde Q}=500$ GeV, $g_{\rm TC}=8$ and with different smearing
parameters: no smearing $\delta E=0$ (dashed line), $\delta
E=\Gamma^{h,\rm SM}_{tot}\simeq 4.03$ MeV (dash-dotted line), and
$\delta E=1$ GeV (solid line). Here and below, $Y_{\tilde Q}=1/3$,
unless noted otherwise.}
 \label{fig:muGG-res}
\end{figure*}

The near-resonance approximation in the $s$-channel production
process (\ref{mu-res}) is valid up to relatively small loop-induced
contributions and higher order correction which may contain extra
loop contributions with technipions, technisigma and technifermions.
These extra contributions can be especially pronounced in the
loop-induced $\gamma\gamma$ and $\gamma Z$ decay channels (see
Fig.~\ref{fig:scalar-to-GG}). Indeed,
\begin{eqnarray} \label{mu-res-GG}
\mu^{\rm res}_{\gamma\gamma}=\frac{\sigma^{mod}(h\to
\gamma\gamma)}{\sigma^{\rm SM}(h\to
\gamma\gamma)}\simeq\frac{1}{c^2_\theta}\frac{\Gamma^{mod}(h\to
\gamma\gamma)}{\Gamma^{\rm SM}(h\to
\gamma\gamma)}\simeq\frac{1}{c^2_\theta}
\frac{|A_W+A_f+A_{\tilde{\pi}}+A_{\tilde{Q}}|^2}{|A^{\rm
SM}_W+A^{\rm SM}_f|^2}
\end{eqnarray}
where $A_{W,f,\tilde{\pi},\tilde{Q}}$ are the amplitudes given by
the SM-like $W,\,f$ loop diagrams (see Fig.~\ref{fig:scalar-to-GG}
(a), (b), (c)), as well as by the new technipion $\tilde{\pi}$ and
technifermion $\tilde{Q}$ loop diagrams (see
Fig.~\ref{fig:scalar-to-GG} (c), (d), (e)). An interference effect
between these contributions may be important. Notably
$A_{\tilde{Q}}\sim s_\theta$ while
$|A_{\tilde{\pi}}|\ll|A_{\tilde{Q}}|$ in general so the interference
effect changes its sign depending on the sign of $s_\theta$ possibly
giving rise to either enhancement or suppression of the
$\gamma\gamma$ signal, or to the SM-like $h\to\gamma\gamma$ signal
strengths in the case of a small mixing angle $s_\theta\ll 1$ (where
the technipion loop contribution disappears as well). Since the
first three diagrams, which are present in the SM, do not exist at
the tree level, their sum is free of divergencies. More precisely,
the divergencies are canceled between diagrams (a) and (b), and the
fermion ($f$ and $\tilde{Q}$) loops are finite individually. We have
found that the sum of technipion loops is finite as well. Also, here
it is reasonable to assume that only heavy top quark loops
contribute to the final result; all other fermions are strongly
suppressed and thus can be neglected.
\begin{figure*}[!h]
\begin{minipage}{0.4\textwidth}
 \centerline{\includegraphics[width=1.0\textwidth]{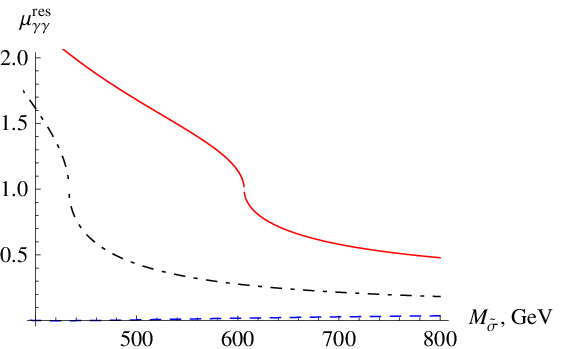}}
\end{minipage}
\hspace{9mm}\vspace{5mm}
\begin{minipage}{0.4\textwidth}
 \centerline{\includegraphics[width=1.0\textwidth]{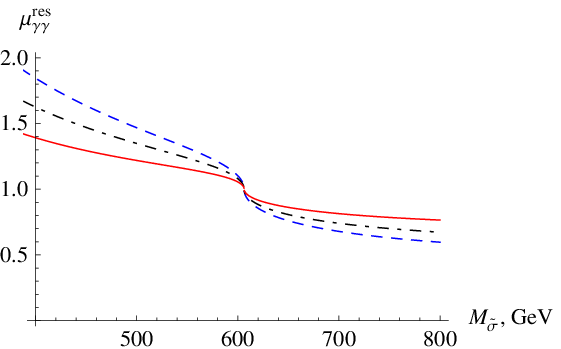}}
\end{minipage}
\begin{minipage}{0.4\textwidth}
 \centerline{\includegraphics[width=1.0\textwidth]{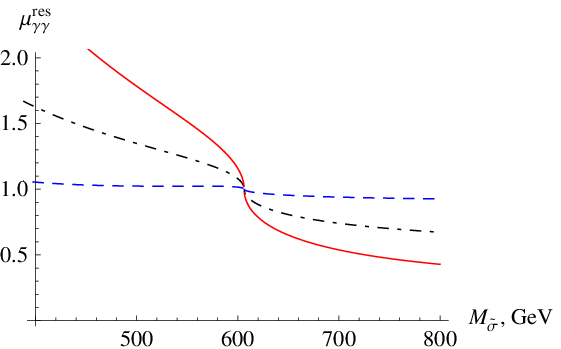}}
\end{minipage}
\hspace{9mm}
\begin{minipage}{0.4\textwidth}
 \centerline{\includegraphics[width=1.0\textwidth]{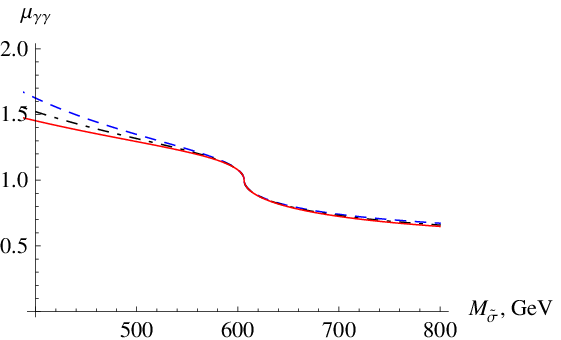}}
\end{minipage}
   \caption{
\small Dependence of the Higgs boson signal strength in the
resonance given by Eq.~(\ref{mu-res-GG}) in the minimal CSTC model
(with scalar $\mu_{S,H}$-terms excluded), $\mu_{\gamma\gamma}^{\rm
res}$, on $M_{\tilde \sigma}$ for different sets of the physical
parameters: (top-left) $g_{\rm TC}=8$, $M_{\tilde Q}=300$ GeV, and
$m_{\tilde \pi}=150,250,350$ GeV, corresponding to dashed,
dash-dotted and solid lines, respectively; (top-right) $g_{\rm
TC}=8$, $m_{\tilde \pi}=350$ GeV, and $M_{\tilde Q}=400,500,700$
GeV, corresponding to dashed, dash-dotted and solid lines,
respectively; (bottom-left) $m_{\tilde \pi}=350$ GeV, $M_{\tilde
Q}=500$ GeV, and $g_{\rm TC}=2,8,15$, corresponding to dashed,
dash-dotted and solid lines, respectively. Finally, bottom-right
figure corresponds to smeared $\mu_{\gamma\gamma}(\delta E)$ given
by Eq.~(\ref{mu-exp}) as a function of $M_{\tilde \sigma}$ for fixed
$m_{\tilde \pi}=350$ GeV, $M_{\tilde Q}=500$ GeV, $g_{\rm TC}=8$ and
with different smearing parameters: no smearing $\delta E=0$ (dashed
line), $\delta E=\Gamma^{h,\rm SM}_{tot}\simeq 4.03$ MeV
(dash-dotted line), and $\delta E=1$ GeV (solid line).}
 \label{fig:muGG-res-spec}
\end{figure*}

A straightforward calculation lead to the following Higgs partial
decay width in the $\gamma\gamma$ channel
\begin{eqnarray} \label{Gamma-mod}
 \Gamma^{mod}(h\to \gamma\gamma) = \frac{\alpha^2 M_h}{16 \pi^3}\cdot |F_W+F_{top}+F_{\tilde{\pi}}+F_{\tilde{Q}}|^2
\end{eqnarray}
where $\alpha=\alpha(M_Z)=1/127.93$ is the fine structure constant
adopted in all numerical calculations, and the individual
contributions from $W$, top-quark, $\tilde{\pi}$ and $\tilde{Q}$
loops read, respectively,
\begin{eqnarray} \nonumber
 && F_W = \frac18\, g\, c_{\theta}\, \frac{M_h}{M_W}\cdot \Big[2 + 3\beta_W + 3\beta_W(2-\beta_W)f(\beta_W)\Big]\,,\\
 && F_{top} = -\frac43\, g\, c_{\theta}\, \frac{m_{top}^2}{M_h M_W}\, \Big[1 + (1-\beta_{top})f(\beta_{top})\Big]\,, \nonumber\\
 && F_{\tilde{\pi}} = -\frac{g_{h\tilde{\pi}}}{2M_h}\, \Big[1 - \beta_{\tilde{\pi}}f(\beta_{\tilde{\pi}})\Big], \qquad
 g_{h\tilde{\pi}} = -2(\lambda_{\rm{TC}}\, u s_{\theta} - \lambda\, v
 c_{\theta})\,,   \label{F-mod} \\
 && F_{\tilde Q} = -2N_{\rm TC}(q_U^2+q_D^2)\, g_{\rm TC}\, s_{\theta}\, \frac{M_{\tilde Q}}{M_h}\,
 \Big[1 + (1-\beta_{\tilde Q})f(\beta_{\tilde Q})\Big]\,, \nonumber
\end{eqnarray}
where we take the number of technicolors $N_{\rm TC}=3$ in numerical
calculations below, $q_{U,D}$ are the techni-up and techni-down
fermion charges, and
\begin{eqnarray}
 f(\beta)=\arcsin^2 \frac{1}{\sqrt{\beta}}\,\qquad
 \beta_X=\frac{4m_X^2}{M_h^2}\,,\qquad X=W,\,{\rm top},\,{\tilde{\pi}},\,{\tilde{Q}}\,.
\end{eqnarray}
The non-minimal case with scalar $\mu$-terms included, the relation
\begin{eqnarray}
g_{h\tilde{\pi}} = -g_{\rm TC}s_{\theta}\frac{M_h^2-m_{\tilde
\pi}^2}{M_{\tilde Q}}
\end{eqnarray}
can be used (cf. Eq.~(\ref{L-hVVpipi})), whereas in the special case
with $\mu_{S,{\rm H}}\to 0$ the relations (\ref{spec-lam}) have to
be employed for calculation of the $g_{h\tilde{\pi}}$ coupling. In
the limit of small $h\tilde{\sigma}$-mixing, the constituent
technifermion and technipion loop contributions to the Higgs boson
width are suppressed by a factor of $s_{\theta}^2\ll 1$, so the
whole expression (\ref{Gamma-mod}) turns to the SM result:
\begin{eqnarray} \label{Gamma-mod-SM}
 \Gamma^{\rm SM}(h\to \gamma\gamma) = \frac{\alpha^2 M_h}{16 \pi^3}\cdot |F^{\rm SM}_W+F^{\rm
 SM}_{top}|^2\,,
\end{eqnarray}
where $F^{\rm SM}_{W,top}$ can be obtained from Eq.~(\ref{F-mod})
with $c_\theta=1$.
\begin{figure*}[!h]
\begin{minipage}{0.4\textwidth}
 \centerline{\includegraphics[width=1.0\textwidth]{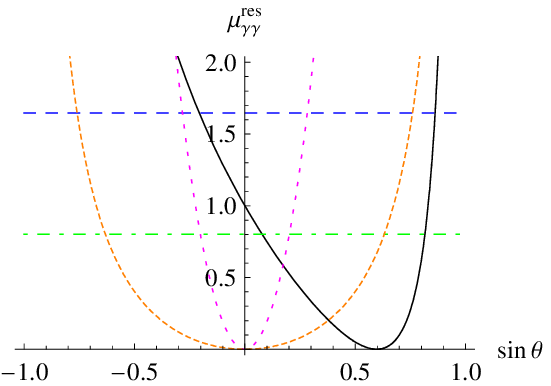}}
\end{minipage}
\hspace{9mm}\vspace{5mm}
\begin{minipage}{0.45\textwidth}
 \centerline{\includegraphics[width=1.0\textwidth]{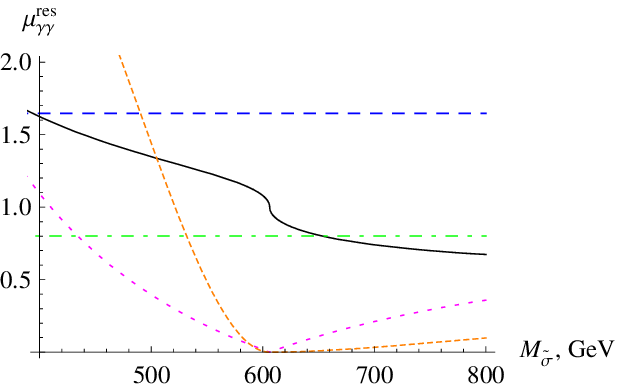}}
\end{minipage}
   \caption{
\small  Partial contributions to the $\mu_{\gamma\gamma}^{\rm res}$
in the non-minimal CSTC (with scalar $\mu_{S,H}$-terms included) as
functions of $s_\theta$ with $m_{\tilde \pi}=150$ GeV, $M_{\tilde
Q}=500$ GeV and $g_{\rm TC}=8$ (left panel) and in the minimal CSTC
without scalar $\mu_{S,H}$-terms as functions of $M_{\tilde \sigma}$
with $m_{\tilde \pi}=350$ GeV, $M_{\tilde Q}=500$ GeV and $g_{\rm
TC}=8$ (right panel), corresponding to $W$-loop (dashed lines), top
quark loop $\times$ 10 (dash-dotted lines), technifermion loop
$\times$ 10 (dotted lines), technipion loop $\times$ 1000
(short-dashed lines). At the both panels, solid lines correspond to
the total Higgs boson signal (resonant) strengths shown for
comparison. The rescaling of the curves is made for better
visibility and comparison.}
 \label{fig:muGG-res-partial}
\end{figure*}

The Higgs boson decay widths in the loop-induced $\gamma\gamma$ and
also in the $\gamma Z$ channels in the non-minimal CSTC with scalar
$\mu_{S,H}$-terms included are shown in Fig.~\ref{fig:Gamma-HGG-VV}
as functions of physical parameters of the model. This figure covers
only $s_\theta>0$ region and is complimentary to
Fig.~\ref{fig:muGG-res-spec}. One notices the regions where the
$\gamma\gamma$ and $\gamma Z$ widths can be very different from the
SM predictions, or close to them, or even turn to zero due to a
specific interference pattern. Also, the relation between
$\gamma\gamma$ and $\gamma Z$ widths strongly depends on parameters.
It is, however, more instructive to look directly at the Higgs
signal strengths in the respective decay channels as functions of
parameters, and we will primarily study the $\gamma\gamma$ channel
in detail here.

In particular, let us investigate to what extent the
$h\tilde{\sigma}$-mixing and the presence of the extra new
$\tilde{\pi}$ and $\tilde{Q}$ states in loops affects the resonance
Higgs signal strength in the $\gamma\gamma$ channel $\mu^{\rm
res}_{\gamma\gamma}$ and its smearing, given by
Eqs.~(\ref{mu-res-GG}) and (\ref{mu-exp}), respectively. For this
purpose, in Fig.~\ref{fig:muGG-res} we show the Higgs boson signal
strength in the $\gamma\gamma$ channel in the resonance region
$\mu_{\gamma\gamma}^{\rm res}(s_\theta)$ given by
Eq.~(\ref{mu-res-GG}) in the non-minimal case of the CSTC model with
scalar $\mu_{S,H}$-terms included. The $\mu_{\gamma\gamma}^{\rm
res}(s_\theta)$ weakly depends on $m_{\tilde \pi}$ value. It also
turns into zero at some $s_\theta^*>0$, which increases with
$M_{\tilde Q}$ and decreases with $g_{\rm TC}$. Note that there is
no symmetry $s_\theta\to -s_\theta$. In general, for $s_\theta<0$,
we always have in the resonance $\mu_{\gamma\gamma}^{\rm
res}(s_\theta)>1$, while smearing over the resonance can change
this. Also, smearing does not change significantly
$\mu_{\gamma\gamma}^{\rm res}(s_\theta)$ at small smearing angles
$s_\theta\to0$. The signal strength is close to unity for two
different cases of the mixing angle: in the no
$h\tilde{\sigma}$-mixing limit $s_\theta\to 0$ and for $s_\theta\sim
0.5-0.7$, while the latter is much more fine-tuned do to a sharp
behavior of $\mu_{\gamma\gamma}^{\rm res}(s_\theta)$; the third
configuration at negative $s_\theta$ appears due to a resonance
smearing described above. Note, that any relatively large mixing
configurations with $s^2_\theta>0.4$ are excluded by EW precision
constraints on $T$-parameter (see above).
\begin{figure*}[!h]
\begin{minipage}{0.7\textwidth}
 \centerline{\includegraphics[width=1.0\textwidth]{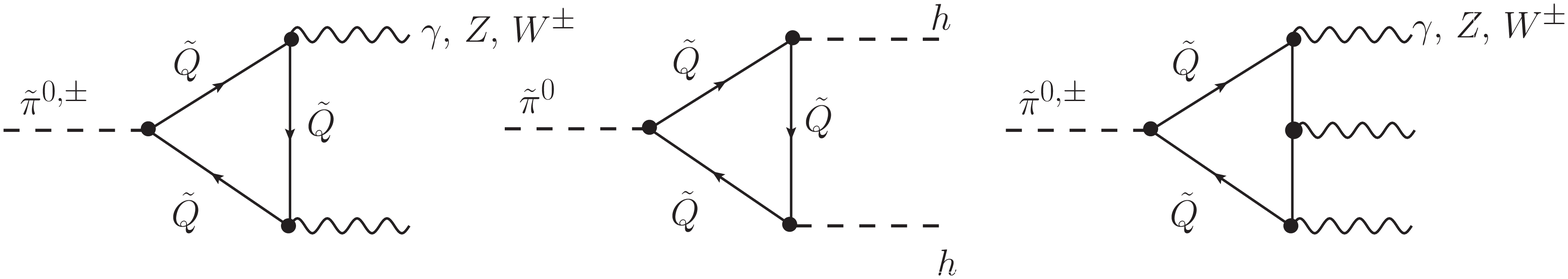}}
\end{minipage}
   \caption{
\small Light technipion loop-induced (2- and 3-body) decay modes in
the leading order through constituent technifermion loops.}
 \label{fig:pi-decay}
\end{figure*}

In Fig.~\ref{fig:muGG-res-spec} we show the same observable
$\mu^{\rm res}_{\gamma\gamma}$, but in the minimal CSTC scenario
without $\mu_{S,H}$-terms, as a function of $M_{\tilde \sigma}$. In
opposition to the non-minimal CSTC, in this case there is a very
strong dependence on $m_{\tilde \pi}$ parameter. Also, in the no
mixing limit $s_\theta\to 0$ which corresponds to $M_{\tilde
\sigma}\to\sqrt{3}m_{\tilde \pi}$, the strength turns to unity
$\mu^{\rm res}_{\gamma\gamma}\to 1$, as expected, and smearing does
not affect this. The current LHC data, in fact, prefer relatively
large technipion mass $m_{\tilde \pi}\gtrsim 250$ GeV and the small
$h\tilde{\sigma}$-mixing configuration in the parameter space, and a
small vicinity around the ``no $h\tilde{\sigma}$-mixing'' limit is
the only region of parameter space which satisfies the data in the
minimal CSTC and the Higgs boson looks as the standard one.

At last, in Fig.~\ref{fig:muGG-res-partial} we show partial
contributions to the Higgs signal strength in the resonance
$\mu^{\rm res}_{\gamma\gamma}$ coming from $W$-loop (dashed lines),
top quark loop $\times$ 10 (dash-dotted lines), technifermion loop
$\times$ 10 (dotted lines), technipion loop $\times$ 1000
(short-dashed lines), where the rescalings of the curves are made to
increase visibility. The shapes of the curves in the minimal and
non-minimal CSTC scenarios are very different, but in both cases
there is a strong interference pattern.

\subsection{Technipion and technisigma phenomenology}

\subsubsection{Technipion decay}

Besides the Higgs boson decay properties studied above, another
important phenomenological implication of the CSTC scenario concerns
possible technipion and technisigma signatures at the LHC. Since
technipions are pseudoscalar particles, at tree level they can be
produced only in pairs $\tilde{\pi}^+\tilde{\pi}^-$ or
$\tilde{\pi}^0\tilde{\pi}^0$, which have rather high invariant
masses $M_{\tilde{\pi}\tilde{\pi}}\gtrsim 300$ GeV, whereas
one-technipion production can be loop induced only (see below). In
order to define the phenomenological signatures of
technisigma/technipion production at colliders, one has to study
primarily the decay modes of produced technipions. In particular, an
identification of the produced $\tilde \pi$ mesons is important for
e.g. studies of the $\tilde \sigma$ meson properties at the LHC,
Yukawa and gauge couplings as well as constituent masses and
degeneration of the mass spectrum of the technifermions, etc.
\begin{figure*}[!h]
\begin{minipage}{0.325\textwidth}
 \centerline{\includegraphics[width=1.0\textwidth]{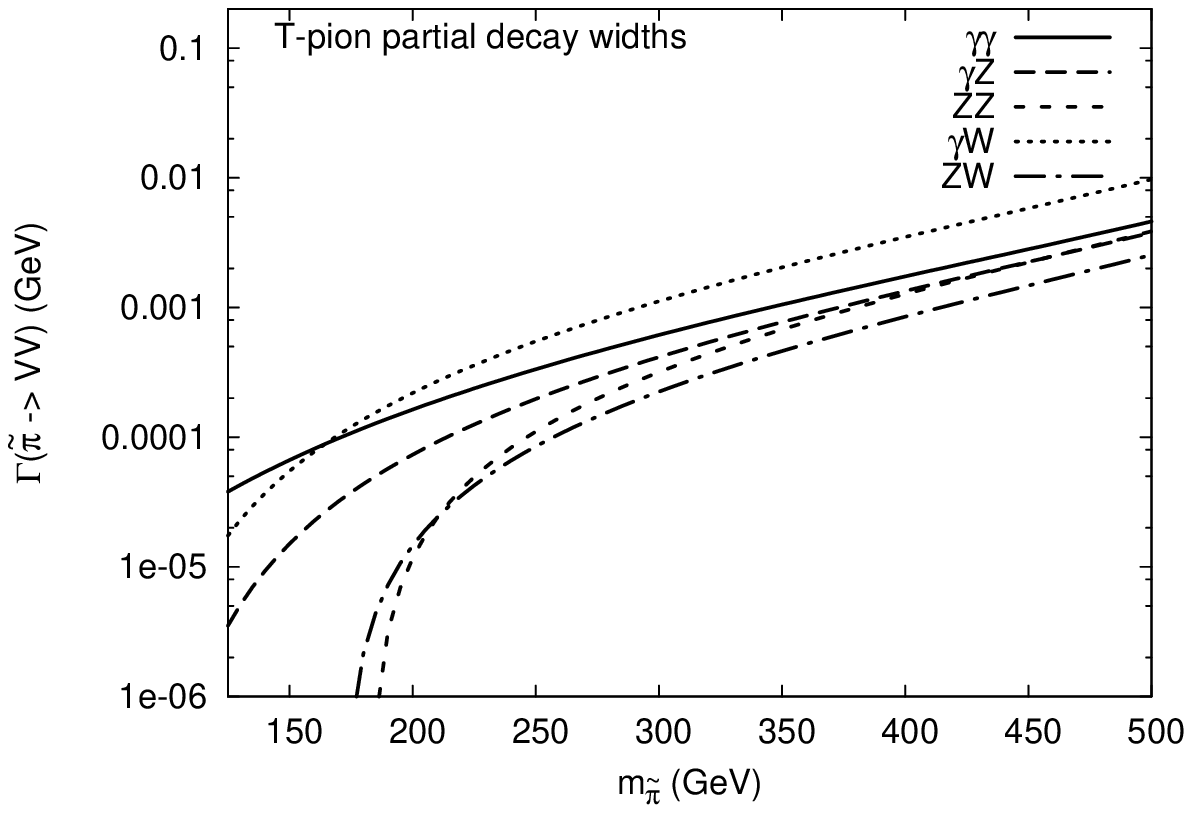}}
\end{minipage}
\begin{minipage}{0.325\textwidth}
 \centerline{\includegraphics[width=1.0\textwidth]{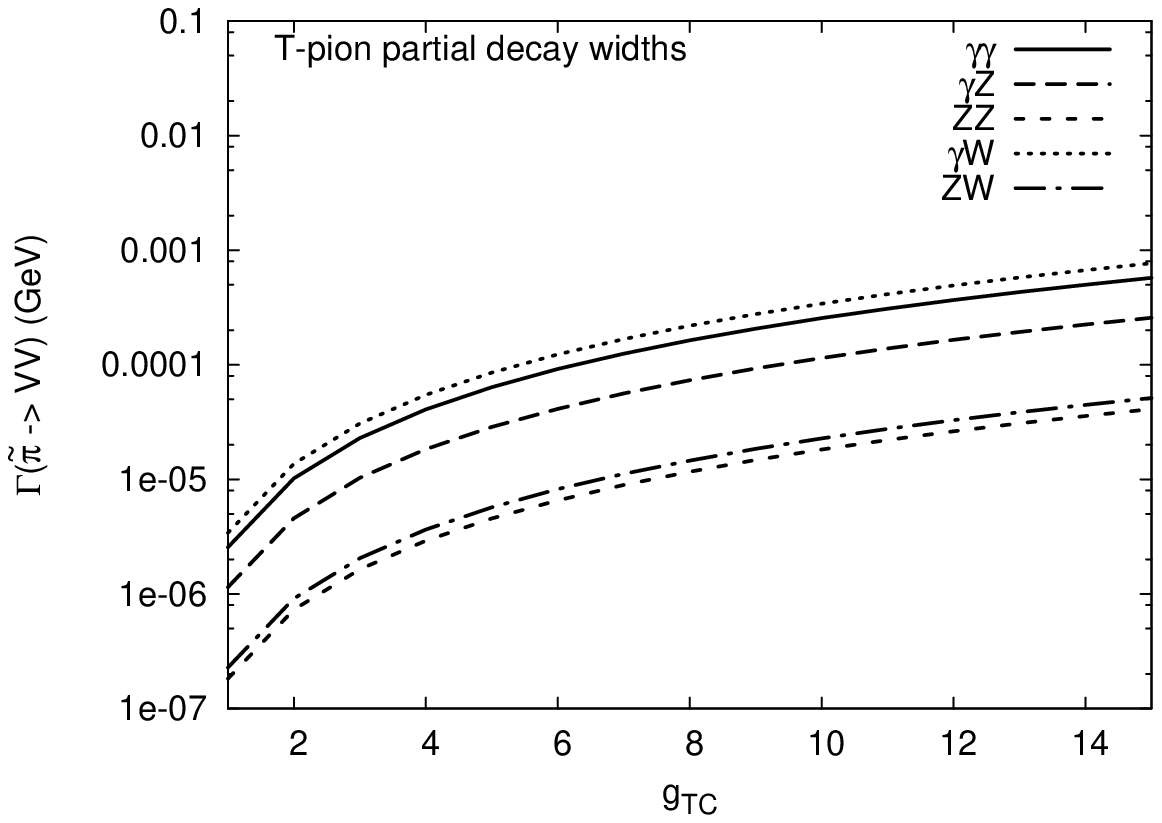}}
\end{minipage}
\begin{minipage}{0.325\textwidth}
 \centerline{\includegraphics[width=1.0\textwidth]{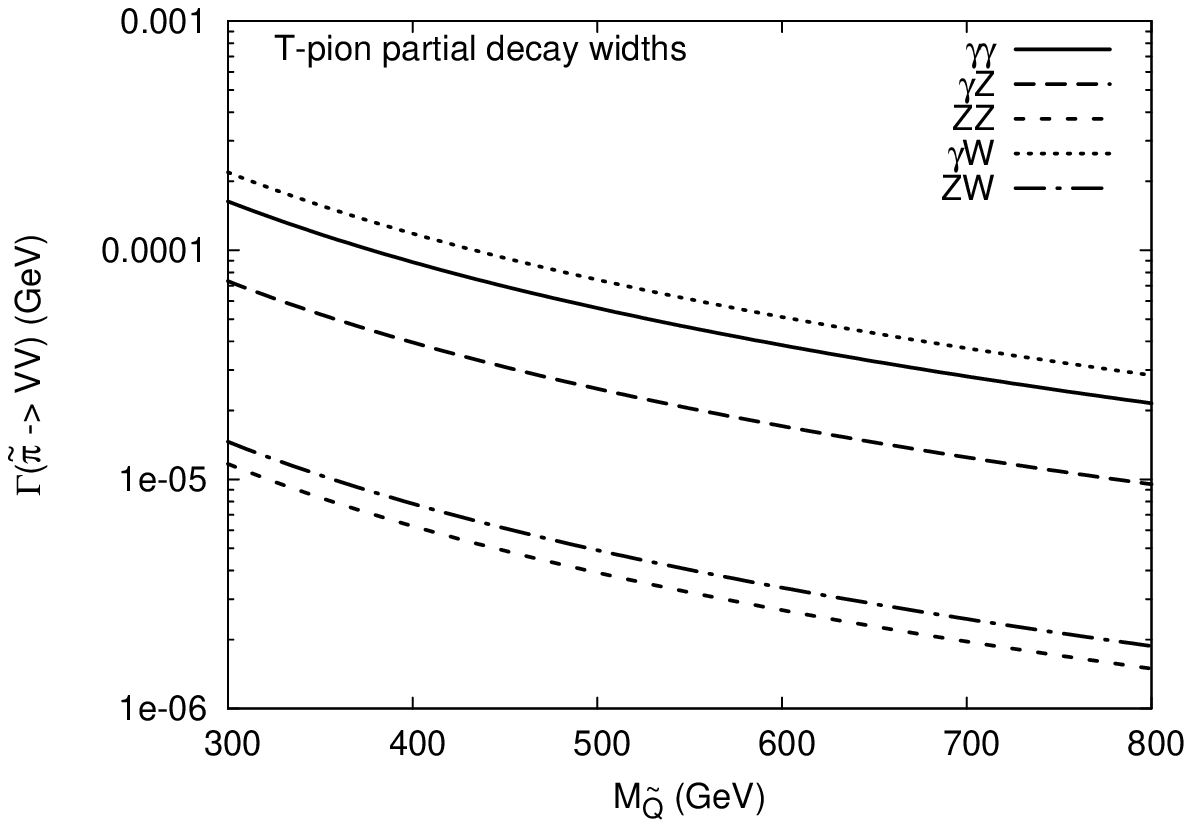}}
\end{minipage}
   \caption{
\small The technipion decay widths in the loop-induced
$\gamma\gamma$, $\gamma Z$, $\gamma W$, $ZZ$ and $ZW$ channels in
the non-minimal CSTC (with scalar $\mu_{S,H}$-terms included) as
functions of physical parameters of the model. The parameters in
each figure are set as follows: (left) $M_{\tilde Q}=300$ GeV,
$c_\theta^2=0.8$, and $g_{\rm TC}=8$; (middle) $M_{\tilde Q}=300$
GeV, $m_{\tilde \pi}=200$ GeV, and $c_\theta^2=0.8$; (right)
$m_{\tilde \pi}=200$ GeV, $c_\theta^2=0.8$, and $g_{\rm TC}=8$.
These results do not depend on $M_{\tilde \sigma}$.}
\label{fig:Gamma-Pi}
\end{figure*}
\begin{figure*}[!h]
\begin{minipage}{0.4\textwidth}
 \centerline{\includegraphics[width=1.0\textwidth]{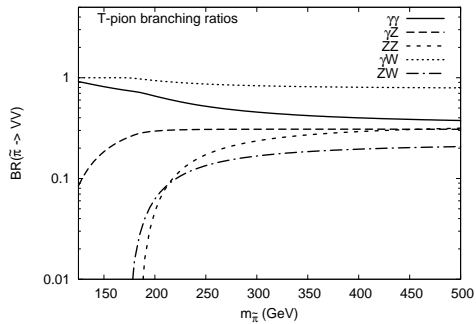}}
\end{minipage}
   \caption{
\small  The neutral and charged technipion branching ratios of the
loop-induced $\gamma\gamma$, $\gamma Z$, $\gamma W$, $ZZ$ and $ZW$
channels in the non-minimal CSTC (with scalar $\mu_{S,H}$-terms
included) as functions of $m_{\tilde\pi}$ for fixed $M_{\tilde
Q}=300$ GeV, $c_\theta^2=0.8$, and $g_{\rm TC}=8$.}
 \label{fig:Pi-BR}
\end{figure*}

It is of special interest for collider phenomenology to study
$\tilde \pi$ decays into vector bosons and, in principle, into a
pair Higgs bosons whose diagrams are represented as generic 2- and
3-body technifermion loop-induced processes in
Fig.~\ref{fig:pi-decay}. In the case of the mass-degenerated
technifermion doublet, it turns out that in the simplest case with
$Y_{\tilde Q}=0$ the 2-body technipion vector boson decay modes are
always forbidden by symmetry encoded in the structure of vertices,
whereas allowed for generic $Y_{\tilde Q}\not=0$ cases. The $\tilde
\sigma$ decays would manifest themselves as multi-lepton final
states with a large lepton multiplicity -- up to twelve leptons from
technipion pair decay in the case of $Y_{\tilde Q}=0$ or up to eight
leptons for $Y_{\tilde Q}=1/3$ in the final state from technisigma
decay (six and four leptons coming from each technipion in the above
cases, respectively), which would be rather challenging but very
interesting to study.
\begin{figure*}[!b]
\begin{minipage}{0.325\textwidth}
 \centerline{\includegraphics[width=1.0\textwidth]{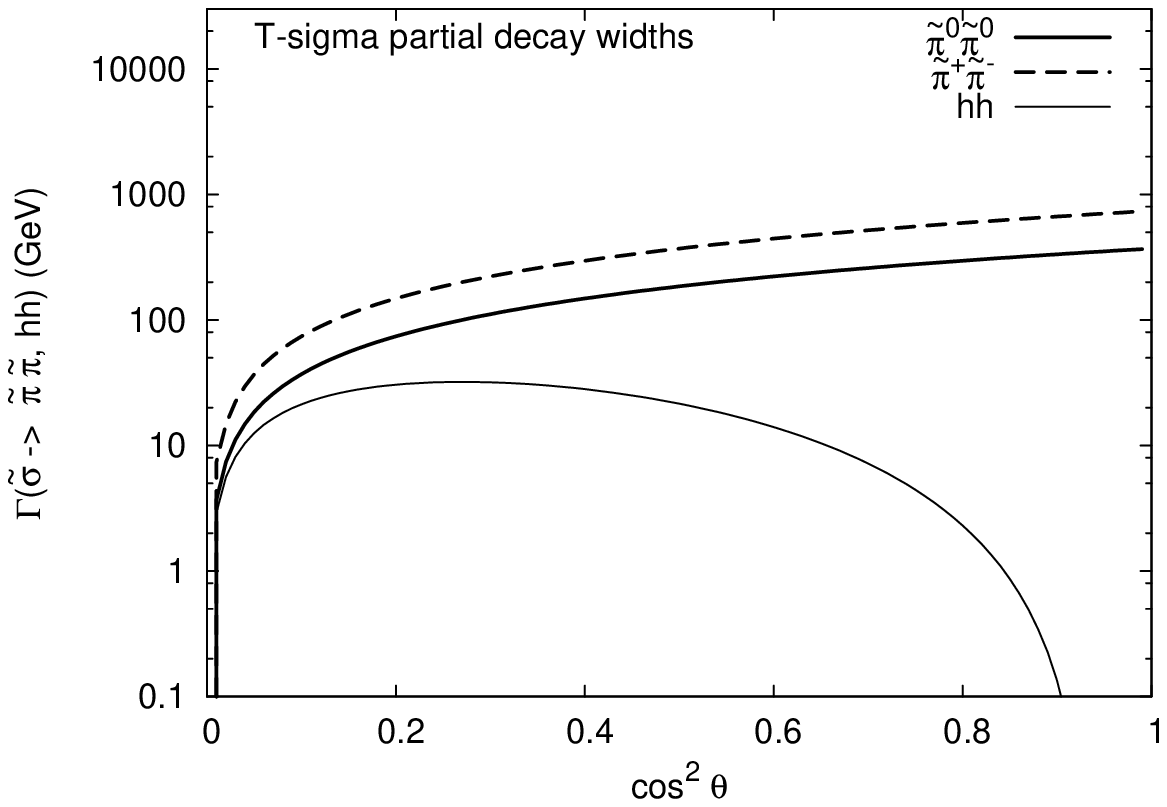}}
\end{minipage}
\begin{minipage}{0.325\textwidth}
 \centerline{\includegraphics[width=1.0\textwidth]{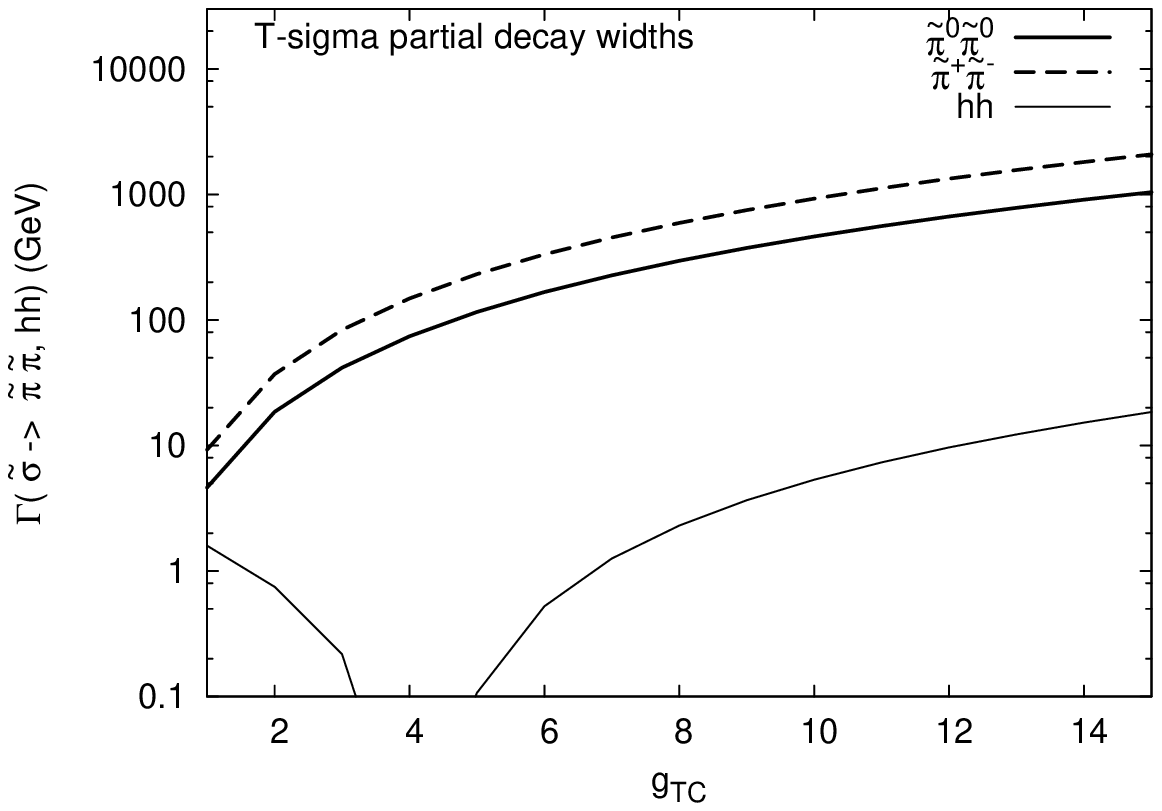}}
\end{minipage}
\begin{minipage}{0.325\textwidth}
 \centerline{\includegraphics[width=1.0\textwidth]{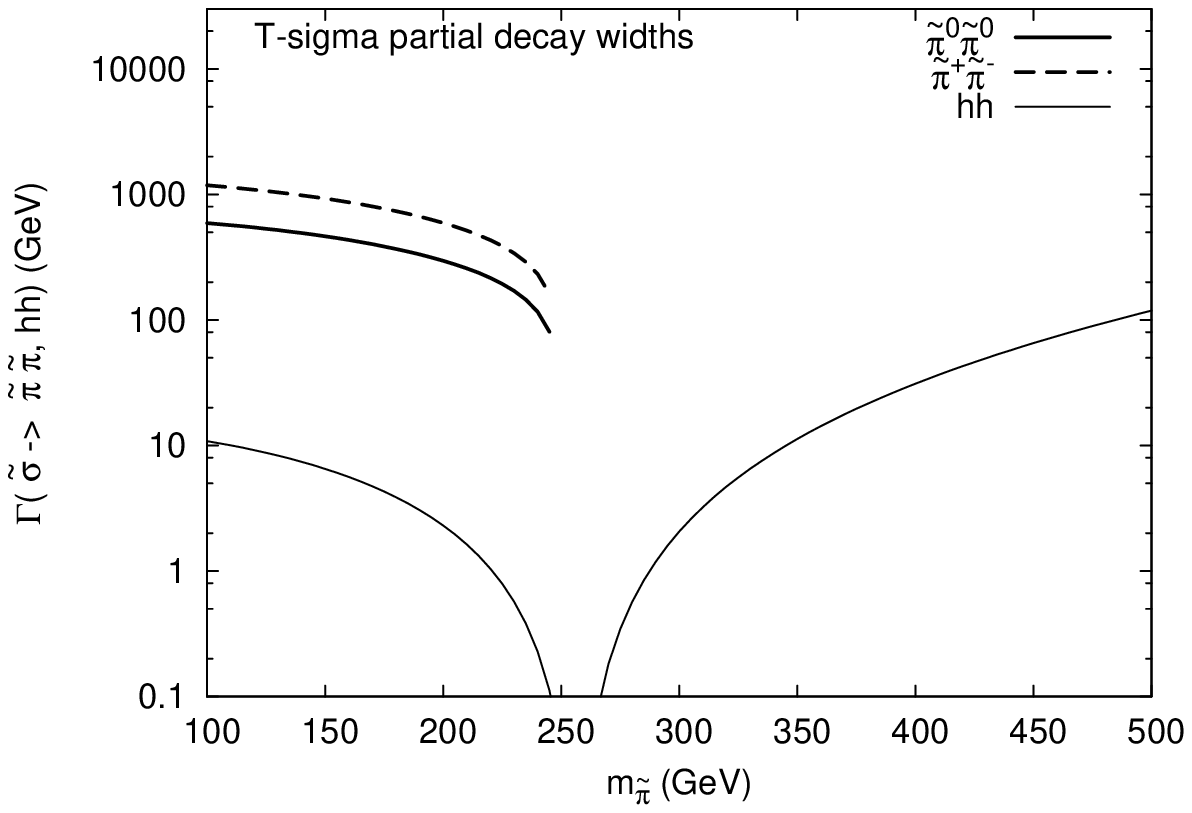}}
\end{minipage}
\begin{minipage}{0.325\textwidth}
 \centerline{\includegraphics[width=1.0\textwidth]{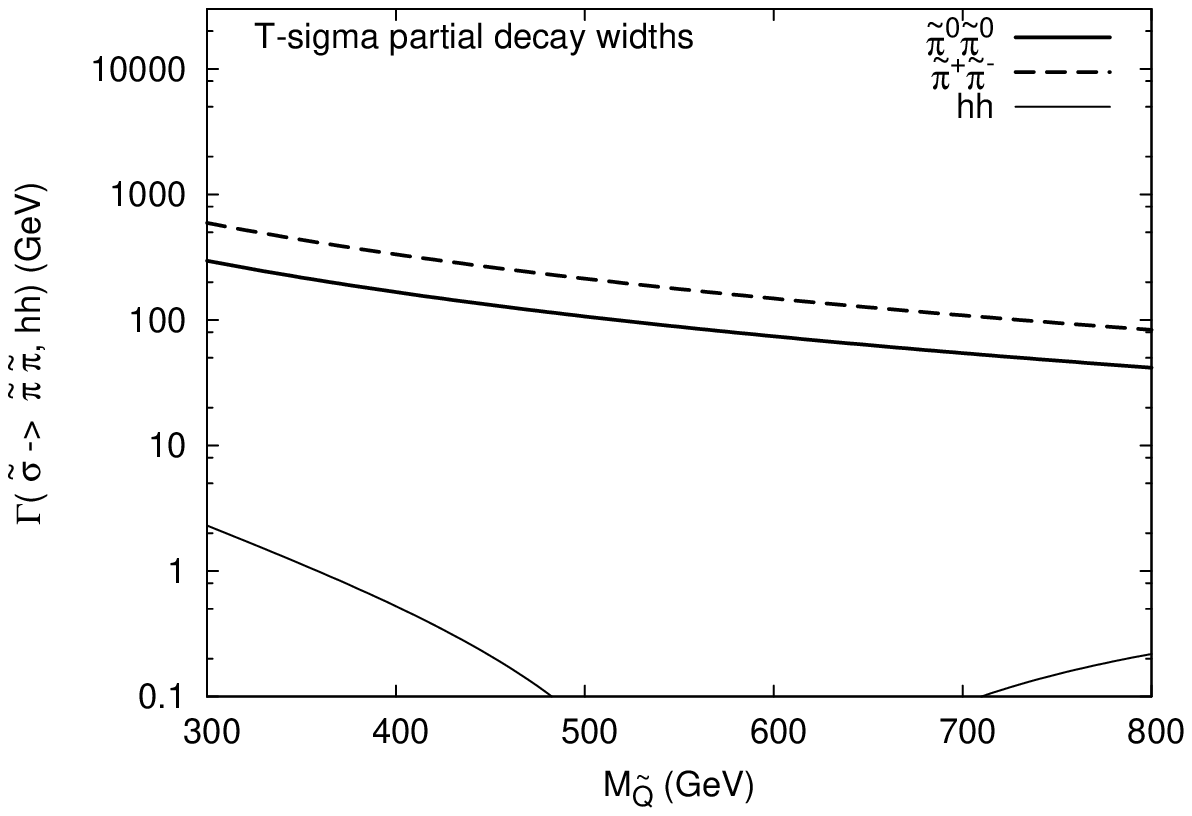}}
\end{minipage}
\begin{minipage}{0.325\textwidth}
 \centerline{\includegraphics[width=1.0\textwidth]{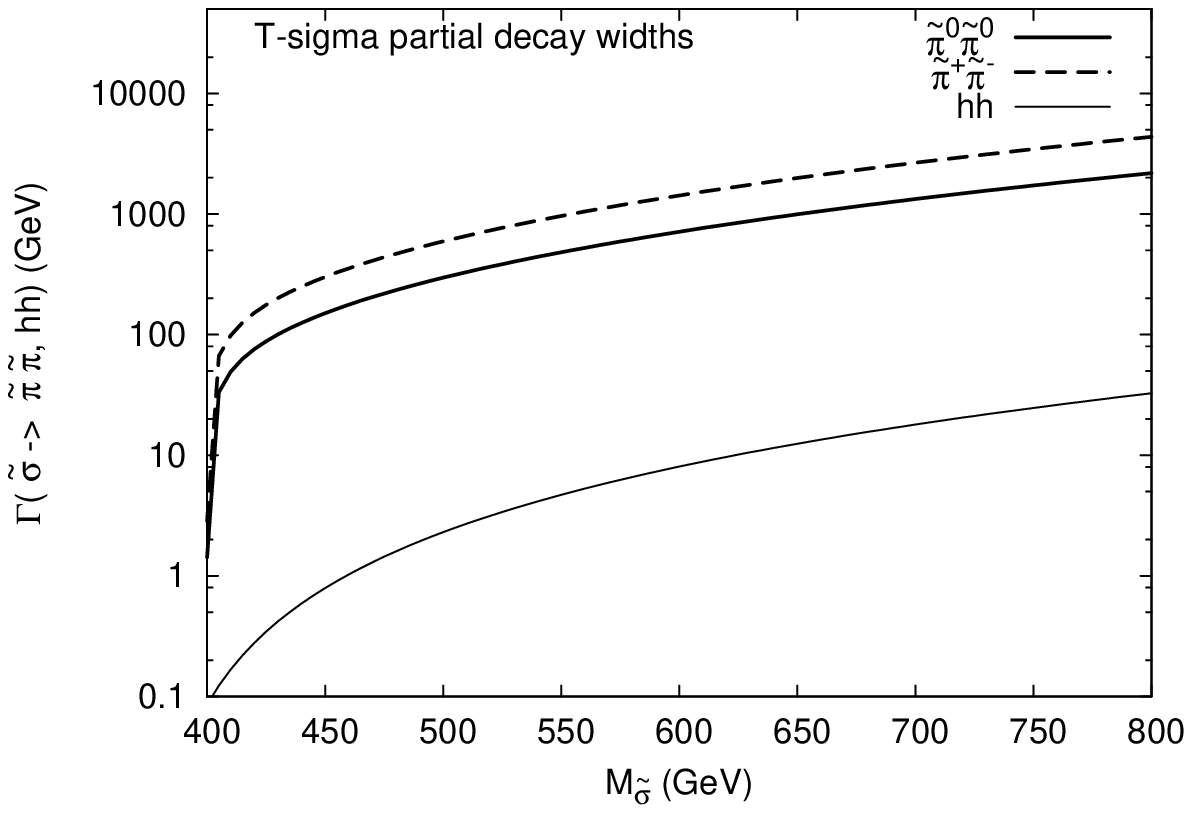}}
\end{minipage}
   \caption{
\small The technisigma tree-level decay widths in the
$\tilde{\pi}\tilde{\pi}$ and $hh$ channels in the non-minimal CSTC
(with scalar $\mu_{S,H}$-terms included) as functions of physical
parameters of the model. The parameters in each figure are fixed as
$m_{\tilde \pi}=200$ GeV, $M_{\tilde Q}=300$ GeV, $M_{\tilde
\sigma}=500$ GeV, $c_\theta^2=0.8$, $g_{\rm TC}=8$, such that in
each figure one drops off a variable from this list corresponding to
the one at the respective $x$-axis while keeping others fixed.}
 \label{fig:Gamma-Sig-scalar}
\end{figure*}

In general, one would deal with many possible four-vector $VVVV$,
four-Higgs $hhhh$ or mixed $hhVV$ final states in order to
reconstruct the technisigma mass, and this procedure gets even more
complicated due a very large $\tilde \sigma$ width. If there are no
visible deviations of the Higgs boson properties from the SM ones,
the technipion/technisigma phenomenology, as well as Higgs-scalar
self-couplings and studies of various loop-induced processes with
the Higgs boson participation, even though very challenging, would
be the only source of information about the CSTC sector possibly
available at the LHC. The technipion 2-body decay modes into the
on-shell gauge bosons, namely, into the $\gamma\gamma$, $\gamma Z$,
$\gamma W$, $ZZ$ and $ZW$ final states (above the corresponding
thresholds), in the case with $Y_{\tilde Q}=1/3$ are given by:
\begin{eqnarray*}
&&\Gamma(\tilde{\pi}^0\to\gamma\gamma)=\frac{\alpha^2 g_{\rm
TC}^2}{4\pi^3}\frac{M_{\tilde Q}^2}{m_{\tilde
\pi}}\,\arcsin^4\Bigl(\frac{m_{\tilde \pi}}{2M_{\tilde Q}}\Bigr)\,,
\qquad \frac{m_{\tilde \pi}}{2M_{\tilde
Q}}<1\,, \\
&&\Gamma(\tilde{\pi}^0\to\gamma Z)=\frac{\alpha^2 g_{\rm
TC}^2}{2\pi^3}\frac{M_{\tilde Q}^2}{m_{\tilde
\pi}}\,\cot^22\theta_W\,\Bigl(1-\frac{M_Z^2}{m_{\tilde
\pi}^2}\Bigr)\,\left[\arcsin^2\Bigl(\frac{m_{\tilde \pi}}{2M_{\tilde
Q}}\Bigr)-\arcsin^2\Bigl(\frac{M_Z}{2M_{\tilde
Q}}\Bigr)\right]^2\,,\\ &&\Gamma(\tilde{\pi}^\pm\to\gamma
W^\pm)=\frac{\alpha^2 g_{\rm TC}^2}{2\pi^3s_W^2}\frac{M_{\tilde
Q}^2}{m_{\tilde \pi}}\,\Bigl(1-\frac{M_W^2}{m_{\tilde
\pi}^2}\Bigr)\,\left[\arcsin^2\Bigl(\frac{m_{\tilde \pi}}{2M_{\tilde
Q}}\Bigr)-\arcsin^2\Bigl(\frac{M_W}{2M_{\tilde
Q}}\Bigr)\right]^2\,,\\ &&\Gamma(\tilde{\pi}^0\to ZZ)=\frac{\alpha^2
g_{\rm TC}^2}{16\pi^3}\,M_{\tilde Q}^2m_{\tilde
\pi}^3\,\bar{\lambda}^3(M_Z^2,M_Z^2,m_{\tilde
\pi}^2)\,C_0^2(M_Z^2,M_Z^2,m_{\tilde \pi}^2;M_{\tilde Q}^2)\,,\\
&&\Gamma(\tilde{\pi}^\pm\to ZW^\pm)=\frac{\alpha^2 g_{\rm
TC}^2}{32\pi^3c_W^2}\,M_{\tilde Q}^2m_{\tilde
\pi}^3\,\bar{\lambda}^3(M_Z^2,M_W^2,m_{\tilde
\pi}^2)\,C_0^2(M_Z^2,M_W^2,m_{\tilde \pi}^2;M_{\tilde Q}^2)\,,
\end{eqnarray*}
respectively, where the normalized K\"allen function is defined in
Eq.~(\ref{Kallen}), and $C_0(m_1^2,m_2^2,m_3^2;m^2)\equiv
C_0(m_1^2,m_2^2,m_3^2;m^2,m^2,m^2)$ is the standard finite
three-point function. Note, the $\tilde \pi^0\to WW$ decay mode is
forbidden by symmetry. The complete set of $\tilde \pi$ decay rates
(the $\tilde{\pi}^0\to hh$ decay rate which, in principle, exists
for heavy technipions vanishes in the ``no $h\sigma$-mixing'' limit
and not included into the analysis) is shown for the non-minimal
CSTC scenario in Fig.~\ref{fig:Gamma-Pi} as functions of the model
parameters. The branching ratios as functions of $m_{\tilde\pi}$ at
a fixed point in the parameter space as an example are shown in
Fig.~\ref{fig:Pi-BR}. Interestingly enough, the total technipion
decay width is dominated by the $\gamma W^\pm$ channel in the
$\tilde \pi^\pm$ decay, and by the $\gamma\gamma$ channel in the
$\tilde \pi^0$ decay, although other decay modes are not negligible
in general.

\subsubsection{Technisigma decay}

The tree-level 2-body $\tilde \sigma$ decay widths into
$\tilde{\pi}\tilde{\pi}$, $f{\bar f}$, $ZZ$ and $WW$ are given by
the following expressions:
\begin{eqnarray*}
&&\Gamma(\tilde{\sigma}\to
\tilde{\pi}\tilde{\pi})=\frac{3g_{\tilde{\sigma}\tilde{\pi}}^2}{8\pi
M_{\tilde
\sigma}}\sqrt{1-\frac{4m_{\tilde{\pi}}^2}{M_{\tilde{\sigma}}^2}}\,,\qquad
g_{\tilde{\sigma}\tilde{\pi}}=-\lambda_{\rm TC}\,u
c_\theta-\lambda\,v s_\theta\,,\\ &&\Gamma(\tilde{\sigma}\to
\bar{f}f) = \frac{g^2s_{\theta}^2}{32\pi}\, M_{\tilde{\sigma}}\,
\frac{M_f^2}{M_W^2}\left(1-\frac{4M_f^2}{M_{\tilde{\sigma}}^2}\right)^{3/2}\,,
\end{eqnarray*}
\begin{eqnarray*}
&&\Gamma(\tilde{\sigma}\to ZZ) =
\frac{g^2s_{\theta}^2}{16\pi}\,\frac{M_Z^2}{M_{\tilde{\sigma}}c_{W}^2}
\left(1-\frac{4M_Z^2}{M_{\tilde{\sigma}}^2}\right)^{1/2}\cdot
\left[1+\frac{(M_{\tilde{\sigma}}^2-2M_Z^2)^2}{8M_Z^4}\right]\,,\\
&&\Gamma(\tilde{\sigma}\to WW) =
\frac{g^2s_{\theta}^2}{8\pi}\,\frac{M_W^2}{M_{\tilde{\sigma}}}
\left(1-\frac{4M_W^2}{M_{\tilde{\sigma}}^2}\right)^{1/2}\cdot
\left[1+\frac{(M_{\tilde{\sigma}}^2-2M_W^2)^2}{8M_W^4}\right]\,,
\end{eqnarray*}
\begin{figure*}[!h]
\begin{minipage}{0.325\textwidth}
 \centerline{\includegraphics[width=1.0\textwidth]{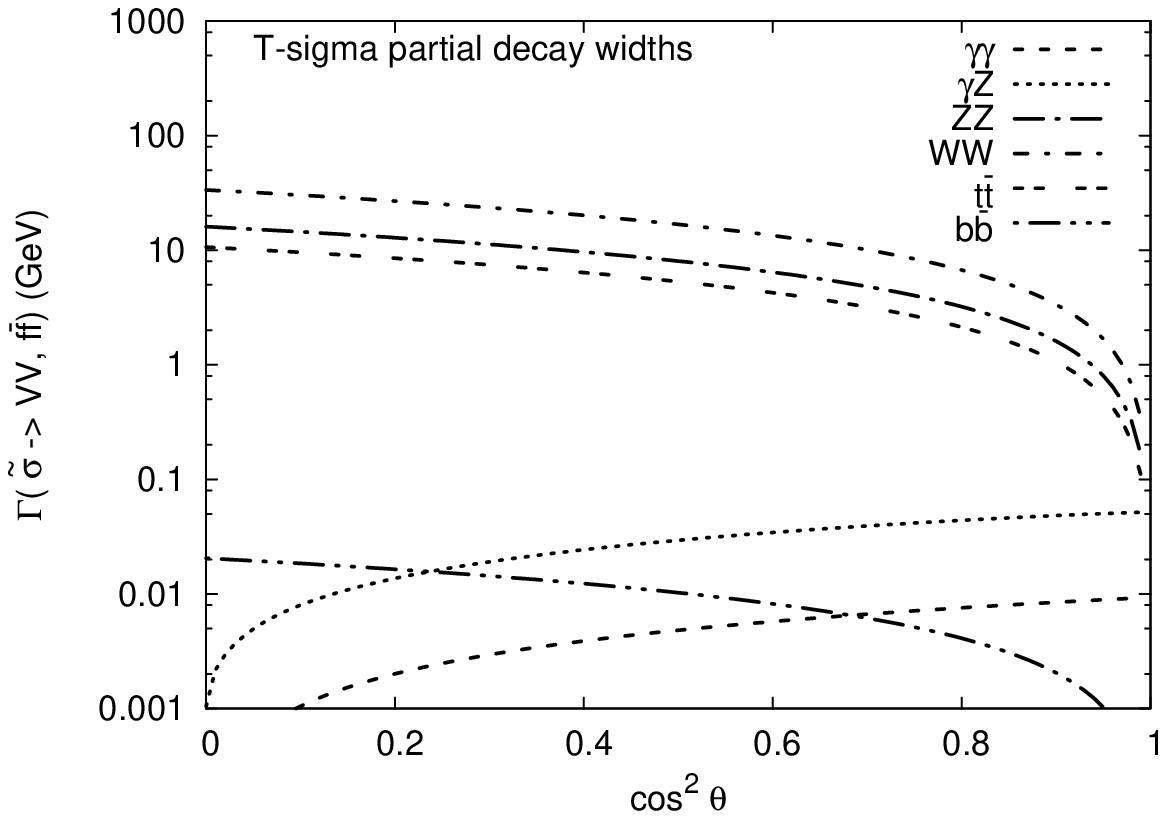}}
\end{minipage}
\begin{minipage}{0.325\textwidth}
 \centerline{\includegraphics[width=1.0\textwidth]{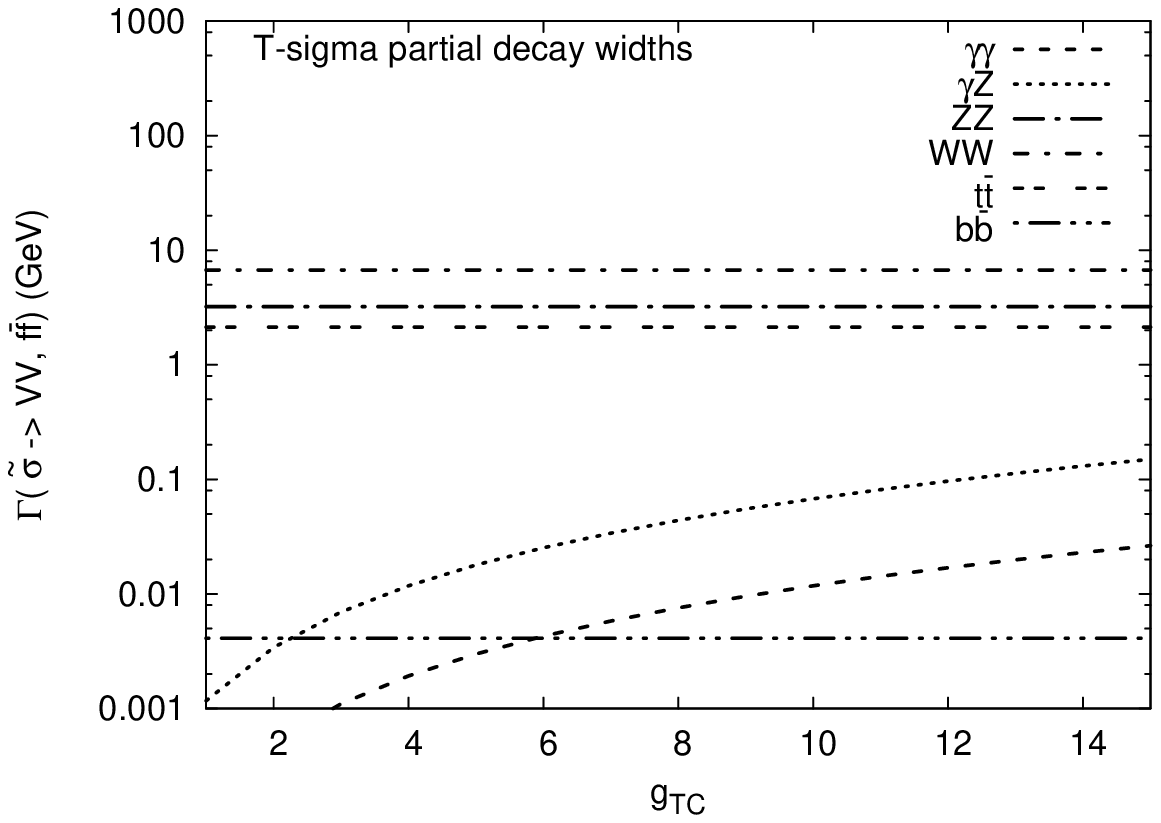}}
\end{minipage}
\begin{minipage}{0.325\textwidth}
 \centerline{\includegraphics[width=1.0\textwidth]{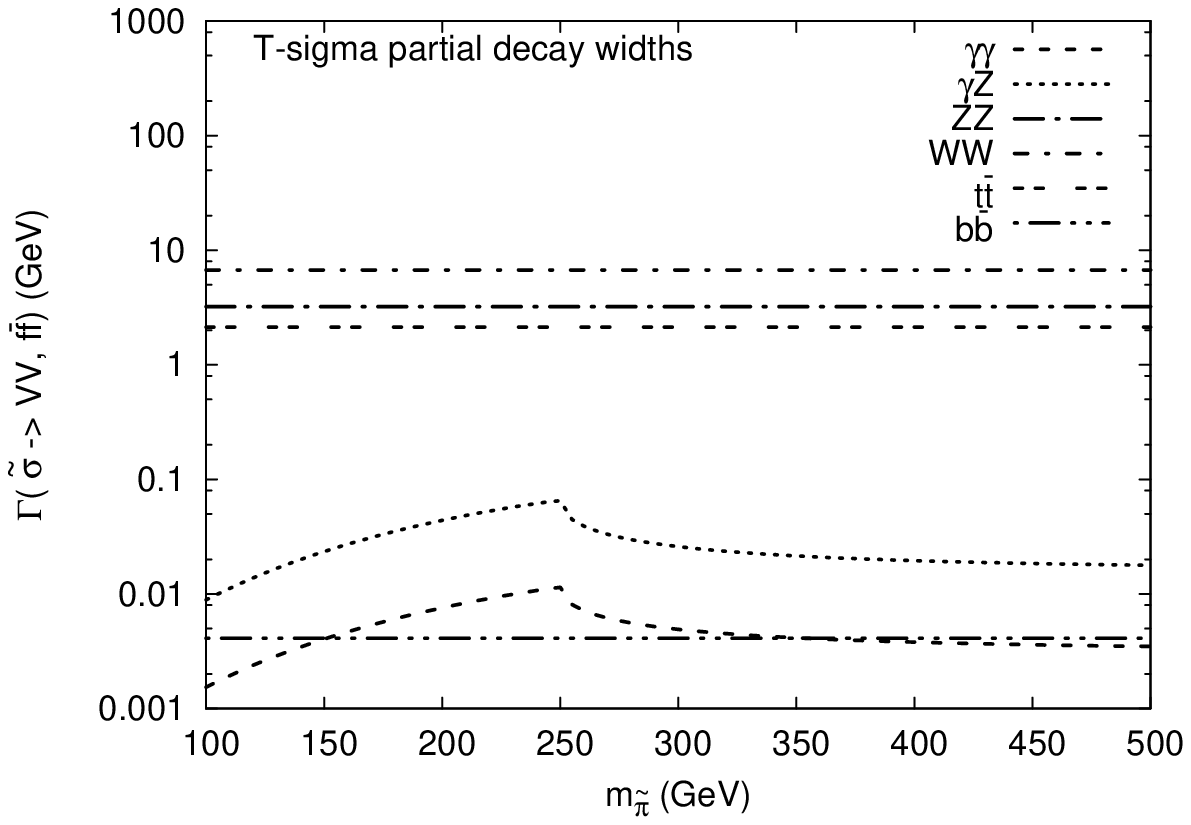}}
\end{minipage}
\begin{minipage}{0.325\textwidth}
 \centerline{\includegraphics[width=1.0\textwidth]{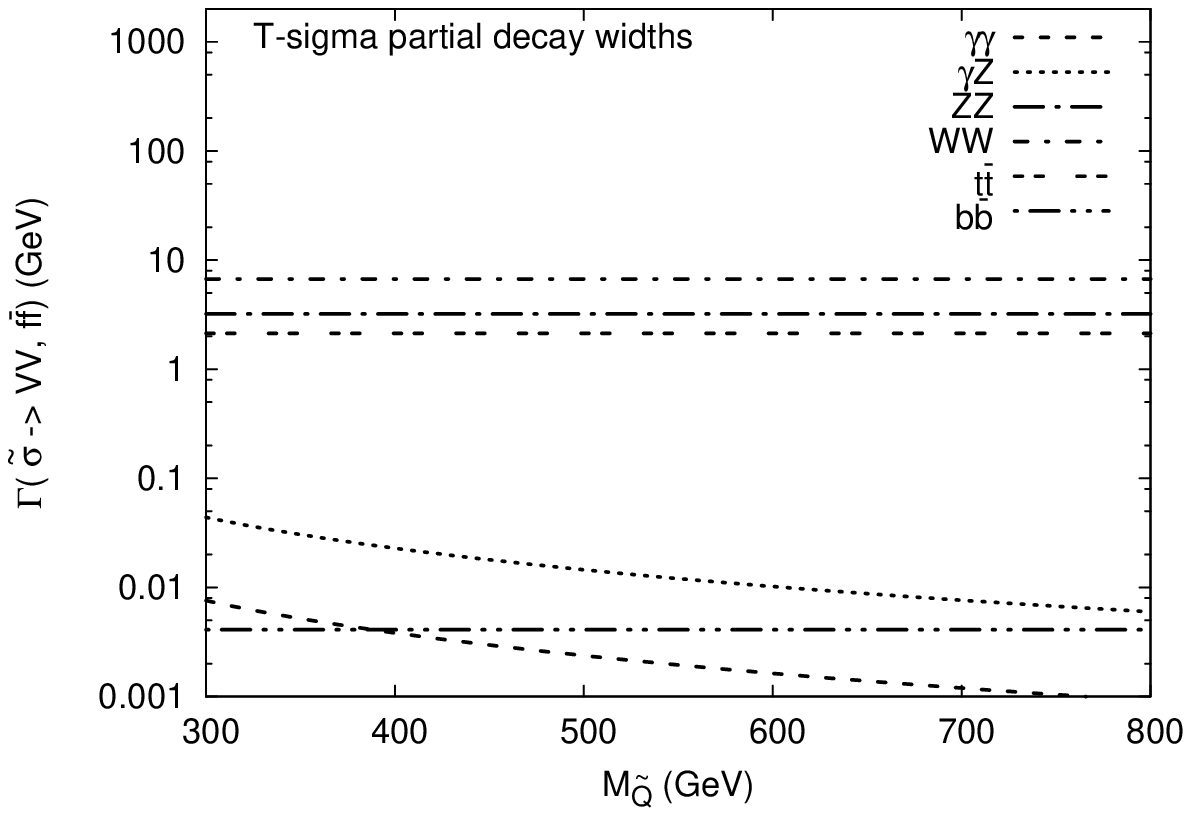}}
\end{minipage}
\begin{minipage}{0.325\textwidth}
 \centerline{\includegraphics[width=1.0\textwidth]{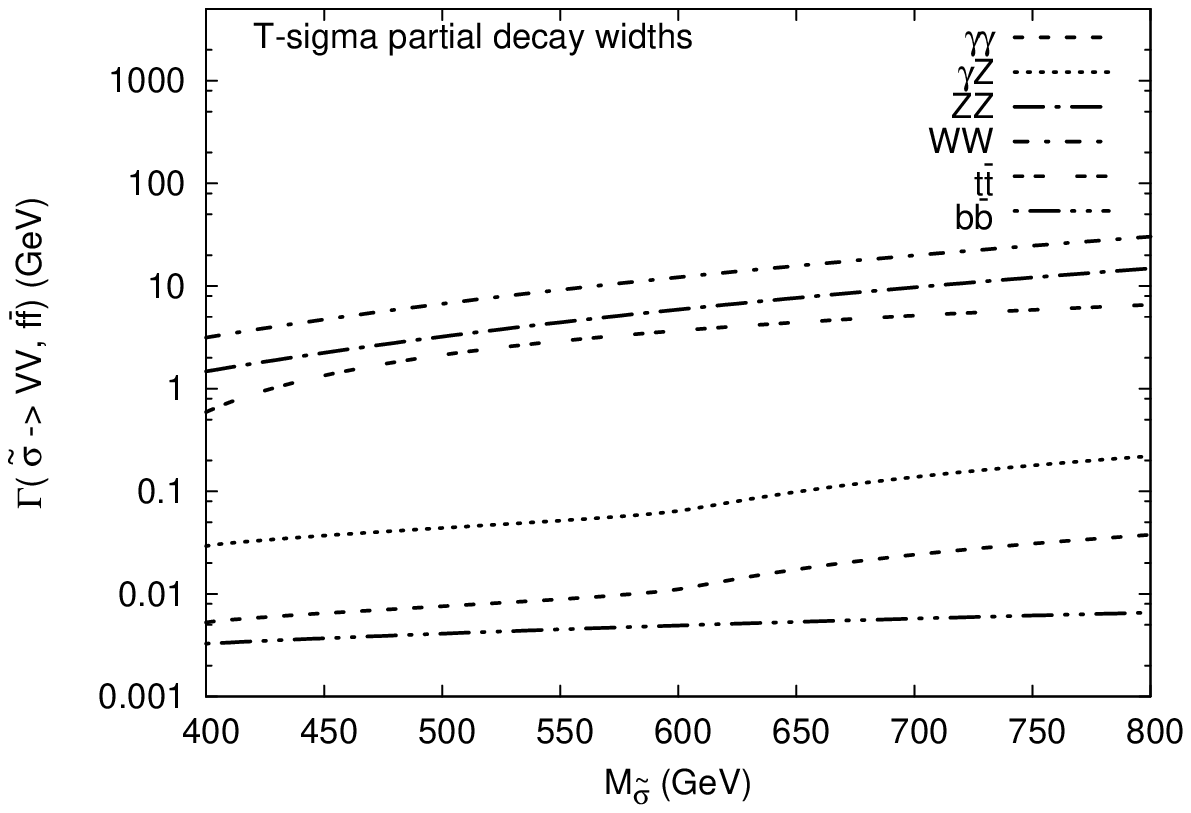}}
\end{minipage}
   \caption{
\small The technisigma tree-level decay widths in the fermion
($t\bar{t}$, $b\bar{b}$) and gauge boson ($\gamma\gamma$, $\gamma
Z$, $ZZ$ and $WW$) channels in the non-minimal CSTC (with scalar
$\mu_{S,H}$-terms included) as functions of physical parameters of
the model. The set-up of parameters is the same as in
Fig.~\ref{fig:Gamma-Sig-scalar}.}
 \label{fig:Gamma-Sig-other}
\end{figure*}
respectively, while the loop-induced $\tilde\sigma$ decay widths in
the $\gamma\gamma$ and $\gamma Z$ channels can be obtained from that
of the Higgs boson by a replacement $c_\theta\to s_\theta,\,M_h\to
M_{\tilde \sigma}$, and thus are not shown here explicitly. The
(pseudo)scalar ($hh$ and $\tilde{\pi}\tilde{\pi}$) decay modes are
shown for the non-minimal CSTC scenario in
Fig.~\ref{fig:Gamma-Sig-scalar} as functions of the model
parameters, while fermion ($t\bar{t}$, $b\bar{b}$) and gauge boson
($\gamma\gamma$, $\gamma Z$, $ZZ$ and $WW$) decay channels are given
in Fig.~\ref{fig:Gamma-Sig-other}. One notices that the technipion
modes of the $\tilde\sigma$ decay strongly dominate the total
$\tilde\sigma$ decay width, and can be as large as a few hundreds
GeV being comparable to $M_{\tilde\sigma}$. Certainly,
$\tilde\sigma$ is a highly unstable and unusually broad state, for
which one cannot use the narrow width approximation, and it is an
open question how to identify it experimentally.

\subsubsection{One-technipion production}

\begin{figure*}[!h]
\begin{minipage}{0.5\textwidth}
 \centerline{\includegraphics[width=1.0\textwidth]{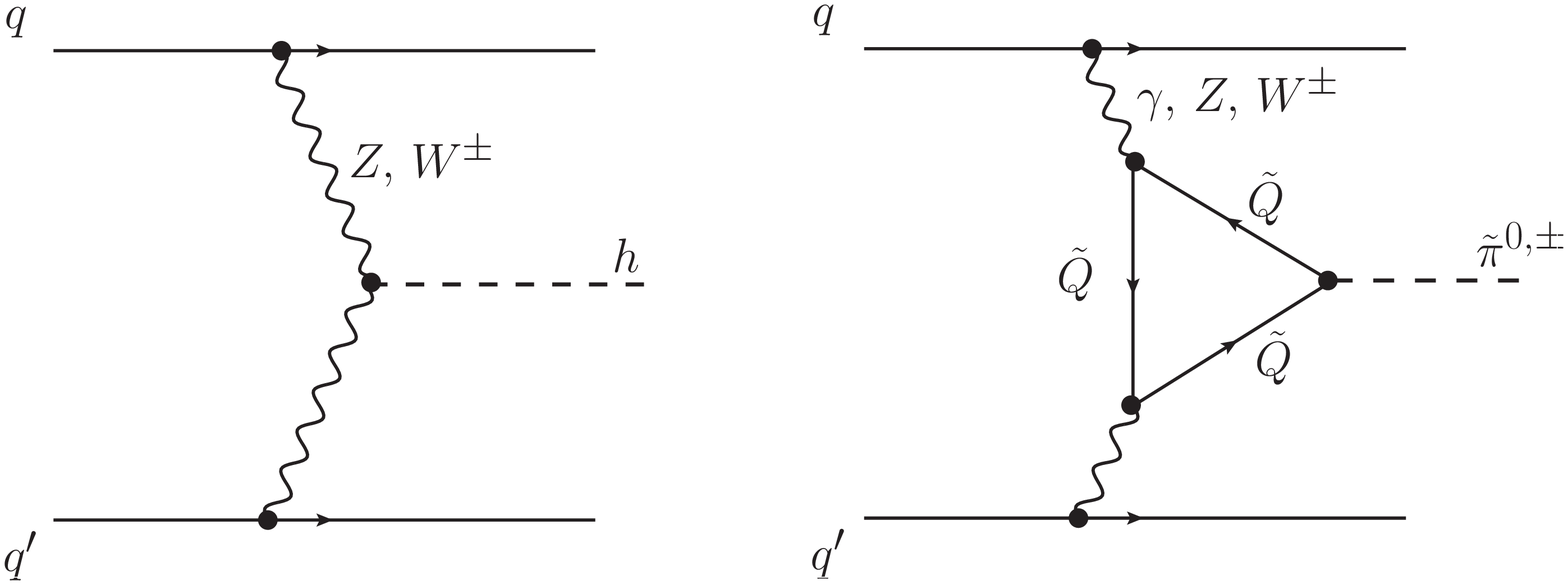}}
\end{minipage}
   \caption{
\small Typical production channels of the Higgs boson at tree level
(left) and technipion via a triangle technifermion loop (right) via
gauge boson fusion in the quark-(anti)quark scattering.}
 \label{fig:PiH-prod-diag}
\end{figure*}

As has been mentioned above, one technipion can be produced only at
the loop level. Let us look into this possibility in more detail
since this channel is especially important for understanding the
discovery potential of Technicolor at the LHC, even in the absence
of any deviations the Higgs boson signal strengths from the SM
predictions. Corresponding typical partonic $2\to3$ hard subprocess
of Higgs boson and $\tilde{\pi}$ production in high energy
hadron-hadron collisions via intermediate vector boson fusion (VBF)
mechanism is shown in Fig.~\ref{fig:PiH-prod-diag}.
The Higgs boson VBF production channel (left panel) shown for
comparison with the technipion channel (right panel) is one of the
key production modes recently studied at the LHC which allowed for
clear discrimination of the Higgs signal and large backgrounds
\cite{ATLAS,CMS}. The Higgs boson has also other production modes
e.g. via gluon-gluon fusion mechanism and the Higgsstrahlung off
gauge bosons and heavy flavor. In opposite to the Higgs boson, one
technipion can only be produced via heavy technifermion triangle
loop in the VBF mechanism. In numerical estimations, it is
explicitly assumed that the incoming quark $q$ and (anti)quark $q'$
loose only a small fraction of their initial energy taken away by
intermediate vector bosons. In this kinematics, the final-state
quarks are seen as forward-backward hard jets, and by measuring
their momenta one accurately reconstructs the invariant mass of the
produced state. An overall one-technipion production rate is
expected to be strongly suppressed compared to the Higgs boson
production rate, which along with extremely narrow technipion
resonance makes it rather hard to measure experimentally but not
impossible.

\begin{figure*}[!h]
\begin{minipage}{0.325\textwidth}
 \centerline{\includegraphics[width=1.0\textwidth]{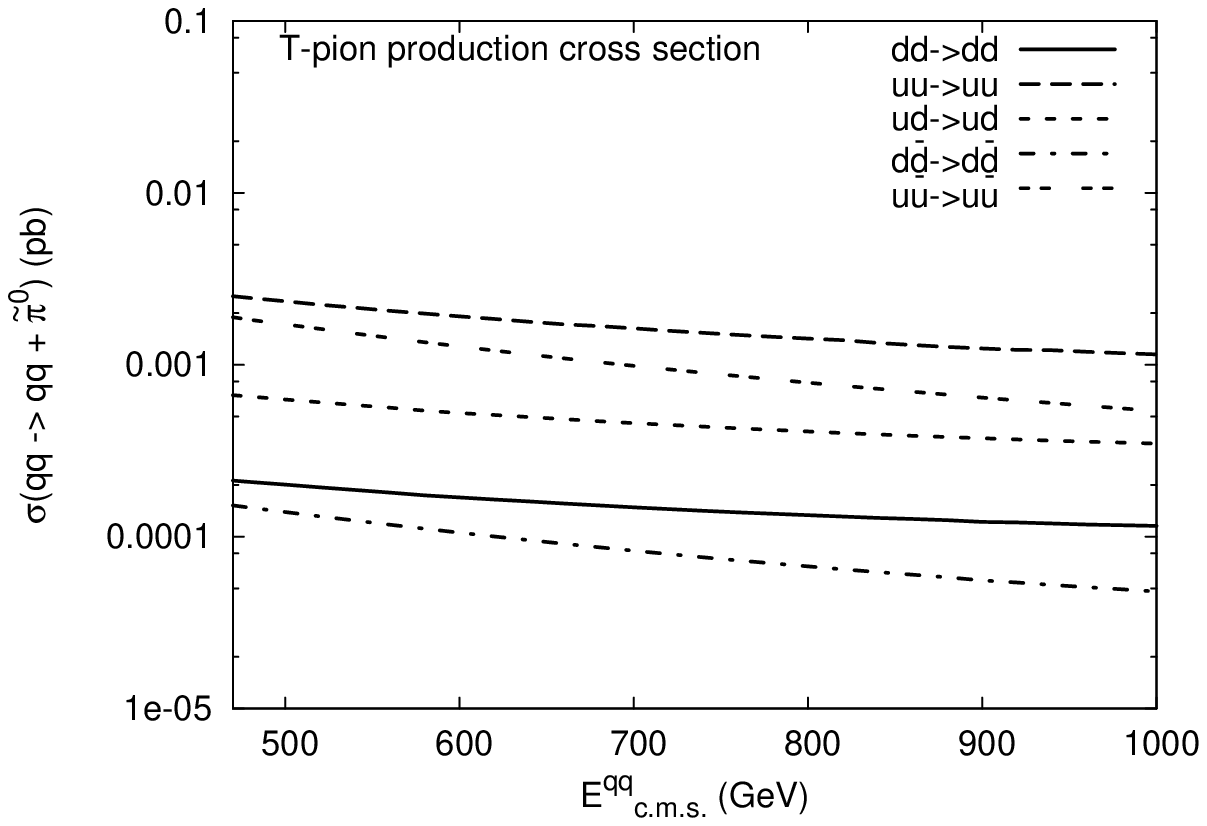}}
\end{minipage}
\begin{minipage}{0.325\textwidth}
 \centerline{\includegraphics[width=1.0\textwidth]{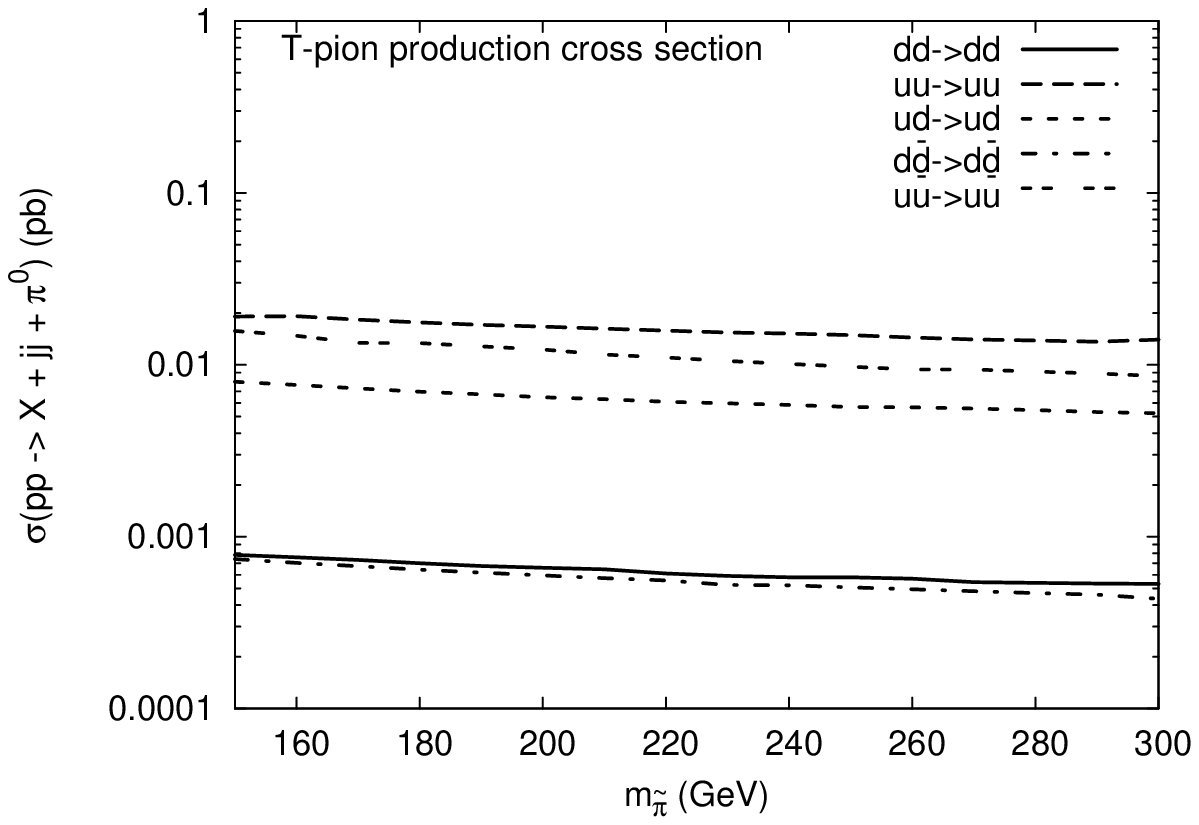}}
\end{minipage}
\begin{minipage}{0.325\textwidth}
 \centerline{\includegraphics[width=1.0\textwidth]{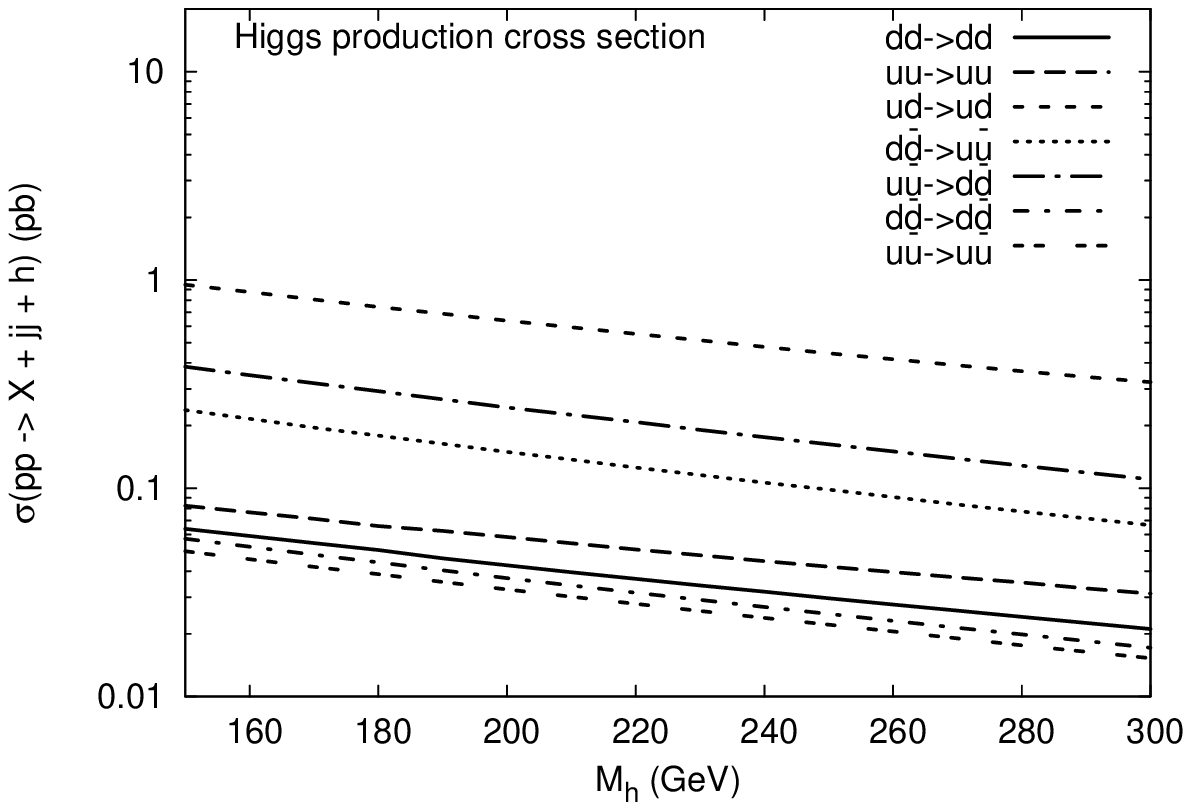}}
\end{minipage}
   \caption{ \small The one-technipion (T-pion) production cross sections via the VBF
   mechanism at the parton level for different
   incoming and outgoing quark $q$ and (anti)quark $q'$ states as functions of $qq'$ invariant
   mass, or c.m.s. energy $E^{qq}_{c.m.s.}=\sqrt{\hat{s}}$ (left), corresponding
   total hadron level cross sections of one technipion production for given
   incoming $qq'$ states in picobarns (before cuts) at the maximal LHC energy
   $\sqrt{s}=14$ TeV as a function of the technipion mass $m_{\tilde \pi}$ (middle),
   and corresponding VBF hadronic cross sections of the Higgs boson as functions
   of its mass $M_h$ shown for comparison. Here, $g_{\rm TC}=8$ and
   $M_{\tilde Q}=300$ GeV are fixed, and the results do not depend on other
   CSTC parameters. In calculations of the hadronic cross sections in this paper
   we have used quark CTEQ5LO PDFs \cite{Lai:1999wy}.}
 \label{fig:Pi0-prod}
\end{figure*}

In Fig.~\ref{fig:Pi0-prod} we show the one-technipion production
cross sections via the VBF mechanism at the parton level for
different incoming and outgoing quark $q$ and (anti)quark $q'$
states. Both parton-level and hadron-level cross sections at the LHC
with $\sqrt{s}=14$ TeV in the relevant kinematics and mass ranges
along with corresponding Higgs boson cross sections in respective
channels are presented (here and below, CTEQ5LO quark PDFs
\cite{Lai:1999wy} were used in calculations). Only up and down
quarks with at least one valence quark as well as contributions with
maximal Cabibbo-Kobayashi-Maskawa mixing terms are included here. We
have not applied any detector cuts or hadronisation corrections
here, which would be the next crucial step in phenomenological
studies of the CSTC model. All the numerical estimates here are done
for the first time in order to understand the potential of the
suggested model. Even for a rather large technifermion-technipion
coupling $g_{\rm TC}=8$ we observe that the hadronic cross sections
of the technipion production (middle panel) by about two orders of
magnitude smaller than those for the Higgs boson (right panel) in
the same masse range. This suppression will be even stronger for
smaller $g_{\rm TC}$ coupling and does not depend on other CSTC
model parameters. The respective production mechanism is thus one of
the ``golden'' channels for technipion and, in general, new
strongly-coupled sector searches at the LHC in measurements with
high statistics.

The discovery potential depends also on the subsequent decay modes
and branching ratios of technipions. As was demonstrated above, the
decay modes of the neutral technipion are similar to the
vector-boson decay modes of the Higgs boson including
$\gamma\gamma$, $ZZ$ and $\gamma Z$ channels, however, $\tilde\pi^0
\to W^+W^-$ mode is forbidden by symmetry. In the range of
relatively small $m_{\tilde\pi}\lesssim 200$ GeV the strategy for
searches of technipions will be similar to that in the Higgs boson
searches. Moreover, for light technipions it turns out that the
$\gamma\gamma$ signals from the Higgs boson and technipion can be
comparable with each other due to a very small $\gamma\gamma$
branching ratio of the Higgs boson ${\rm BR}(h\to\gamma\gamma)\simeq
10^{-3}$, while corresponding technipion branching is relatively
large ${\rm BR}(\tilde{\pi}\to\gamma\gamma)\simeq 0.5-1.0$ (see
Fig.~\ref{fig:Pi-BR}). The issue with detection of such light
technipions in the $\gamma\gamma$ or $\gamma Z$ channels can arise,
however, due to a very narrow technipion resonance since in the mass
range $\sim 150$ GeV the total technipion decay width amounts to
$\lesssim0.1$ MeV (see Fig.~\ref{fig:Gamma-Pi}). Such an extremely
narrow resonance, in principle, can be missed in the Higgs-type
searches at the LHC, and an additional investigation of this
possibility is necessary. Also, a possibility of a relative
proximity or even an overlap of the Higgs resonance and extremely
narrow technipion resonance is not completely excluded, and remains
to be an interesting opportunity. Further, a more elaborate analysis
and the search for light technipions in the existing LHC data is
required.

At last, heavier technipions $m_{\tilde \pi}\gtrsim 200$ GeV can be
searched for in the $\gamma\gamma$, $\gamma Z$ and $ZZ$ decay
channels which have comparable branchings. The dominant modes for
the heavy Higgs boson searches are typically $WW$ and $ZZ$ ones with
large branchings, whereas $\gamma\gamma$, $\gamma Z$ branchings of
the Higgs decay are practically zeroth. The only common channel for
technipion and Higgs boson in the high mass range is the $ZZ$ one.
However, having comparable branchings, the technipion production
rate is strongly suppressed compared to that of the Higgs boson (see
above). So, the current LHC statistics may not be enough for
establishing significant constraints onto the CSTC model parameter
space for the higher technipion masses, and further studies are
certainly needed.

\subsubsection{Technipion pair production}

\begin{figure*}[!h]
\begin{minipage}{0.7\textwidth}
 \centerline{\includegraphics[width=1.0\textwidth]{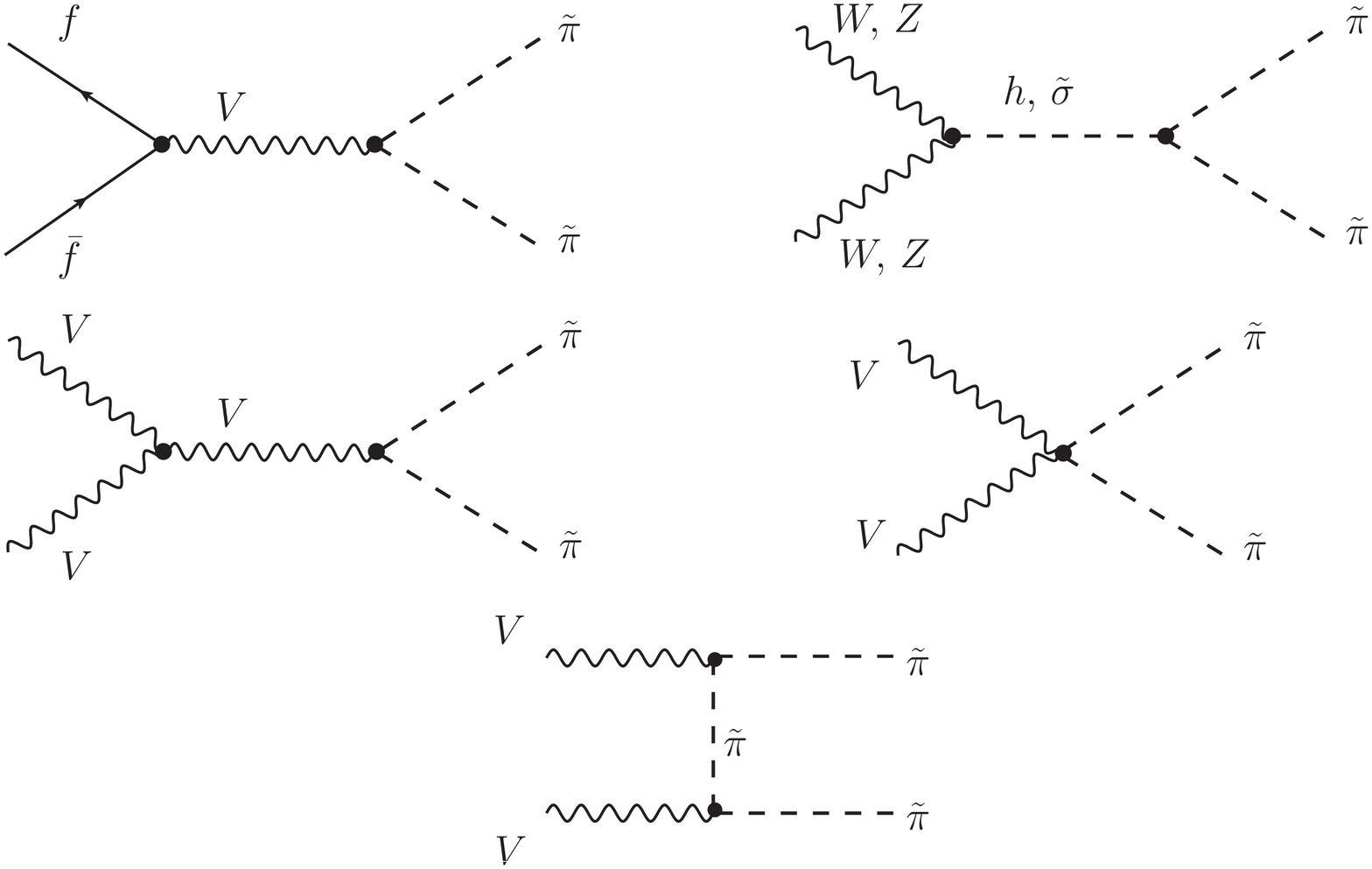}}
\end{minipage}
   \caption{
\small Typical technipion production channels in the leading order,
relevant for collider phenomenology. Here, $V=Z,\,W,\,\gamma$ in
appropriate places. The $ggh$ and $gg\tilde{\sigma}$ couplings are
heavy quark loop-induced ones in the leading order.}
 \label{fig:pi-production}
\end{figure*}

Typical leading-order (tree-level)  processes of the
$\tilde{\pi}$-pair production in $f\bar{f}$ and vector boson fusion
at the LHC are shown in Fig.~\ref{fig:pi-production}. Besides rather
high $\tilde{\pi}\tilde{\pi}$ pair invariant mass
$M_{\tilde{\pi}\tilde{\pi}}\gtrsim 300$ GeV, an additional
suppression in $VV$ and $f\bar{f}$ production channels appear due to
rather weak couplings $g$ and $g^2$ in $\tilde{\pi}\tilde{\pi}V$ and
$\tilde{\pi}\tilde{\pi}VV$ vertices respectively (cf.
Eq.~(\ref{L-piV})), as well as due to a large off-resonant
suppression in $s$-channel subprocesses with intermediate Higgs and
gauge bosons, which are much lightest than
$M_{\tilde{\pi}\tilde{\pi}}$.

Thus, one may naively assume that the largest contribution to the
$\tilde{\pi}^+\tilde{\pi}^-$ and $\tilde{\pi}^0\tilde{\pi}^0$
production rates comes essentially from the intermediate technisigma
resonance with the $\tilde{\pi}\tilde{\pi}\tilde\sigma$ coupling
\begin{eqnarray}
g_{\tilde{\pi}\tilde{\pi}\tilde\sigma} = -g_{\rm
TC}c_{\theta}\frac{M_{\tilde \sigma}^2-m_{\tilde \pi}^2}{2M_{\tilde
Q}}
\end{eqnarray}
which is not suppressed in the small mixing limit (for not very
heavy technifermions). However, in the latter case one encounters
more sources of suppression. Firstly, the production rate of the
$\tilde\sigma$ itself in the SM-like channels most likely to be
suppressed by a small mixing angle, i.e. by the $s_\theta^2\ll 1$
factor in the cross section, compared to the Higgs boson production
rate with $c_\theta^2\sim 1$ (see the previous Section). Secondly,
the $\tilde \sigma$ total decay width dominated by the technipion
channel (in analogy to hadron physics) is typically large, of the
order of a few hundreds of GeV, which means that there will be no
any resonant enhancement in the $\tilde{\pi}\tilde{\pi}$ production
rate associated with the technisigma channel. Thus, overall rates of
the tree-level $\tilde \sigma$ and $\tilde{\pi}\tilde{\pi}$
production are expected to be rather small, similarly to the
loop-induced one technipion rates calculated above. Moreover, in the
small mixing or ``no $h\tilde\sigma$-mixing'' scenario the only
possible $\tilde \sigma$ production channel is through the gauge
boson fusion through the technifermion and technipion triangles
since the $\tilde{Q}\tilde{Q}\tilde{\sigma}$ coupling
(\ref{L-QhSpi}) is finite
\begin{eqnarray}
g_{\tilde{Q}\tilde{Q}\tilde{\sigma}} = - g_{\rm TC}\,c_{\theta}\,,
\end{eqnarray}
and can be rather large due to the ``fat'' TC coupling $g_{\rm
TC}>1$. Besides the dominant technisigma decay mode, the
$\tilde{\pi}\tilde{\pi}$ pair may also be produced at one loop level
via $\tilde{Q}$ box diagrams. These details of the lightest
technihadron dynamics would make the search for new
technipion/technisigma states to be rather challenging at the LHC,
but not impossible.
\begin{figure*}[!h]
\begin{minipage}{0.45\textwidth}
 \centerline{\includegraphics[width=1.0\textwidth]{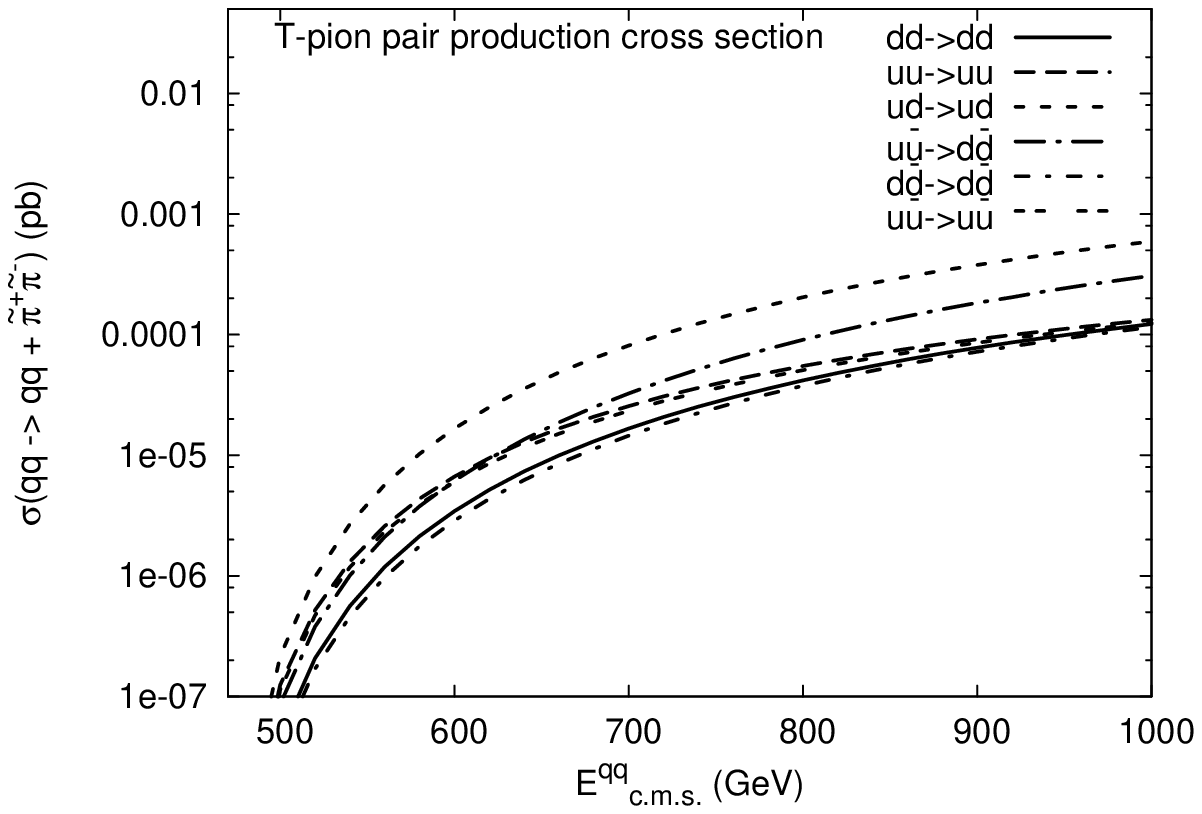}}
\end{minipage}
\begin{minipage}{0.45\textwidth}
 \centerline{\includegraphics[width=1.0\textwidth]{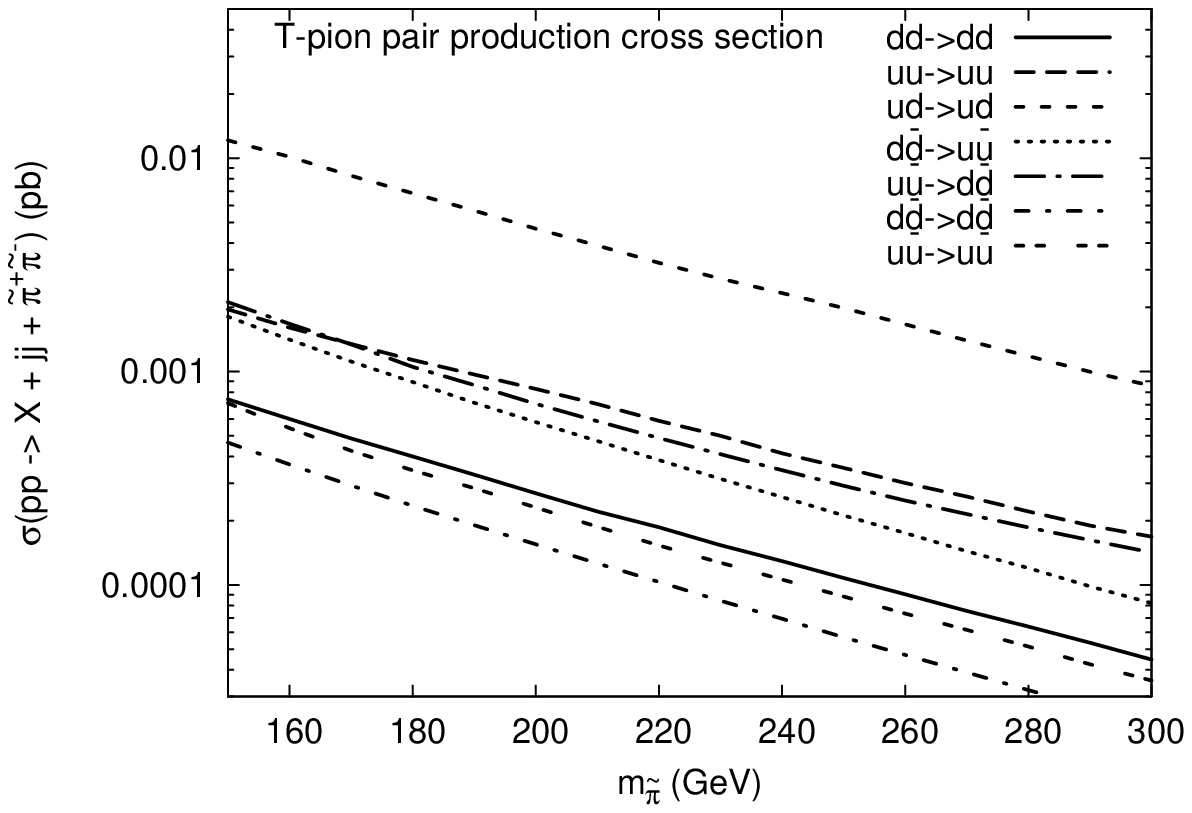}}
\end{minipage}
   \caption{
\small The one-technipion (T-pion) pair production cross sections
via the VBF mechanism at the parton level for different incoming and
outgoing quark $q$ and (anti)quark $q'$ states as functions of $qq'$
invariant mass, or c.m.s. energy $E^{qq}_{c.m.s.}=\sqrt{\hat{s}}$
(left), corresponding total hadron level cross sections of the
technipion pair production for given incoming $qq'$ states in
picobarns (before cuts) at the maximal LHC energy $\sqrt{s}=14$ TeV
as a function of the technipion mass $m_{\tilde \pi}$ (right). Here,
$g_{\rm TC}=8$, $c_\theta^2=0.8$, $M_{\tilde\sigma}=600$ GeV and
$M_{\tilde Q}=300$ GeV are fixed.}
 \label{fig:PiP-PiP-prod}
\end{figure*}

For illustration, in Fig.~\ref{fig:PiP-PiP-prod} we present the
$\tilde{\pi}^+\tilde{\pi}^-$ pair production cross sections at the
parton level in the VBF mechanism as functions of the $qq'$
center-of-mass energy for different initial and final quarks (left)
and the corresponding hadron-level cross sections at the LHC
($\sqrt{s}=14$ TeV) as functions the technipion mass (right). The
quark-antiquark fusion mechanism going via $h$ or $\tilde\sigma$
resonance is assumed to be negligible in the forward/backward jets
kinematics considered here and was not included in this calculation.
In opposite to the one-technipion production cross sections shown in
Fig.~\ref{fig:Pi0-prod}, the parton-level
$\tilde{\pi}^+\tilde{\pi}^-$ pair production cross sections increase
at higher $qq'$ c.m.s. energies (or larger quark fractions $x$) and
can reach the same magnitudes as the one-technipion cross sections
at $E^{qq}_{c.m.s.}\gtrsim 700$ GeV. The hadronic
$\tilde{\pi}^+\tilde{\pi}^-$ cross sections drop faster than
corresponding one-technipion cross sections and have similar
order-of-magnitude values for the light $\tilde\pi$ mass range. This
means that both one- and two-technipion processes should be studied
on the same footing. The latter, however, would be more difficult to
identify experimentally due to a larger multiplicity of leptons and
tiny widths of the technipions.

\section{Summary}

To summarize, in this work we have constructed and investigated in
major details the chiral-symmetric (vector-like) Technicolor
scenario, according to which a new sector of technifermions in
confinement interacts with the SM gauge bosons by means of
vector-like gauge couplings. Our analysis is based upon the gauged
linear $\sigma$-model with initially {\rm global} chiral-gauge
$SU(2)_{\rm L}\otimes SU(2)_{\rm R}$ group broken down to the {\it
local} LR-symmetric SM weak isospin symmetry $SU(2)_{L+R\equiv W}$
group in the technifermion sector.

The Higgs boson in this scenario is considered as a separate
(fundamental or composite) scalar state and introduced in the same
way as in the one-doublet SM. Nevertheless, we have shown that the
electro-weak symmetry breaking at the scale $M_{\rm EW}\sim 100$ GeV
can be initiated {\rm dynamically} by the presence of the confined
vector-like technifermion sector, namely, it is triggered by the
technifermion condensate at the techniconfinement scale,
$\Lambda_{\rm TC}\gtrsim M_{\rm EW}$, together with the chiral
symmetry breaking. This thus leads to the effective SM Higgs
mechanism of dynamical electro-weak symmetry breaking.

Remarkably, this model is well consistent with both EW precision
constraints and, simultaneously, with the recent SM-like Higgs boson
observations at the LHC in the small Higgs-technisigma mixing limit.
At the same time, the model predicts the existence of extra new
lightest technihadron states, namely, physical technipions $\tilde
\pi$ and technisigma $\tilde \sigma$, at the LHC energy scales,
giving rise to rich Technicolor phenomenology at the LHC. Detection
prospects for these new states have also been discussed, and the
most phenomenologically important decay modes of $\tilde \pi$ and
$\tilde \sigma$, as well as technipion production cross sections,
were quantified over physically reasonable regions of parameter
space.

In the absence of noticeable deviations from the SM predictions in
the Higgs signal strengths, the suggested scenario is capable of
explaining of what triggers the SM Higgs mechanism, the nature of
the Higgs vev in the nearly-conformal limit of the new
strongly-coupled dynamics. The proposed vector-like Technicolor
scenario, in its simplest form considered here, does not attempt to
resolve the naturalness problem of the SM, i.e. does not provide a
mechanism protecting the Higgs boson mass itself from becoming
arbitrary large. Nevertheless, this minimal realization of the TC
ideas preserving the effective Higgs mechanism of the SM opens up
new prospects for more elaborated scenarios with extended
chiral-gauge groups possibly predicting the light composite Higgs
boson(s) with well-defined vector-like ultraviolet completion, which
is the subject of our further analysis. At last, as a specific
prediction of this class of models, the lightest neutral heavy
weakly-interacting technibaryon state gives rise to a suitable Dark
Matter candidate making it to be especially attractive opportunity
for astrophysical New Physics searches, and a corresponding analysis
is planned for future studies.

\vspace{1cm}

{\bf Acknowledgments}

Stimulating discussions and helpful correspondence with Johan
Bijnens, Gabriele Ferretti, Stefano Frixione, Christophe Grojean,
Giuliano Panico, Sabir Ramazanov, Johan Rathsman, Slava Rychkov,
Francesco Sannino, Torbj\"orn Sj\"ostrand and Peter Skands are
gratefully acknowledged. This work was supported in part by the
Crafoord Foundation (Grant No. 20120520). R.P. is thankful to the
CERN Theory Group for support and inspiring discussions during his
visit at CERN. V.K. is especially grateful to the Lund THEP Group
for support and hospitality during his visit at Lund University at
the final stage of this work.

\end{document}